\documentclass[traditabstract]{aa}
\usepackage{graphicx}
\usepackage{psfrag}
\usepackage{natbib}
\usepackage[varg]{txfonts}
\usepackage{longtable}

\begin{document}
\def\lsun{L_{\sun}}
\def\msun{M_{\sun}}
\def\dendro{{\tt dendrogram}}
\title{APEX/SABOCA
observations of small-scale structure of infrared-dark clouds }
\subtitle{I. Early evolutionary stages of star-forming cores\thanks{Based on observations carried out with the Atacama Pathfinder Experiment (APEX). APEX is a collaboration between Max Planck Institut f\"ur Radioastronomie (MPIfR), Onsala Space Observatory (OSO), and the European Southern Observatory (ESO).} }

\author{
Sarah E. Ragan,
Thomas Henning, 
Henrik Beuther
}
 
\authorrunning{Ragan et al.} 
\titlerunning{IRDC cores with SABOCA}

\institute{
Max-Planck-Institute for Astronomy, K\"onigstuhl 17, 69117 Heidelberg, Germany \\
\email{ragan@mpia.de} \label{mpia}}

\date{Received 9 May 2013 ; accepted 28 August 2013}

\abstract
{Infrared-dark clouds (IRDCs) harbor the early phases of cluster and high-mass star formation and are comprised of cold ($\sim$20\,K), dense ($n > 10^4$\,cm$^{-3}$) gas. The spectral energy distribution (SED) of IRDCs is dominated by the far-infrared and millimeter wavelength regime, and our initial Herschel study examined IRDCs at the peak of the SED with high angular resolution. Here we present a follow-up study using the SABOCA instrument on APEX which delivers 7.8$''$ angular resolution at 350\,$\mu$m, matching the resolution we achieved with {\em Herschel}/PACS, and allowing us to characterize substructure on $\sim$0.1\,pc scales. Our sample of 11 nearby IRDCs are a mix of filamentary and clumpy morphologies, and the filamentary clouds show significant hierarchical structure, while the clumpy IRDCs exhibit little hierarchical structure.  All IRDCs, regardless of morphology, have about 14\% of their total mass in small scale core-like structures which roughly follow a trend of constant volume density over all size scales.  Out of the 89 protostellar cores we identified in this sample with {\em Herschel}, we recover 40 of the brightest and re-fit their SEDs and find their properties agree fairly well with our previous estimates ($<T> \sim 19$\,K). We detect a new population of ``cold cores'' which have no 70\,$\mu$m counterpart, but are 100 and 160\,$\mu$m-bright, with colder temperatures ($<T> \sim 16$\,K). This latter population, along with SABOCA-only detections, are predominantly low-mass objects, but their evolutionary diagnostics are consistent with the earliest starless or prestellar phase of cores in IRDCs. }

\keywords{catalogs -- stars: formation -- ISM: structure -- submillimeter: ISM}

\maketitle

\section{Background and Motivation}
\label{sec:bg}

Despite the importance of high-mass stars to the energy budget of galaxies, their formation remains a major open question in astronomy \citep{ZinneckerYorke2007}.  A major hindrance to progress is the inherent difficulty in obtaining well-defined initial conditions observationally. Because high-mass stars are rare, they are on average at large distances, thus angular resolution is of paramount importance.  With the advent of {\em Spitzer Space Telescope} and {\em Herschel Space Observatory}  \citep{A&ASpecialIssue-Herschel}, coupled with extensive ground-based survey efforts, we now can approach this observational task statistically.

Ever since their discovery in absorption in mid-infrared Galactic plane surveys, ISO data \citep{Perault1996} and MSX observations \citep{egan_msx}, infrared-dark clouds (IRDCs) have been subject to intense study because they are the cold (T $< 20$\,K), dense ($n > 10^{4}$\,cm$^{-3}$) environments believed to be required for the formation of high-mass stars and clusters \citep[cf.][]{Rathborne2006, Ragan_spitzer, Battersby2010}.  They are located throughout the Galactic plane, concentrated within the spiral arms \citep{Jackson_galdistr_IRDCs}. The formation of IRDCs, their kinematic structure and population of self-gravitating cores, and their final dissipation are all important ingredients on our way to understanding the nature of high-mass star formation.

In {\em Spitzer} continuum bands, IRDCs appear ``dark,'' but only {\em Herschel} observations showcase the transition from dust absorbing the galactic background (shortward of $\sim$100\,$\mu$m) into optically thin emission of cold dust structures (longward of $\sim$100\,$\mu$m).  On small scales, {\em Herschel} has transformed our view of IRDCs by probing their deeply embedded core population, a fact that has been extensively exploited in recent literature \citep[][]{A&ASpecialIssue-Henning,A&ASpecialIssue-Beuther,A&ASpecialIssue-Linz, Bontemps2010_aquila, DWT2010_Polaris, Koenyves2010, Hennemann2010_rosette, Motte2010_hobys, Battersby2011, Beuther_18454, Giannini2012, Ragan2012b, Rygl2013_lupus, Gaczkowski2013, Pitann2013,Stutz2013}.  In these works, the spectral energy distribution (SED) of compact or point-like sources have been analysed, thus giving estimates of the mass, luminosity, and average temperature of the objects.

While {\em Herschel} provides a new look at the star formation in IRDCs, the properties of the cores are not strongly-constrained by {\em Herschel} observations alone. For example, in \citet[][hereafter R12]{Ragan2012b}, we find the presence of a 24\,$\mu$m MIPS and PACS 70\,$\mu$m counterpart to be an important evolutionary indicator. On the long-wavelength ($\lambda >$ 160\,$\mu$m) side of the SED, however, there are very few constraints at matching angular resolution. In order to address this issue, we obtained maps of eleven IRDCs selected from the Earliest Phases of Star Formation (EPoS) {\em Herschel} guaranteed time key program sample of IRDCs (R12) with APEX telescope using SABOCA, a bolometer operating at 850\,GHz, or 350\,$\mu$m. The angular resolution, 7.8$''$ is well-matched to the {\em Herschel}/PACS resolution, so the physical scales can be directly compared. The goal of this study is to examine the cloud structure seen with SABOCA first independently then in concert with the {\em Herschel} dataset. Do we recover all {\em Herschel} cores?  How reliable are the properties derived from our original ``PACS-only'' SEDs? What is the nature of recovered PACS sources and new sources found only with SABOCA?

To this end, we first employ the hierarchical structure identification algorithm, {\tt dendrograms} to quantify the emission structures. Then we correlate the structures with those known from {\em Herschel} observations. We model the SEDs to derive or place (upper limits on) the mass, luminosity, and temperature of each structure to place them in evolutionary context. These observations prove to be critical in characterizing the early pre- and protostellar phases of star formation in IRDCs.

\begin{figure}
\begin{center}
\includegraphics[scale=0.6]{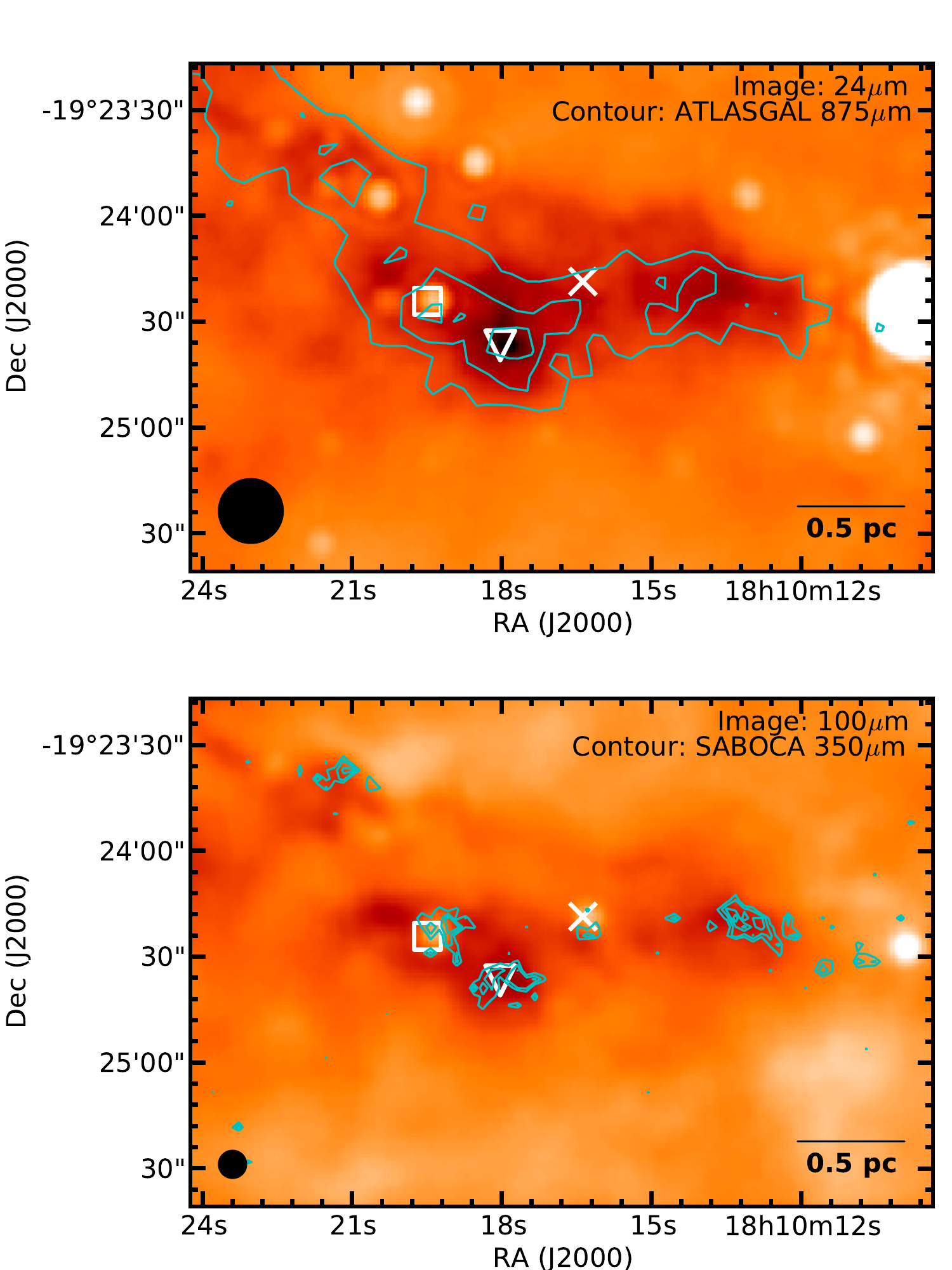} 
\end{center}
\caption{Top: 24\,$\mu$m image of region in IRDC011.11-0.12 with ATLASGAL 875\,$\mu$m contours. The 19.2$''$ ATLASGAL beam at this wavelength is shown in the lower-left corner. Bottom: PACS 100\,$\mu$m image of the same region over-plotted with SABOCA 350\,$\mu$m contours. The 7.8$''$ SABOCA  beam is shown at the lower-left corner.  The white square marks the position of a 24\,$\mu$m-bright protostar; the white $\times$ marks the position of a 24\,$\mu$m-dark protostar; the white triangle marks a IR-dark core. \label{f:g11tiny}}
\end{figure}

\section{Observations and data reduction}
\label{sec:obs}

\begin{table*}
\begin{center}
\caption{Target and observation summary \label{tab:obstable}}
\begin{tabular}{llclcclccc}
\hline \hline
IRDC & RA (J2000)    & Dec (J2000)    & Distance &  $M_{atlasgal}^b$ & map size   & observation  & $t_{int}$  & rms & $N$(H$_2$)$^{sens}$  \\
name$^a$ & ($^{h}$:$^{m}$:$^{s}$) & ($^{\circ}$:$^{'}$:$^{''}$) & (kpc) & ($\msun$) & ($' \times '$)  & date & (sec) &  (Jy beam$^{-1}$)  & (cm$^{-2}$)   \\
\hline
IRDC\,310.39-0.30 & 13:56:04.9 & -62:13:42 & 5.0$\pm$1.2 & 1398 & 6 $\times$ 6  & 2011-04-03  & 7560 & 0.22 & 8.44E+21 \\
IRDC\,316.72+0.07* & 14:44:19.2 & -59:44:29 & 2.8$\pm^{0.4}_{0.5}$ & 3165 & 6 $\times$ 6  & 2011-04-14  & 7560  & 0.26 & 9.97E+21 \\
IRDC\,320.27+0.29 & 15:07:45.0 & -57:54:16 & 2.3$\pm$0.4 & 156 & 6 $\times$ 6  & 2011-04-15  & 7500  &  0.26 & 9.97E+21 \\
IRDC\,321.73+0.05 & 15:18:13.1 & -57:21:52 & 2.3$\pm$0.4 & 564& 7 $\times$ 7  & 2011-08-16 & 7980  &  0.23 & 8.82E+21 \\
IRDC\,004.36-0.06 & 17:55:45.6 & -25:13:51 & 3.3$\pm^{1.0}_{1.6}$ & 327 & 6 $\times$ 5  & 2011-08-22 & 7860 &  0.19 & 7.29E+21 \\
IRDC\,009.86-0.04 & 18:07:37.4 & -20:26:20 & 2.7$\pm^{0.6}_{0.8}$ & 143 & 6 $\times$ 6  & 2011-04-14  & 7440  & 0.19 & 7.29E+21 \\
IRDC\,011.11-0.12* & 18:10:27.7 & -19:20:59 & 3.4$\pm$0.5 & 5045 & 9 $\times$ 23 & 2010-05,11 & 21000 & 0.27 & 1.04E+22\\
IRDC\,015.05+0.09 & 18:17:40.3 & -15:49:10 & 3.0$\pm^{0.4}_{0.5}$ & 160 & 6 $\times$ 6  & 2011-08-16 & 3840 & 0.43 & 1.65E+22\\
IRDC\,18223*  & 18:25:11.5 & -12:49:45 & 3.5$\pm^{0.3}_{0.4}$ & 3501 & 4 $\times$ 16 & 2011-04-19  & 12060 & 0.21 & 8.06E+21\\
IRDC\,019.30+0.07 & 18:25:53.8 & -12:06:18 & 2.4$\pm^{0.4}_{0.5}$ & 624 & 7 $\times$ 7  & 2011-04-19 & 4500 & 0.42 & 1.61E+22\\
IRDC\,028.34+0.06* & 18:42:50.3 & -04:02:17 & 4.5$\pm$0.3 & 15011 & 7 $\times$ 8  & 2011-08-16  & 3780 & 0.40 & 1.54E+22 \\
\hline
\end{tabular}
\end{center}

\tablefoottext{a}{Filamentary IRDCs denoted with asterisk.}

\tablefoottext{b}{Mass above $N \sim$10$^{21}$\,cm$^{-2}$ from ATLASGAL survey \citep[see ][]{Ragan2012b}.}

\end{table*}

\subsection{Sample Selection}
\label{ssec:sample}

As part of the {\em Herschel} Earliest Phases of Star formation (EPoS) guaranteed time key program, we  investigated the protostellar core population in 45 massive regions (R12) From this sample, we selected eleven IRDCs which exhibit a range in morphology and protostellar populations. The clouds, their positions and kinematic distances are listed in Table~\ref{tab:obstable}.  The sample contains a mix of filamentary and clumpy clouds, where ``filamentary'' is defined as a cloud with dense ($N>10^{21}\,\mathrm{cm}^{-2} $) material elongated preferentially along one axis by at least an aspect ratio of 3:1. Clouds that qualify as ``filaments'' are IRDC\,011.11-0.12, IRDC\,028.34+0.06, IRDC\,316.72+0.07, and IRDC\,18223. These are also the most massive clouds (see Table~\ref{tab:obstable}), but they have similar distances to the seven clumpy clouds in the sample.

We have selected well-studied clouds, which has the added advantage that existing molecular line surveys \citep[e.g.][Tackenberg et al. (submitted)]{Ragan_msxsurv, Vasyunina2011} have already verified that the IRDCs contained in our selected fields are coherent in velocity space. Thus, we are confident that the clouds are each genuine cold, absorbing entities at a common distance (all on the near side of the Galaxy) and not chance-aligned clouds.

\subsection{APEX observations and data reduction}
\label{ssec:apex}

We surveyed our sample at the Atacama Pathfinder Experiment (APEX) 12-m telescope using the Submillimetre APEX Bolometer Camera \citep[SABOCA,][]{Siringo_saboca}, a 37-element array operating at 350\,$\mu$m with 7.8$''$ angular resolution. Our pilot study to map IRDC011.11-0.12 was completed in 2010 (project M085-0029), and the remaining ten sources were observed in April and August of 2011 (project M087-0021). Throughout the observations, the precipitable water vapor (PWV) was required to be below 0.5\,mm and ranged from 0.2\,mm to 0.5\,mm throughout the observations. The atmospheric opacity at zenith was calculated using the {\tt skydip} procedure and was found to be less than 1. 

The data were reduced using the BOA software \citep{Schuller_boa} using the standard iterative source-masking procedure. Each map is comprised of a series of spiral scans. In the first iteration, the scans are calibrated based on the estimated opacity, noisy channels are then flagged, and correlated noise is removed. This produces the initial map.  In subsequent iterations, we use a smoothed version of the initial map that serves as a model for the actual emitting structure. This model is subtracted from the initial map in order to model (``flat-field'') the residual flux. The model is then added back to the image. The process is repeated on the new image until the flux levels plateau and the artefacts (e.g. ``bowls'' surrounding strong emssion regions) are minimized. 

A summary of the observations is given in Table~\ref{tab:obstable}, which notes the size of our bolometer maps, the total integration time, and the rms that we achieved in each map. We compute the rms from $1' \times 1'$ sized emission-free regions in each map.  An example map of a small region in IRDC\,011.11-0.12 is shown in Figure~\ref{f:g11tiny}, and the full set of maps are presented in Appendix \ref{s:imgal}.  

\section{Structure Identification}
\label{sec:dendro}

\subsection{Method}
\label{sec:method}

The defining hierarchical nature of molecular clouds has given rise to many characterization techniques which attempt to assign gas, traced either by molecular line emission or dust, to structure on various scales.  For consistency, we adopt the size-based terminology given to these scales in \cite{BerginTafalla_ARAA2007}: a ``core'' ranges in size between 0.03 and 0.2\,pc and a ``clump'' from 0.3 to 3\,pc, and both entities are found within the boundaries of several parsec sized ``clouds''.  The SABOCA angular resolution accesses the ``core'' scale in our sample of IRDCs.

One method that quantitatively describes this hierarchy is a structure tree algorithm, popularly implemented in astronomy with the \dendro~representation \citep{dendrograms}. An advantage to using this technique is that it imposes no assumptions about the shape or emission profile of the tree structures, and it operates on both two and three dimensional datasets.  We use the {\tt python} implementation, {\tt astrodendro}\footnote{http://github.com/bradenmacdonald/astrodendro}.  Our motivation for using such an algorithm is to quantify the relative flux contributions at 350\,$\mu$m from the core and from the cloud. Our methodology is described below.

A \dendro~tree has three main features: a trunk, branches, and leaves, which amount to surfaces within increasing contour levels in our two-dimensional dataset.  We define our lowest contour level, or the intensity associated with the trunk of the \dendro~ tree, to be 3-$\sigma$ (roughly corresponding to a mean minimum detectable column density of 10$^{22}$\,cm$^{-2}$, see Table~\ref{tab:obstable} and Section~\ref{ssec:structure}), and we require 1-$\sigma$ steps and eight contiguous pixels (selected to match the number of pixels per APEX beam) to define further branches up the intensity tree.  The peaks comprise the so-called ``leaf'' population.  We show an example \dendro~tree in Figure~\ref{f:tree_diagram} for IRDC\,015.05+0.09.  In this example, there are four leaves with no parent structure (3, 6, 7, and 8) and four leaves which are nested in a parent structure. The minimum contour level is shown with the dashed line.  

\begin{figure}
\includegraphics[width=\linewidth]{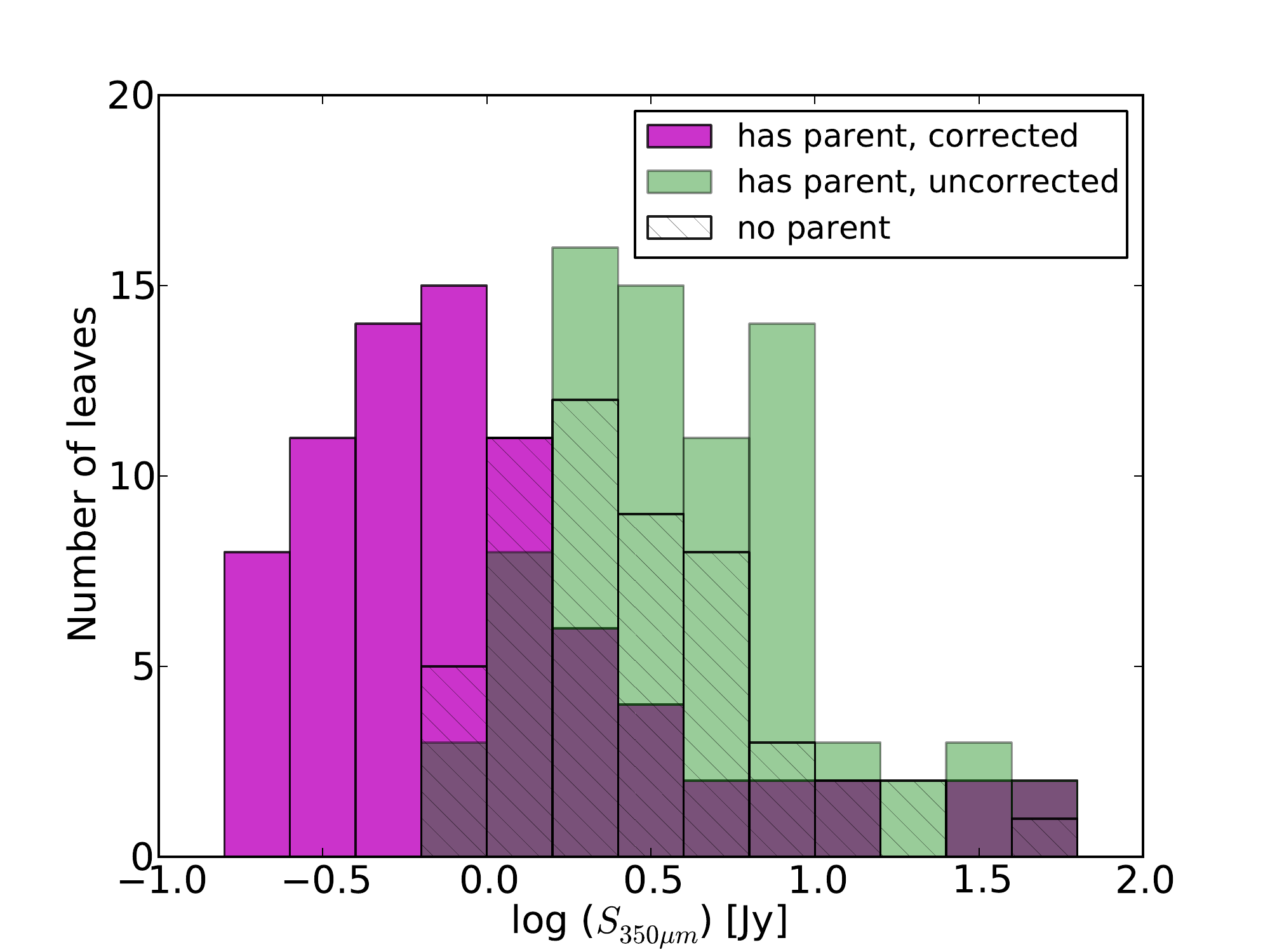}
\caption{\label{f:forphans} Distribution of $S_{350{\mu}m}$ for corrected (magenta) and uncorrected (green) leaves, compared with leaves with no parent structure (black hatched).}
\end{figure}

\begin{figure}
\includegraphics[width=\linewidth]{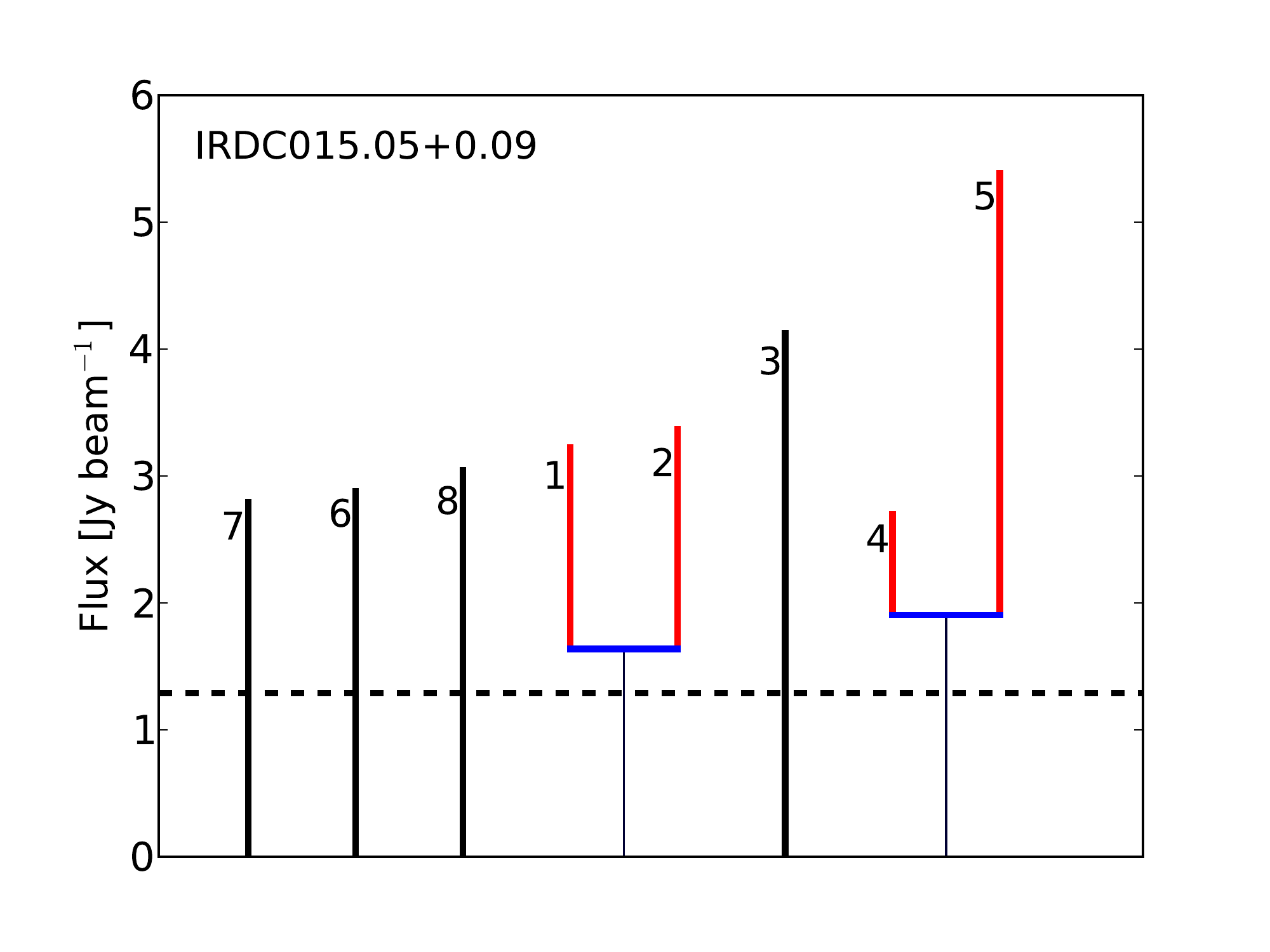}
\caption{\label{f:tree_diagram} Example of \dendro~tree structure for IRDC\,015.05+0.09. The indices are those listed in Table~\ref{tab:newcores} in order of increasing right ascension. The black leaves (3, 6, 7, and 8) have no parent tree structure, and the red leaves are nested in separate trunks. Their flux values were corrected for the local merge level, which is marked in blue. The horizontal black dashed line shows the 3-$\sigma$ minimum flux requirement for this cloud. }
\end{figure}

\begin{figure}
\includegraphics[width=\linewidth]{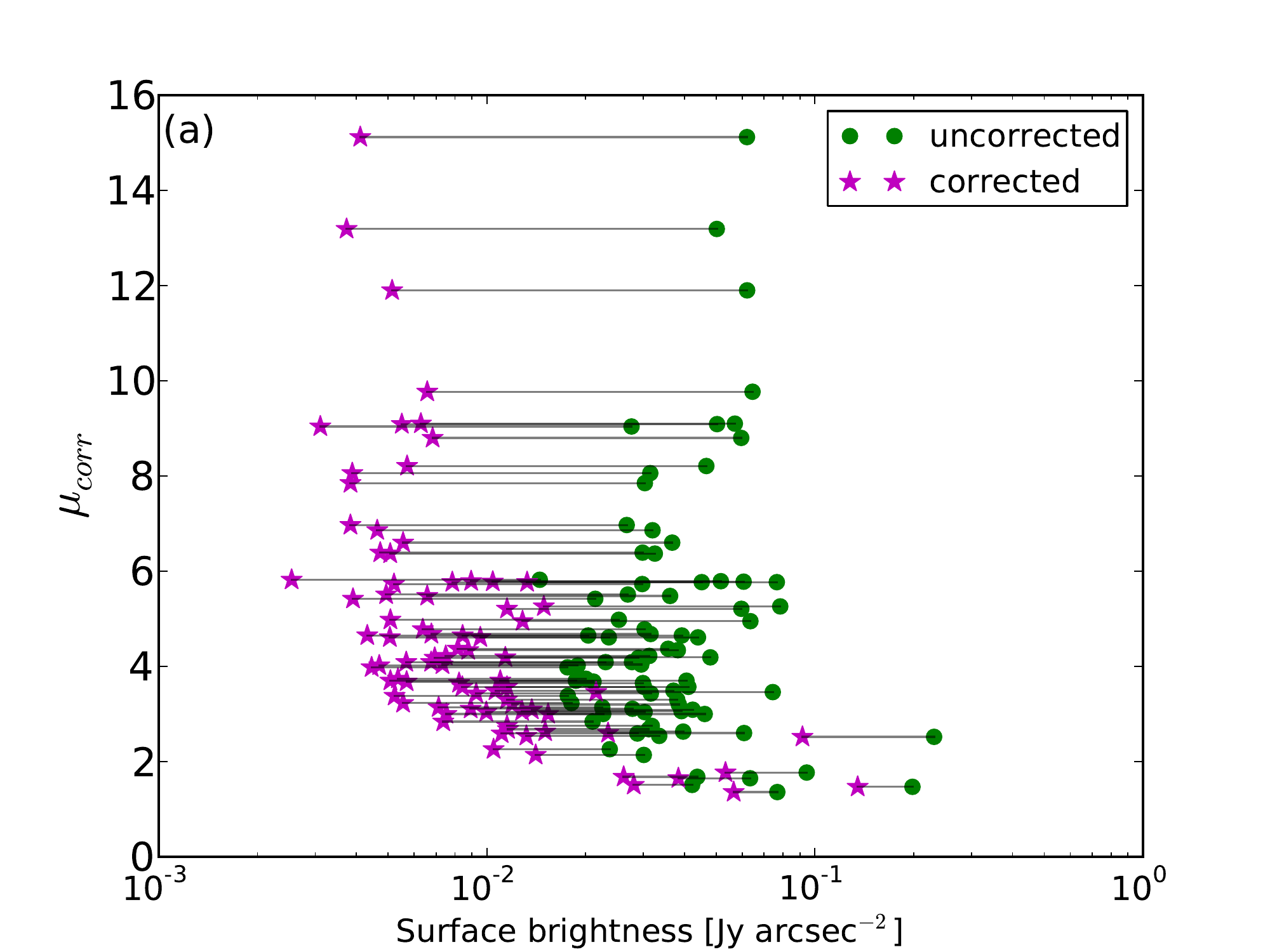}
\includegraphics[width=\linewidth]{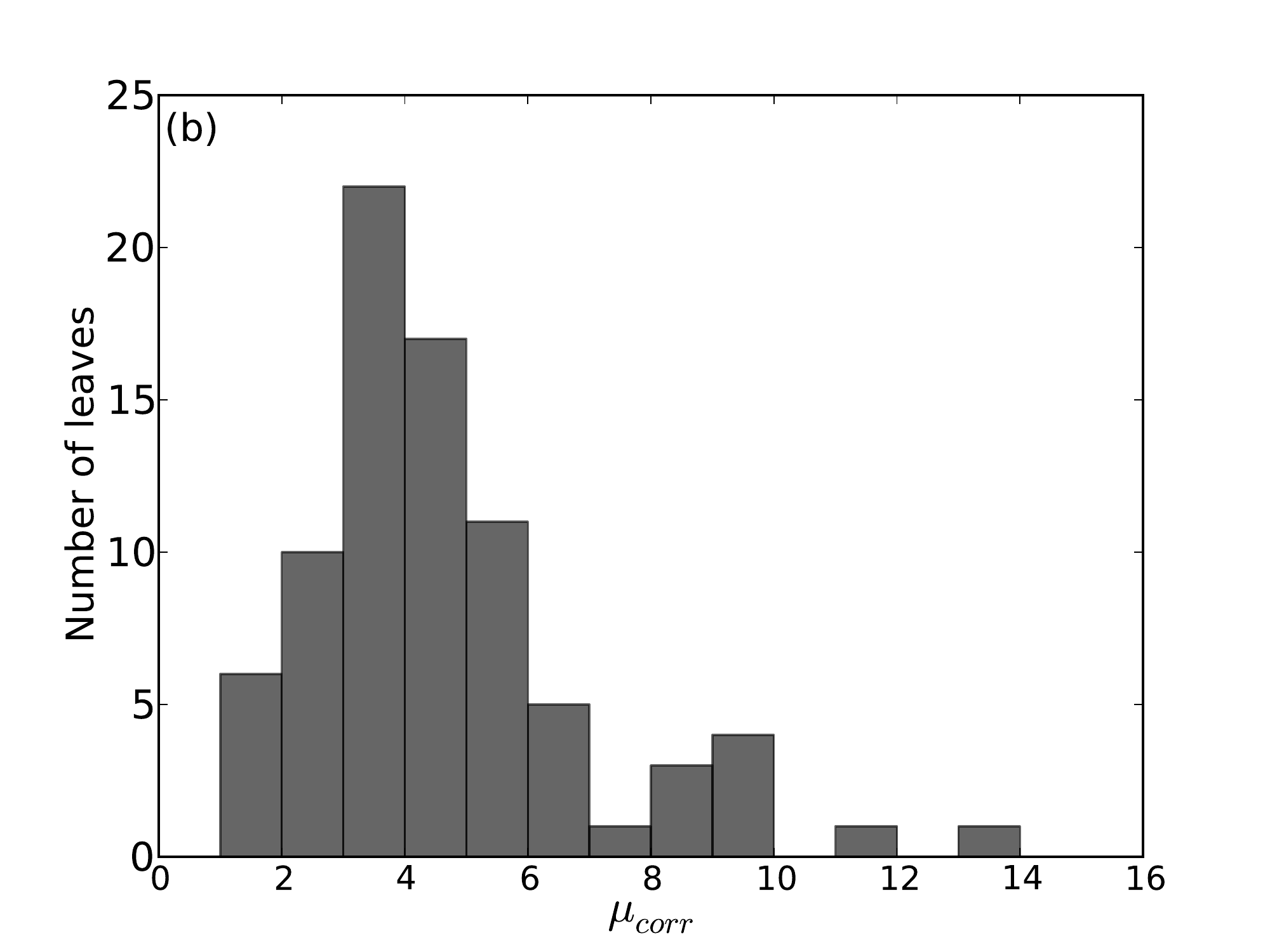}
\caption{\label{f:flux_corr} (a) Flux correction factor, $\mu_{corr}$ (see Section~\ref{sec:method}), as a function of the leaf surface brightness in Jansky per square arcsecond. The green dots are the uncorrected measurements, and the magenta stars are the corrected surface brightness. (b) Histogram showing $\mu_{corr}$ for the leaves with parent flux contribution. The mean value of $\mu_{corr}$ is 4.8.}
\end{figure}

Table~\ref{tab:newcores} lists the positions of all leaves and the total flux of each leaf, 135 in total in the 11 IRDCs.  If the leaf stems from a trunk or branch (i.e. a parent structure, the average flux of which is known within the algorithm as the ``merge level''), we correct that flux by subtracting the merge level flux from each leaf pixel. In Figure~\ref{f:forphans} we show the flux distribution for leaves with no parent structures and both the corrected and uncorrected flux distributions for leaves stemming from parent structures\footnote{We note that this method of separating leaf flux from branch flux will give lower flux values than would have otherwise been estimated. Throughout the following calculations, we adopt the corrected value but note the upper limit had the parent flux been included.}. We define the correction factor, $\mu_{corr}$, as the ratio of uncorrected to corrected total leaf flux. We plot $\mu_{corr}$ as a function of leaf surface brightness in Figure~\ref{f:flux_corr}a. The mean value for $\mu_{corr}$ in child leaves is 4.8, but ranges between 1.4 and 15 (see Figure~\ref{f:flux_corr}b). We find no trend in the needed correction as a function of surface brightness, and the implications of this will be further discussed in Section~\ref{ss:nature}.

This method is designed to isolate the flux of the core structures by removing the contribution of the less dense surrounding cloud. The SABOCA observations do filter out some large-scale structure, but here we take advantage of the superb angular resolution of these data and assume that at the locations of the emission peaks, the small scale core is the dominant source. We will address the hierarchical cloud structure in a forthcoming publication. In this paper, we concentrate on the core population.

\subsection{Core recovery rates}
\label{sec:corerecovery}

We cross-matched the positions of the SABOCA leaves with the {\em Herschel}/PACS point source catalog of R12 using the {\tt TOPCAT} software package requiring a match within a 10$\arcsec$ radius.  We note which of the SABOCA leaves have a {\em Herschel}/PACS counterpart in Table~\ref{tab:newcores}.  For the remainder of the paper, we distinguish the types of objects based on in which wavelengths they are detected.  All SABOCA detections are referred to as ``leaves'' regardless of whether they have {\em Herschel} counterparts.  We refer to {\em Herschel} sources as ``cores'' and distinguish the sub-sample that we detect with SABOCA as ``recovered cores.'' Any core appearing in the R12 catalog (with counterparts at 70, 100, and 160\,$\mu$m) is a ``PACS core'' and may or may not have a 24\,$\mu$m counterpart. A ``cold core'' is detected only in the 100 and 160\,$\mu$m bands of PACS (not 24 or 70\,$\mu$m), but is also a recovered core with SABOCA\footnote{We again emphasize that the use of the term ``core'' reflects the compact size of the structures (see Section~\ref{sec:method}), but given our present resolution limitations and the range of masses, the ``cores'' could be the precursors to individual stars, binaries, or bound multiples.}.

The statistics of the PACS core recovery rate per cloud are given in Table~\ref{tab:corepop}.  For these 11 IRDCs, R12 catalogued a population of 89 PACS cores and plus another 76 candidate cold cores. This latter population could not be modeled using {\em Herschel} data alone due to poor spectral coverage at the relevant size scale and were therefore not cataloged at that time.  We recover 52 (58\%) PACS cores in SABOCA emission, 40 (45\%) of which are in the form of leaf structures, otherwise they coincide with branch level emission. We recover 14 (18\%) candidate cold cores, 12 (16\%) of which are leaves.  The figures in Appendix~\ref{s:imgal} show filled circles for recovered cores (i.e. {\em Herschel} point sources with a SABOCA leaf counterpart) and empty circles for unrecovered cores. 

In three cases, a leaf overlaps two or more {\em Herschel} cores.  The affected objects are noted in Table~\ref{tab:newcores}. Since the leaf flux has already been corrected for parent contribution, we assume that similar to PACS wavelengths, the flux of each core scales with the (predicted) luminosity from previous fits (see Figure 8 in R12).  Therefore, only for the purposes of SED-fitting, we use the luminosity ratio of the cores that fall in the leaf area to divide the SABOCA flux between these core contributions.  The factors by which the fluxes are scaled are detailed in Section~\ref{ss:individual}. We then consider the scaled 350\,$\mu$m flux with the closest-matching {\em Herschel} core in the SED-fitting analysis that follows. 

\begin{figure}
\includegraphics[width=\linewidth]{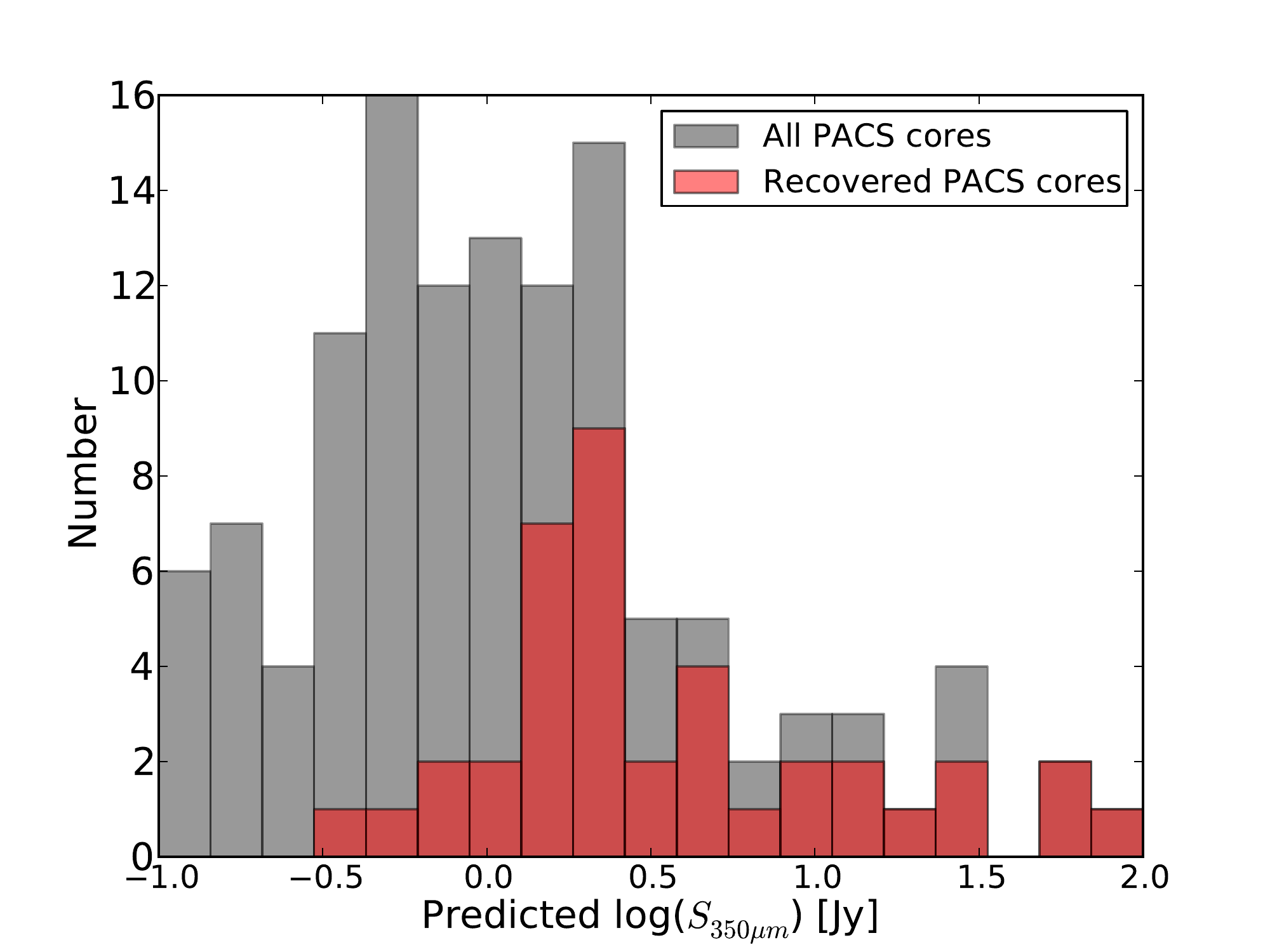}
\caption{\label{f:predicted350} The distribution of flux density at 350\,$\mu$m from all PACS cores in R12 for the IRDCs in this sample (gray). The red histogram shows the portion of these cores which were recovered by our SABOCA observations.}
\end{figure}

The recovery rate of PACS cores is mainly governed by the sensitivity of our SABOCA observations. In Table~\ref{tab:obstable}, we list the rms levels of each map, which range from 0.19 to 0.43 Jy beam$^{-1}$.  Based on the original SED fits in R12, the median predicted flux at 350\,$\mu$m is 0.58 Jy (in contrast to 2.4\,Jy median predicted flux for recovered cores), which does not meet the 3-$\sigma$ detection requirement.  Figure~\ref{f:predicted350} shows the distributions of predicted 350\,$\mu$m flux based on the best-fit SED to the PACS data from R12, showing that we clearly recover the brightest cores.  In other cases, IRDC\,028.34 and IRDC\,316.72 especially, source confusion plays a role in some bright PACS cores not being recovered. Clearly, our recovery rate is dominated by our SABOCA sensitivity.

The recovery rate for cold cores is subject to the same limitations as above, with the added difficulty that the original catalog of 73 cold cores was based on the criterion of only two detections at 100 and 160\,$\mu$m, rather than the three required for PACS cores. The 100 and 160\,$\mu$m bands had the highest uncertainties and greater large scale variation, thus increasing the propensity for contamination of the candidate catalog. Therefore their poor recovery rate is unsurprising.

\begin{figure}
\includegraphics[width=\linewidth]{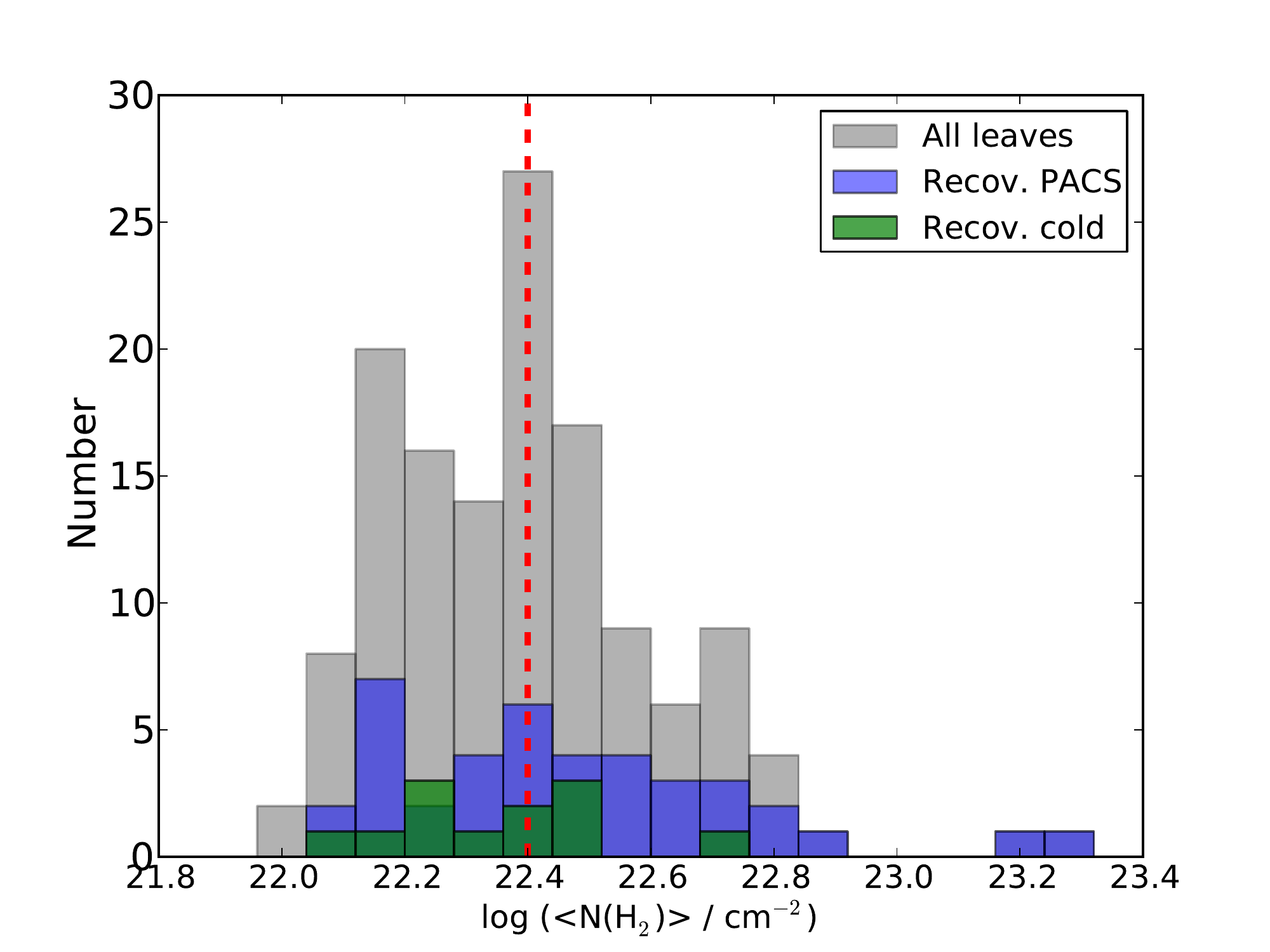}
\caption{\label{f:nhhist} Distribution of mean column density of all leaf structures. Recovered PACS cores are shown in blue, and recovered cold cores are shown in green. The median column density of 2.4 $\times$ 10$^{22}$ cm$^{-2}$ is shown in the vertical dashed line.}
\end{figure}

\section{Results}
\label{sec:results}

\subsection{IRDC small-scale structure}
\label{ssec:structure}

For each leaf extracted from the SABOCA map, we calculate the mass via the standard formulation, 

\begin{equation}
M_\mathrm{350{\mu}m} =\frac{ S_{350{\mu}m}~d^2~R_\mathrm{gd}  }{ \kappa_{350{\mu}m}~B_{\nu}(T_\mathrm{d})}.
\label{eq:m350}
\end{equation}

\noindent where $S_{350{\mu}m}$ is the specific flux, $d$ is the distance, $R_\mathrm{gd}$ is the gas-to-dust ratio assumed to be 100, and $B_{\nu}$ is the Planck function at dust temperature, $T_\mathrm{d}$ that we first assume here to be uniformly 20\,K, and the column density

\begin{equation}
N_{H_2} = S_{350{\mu}m}~R_\mathrm{gd}~[ \Omega_\mathrm{beam}~\mu~m_\mathrm{H}~\kappa_{350{\mu}m}~B_{\nu}(T_\mathrm{d}) ]^{-1}
\end{equation}

\noindent where $\Omega_\mathrm{beam}$ is the size of the beam, $\mu$ is the mean molecular weight (2.36), and m$_\mathrm{H}$ is the mass of hydrogen. We assume the dust mass opacity at 350~$\mu$m, $\kappa_{350{\mu}m}$, to be 10.1\,cm$^2$ g$^{-1}$, which is taken from \citet{ossenkopf_henning} for dust grains with thin ice mantles at $n = 10^6$ cm$^{-3}$ (column 5, hereafter OH5), which is supported for similar environments in the recent literature \citep[e.g.][]{Shirley2011b, Kainulainen2013}. Both the mass and mean column density for each leaf are reported in Table~\ref{tab:newcores}.  Given the flux limits discussed in Section~\ref{sec:corerecovery}, our mass sensitivity is roughly 1-2\,$\msun$ depending on the leaf's environment. Our typical column density sensitivity is $\sim$10$^{22}$\,cm$^{-2}$.

We show the distribution of mean leaf column densities in Figure~\ref{f:nhhist}, which are left uncorrected for parent flux. The median value is 2.4 $\times$ 10$^{22}$\,cm$^{-2}$.  leaves of greater column density are more likely to be seen also as a {\em Herschel} core. While most of the cold cores are below the ensemble median column density, 24 of the 40 PACS cores are above the median value.  The column density values and the tendency for the higher column density structures to have star formation (appearing as a PACS core) are consistent with what is observed in local clouds \citep{Enoch2008}.

For more accurate mass estimates, we evaluate Equation~\ref{eq:m350} at the best-fit or upper-limit temperature, for recovered cores and leaves, respectively. The best-fit temperature is given in Table~\ref{tab:coreprops}, and the upper-limit temperature for leaves with no {\em Herschel} counterpart is calculated by fitting an SED to the 3-$\sigma$ detection limit of PACS fluxes in all bands plus the corrected 350\,$\mu$m flux and is used to determine the leaf mass, giving a lower limit.  With only a few exceptions, the high-mass leaves ($M_{350{\mu}m} > 40\,\msun$) correspond to PACS cores. The mean and median masses for PACS cores are 239 and 16\,$\msun$, respectively, while for the remaining leaves, we find 28 and 8\,$\msun$. 

\begin{figure*}
\begin{center}
\centerline{
\includegraphics[width=3.8in]{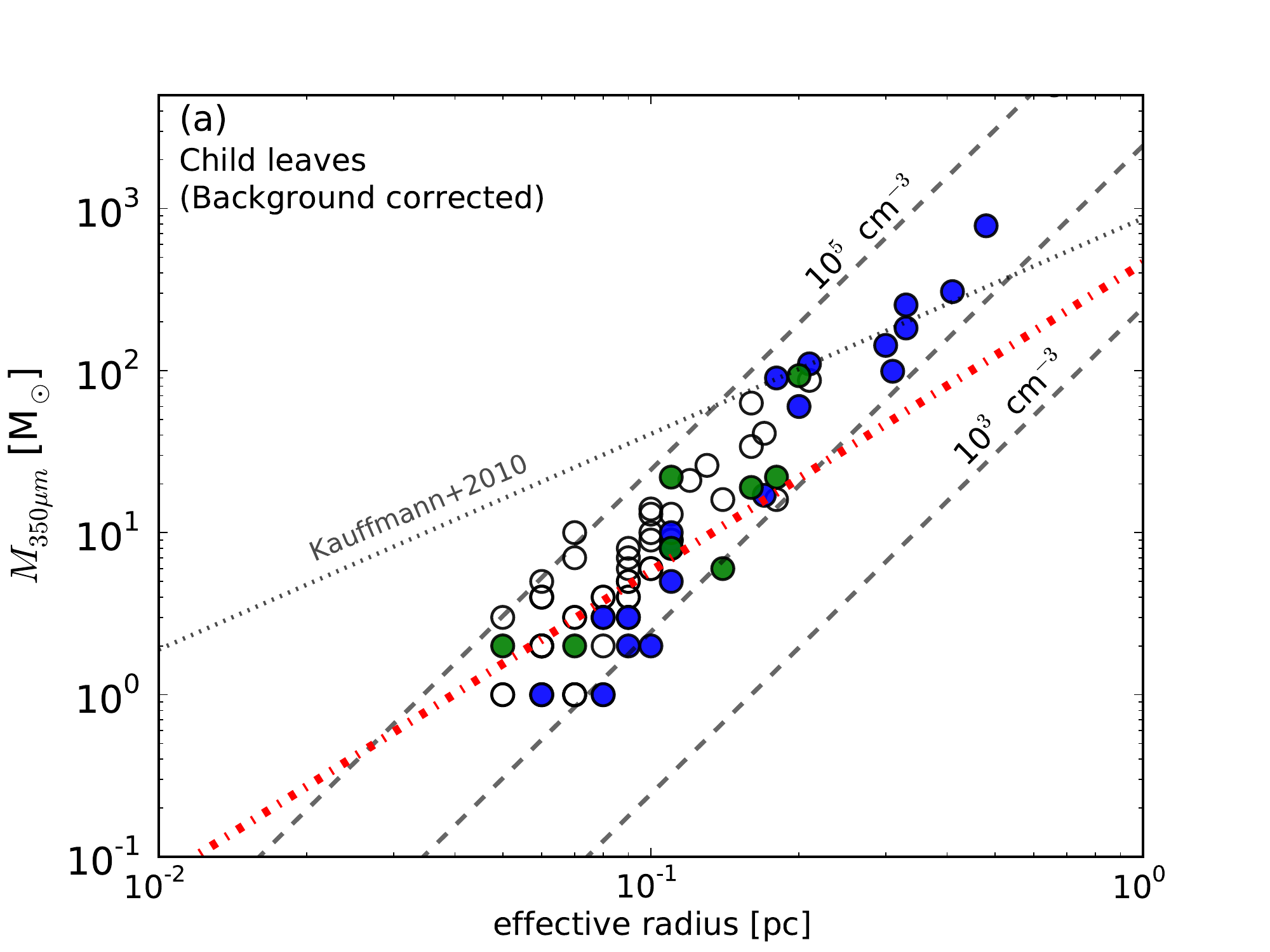}
\includegraphics[width=3.8in]{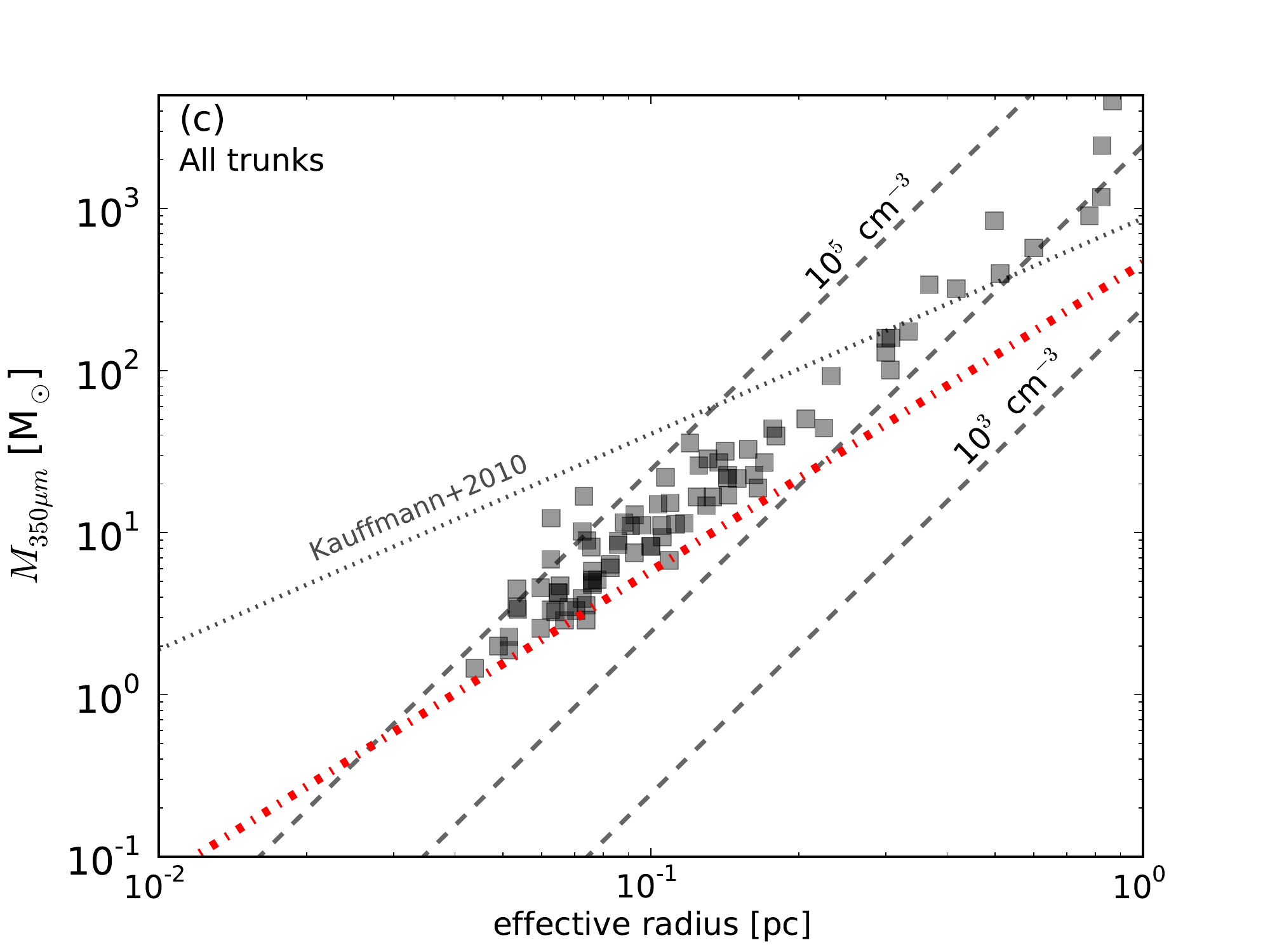}
}
\centerline{
\includegraphics[width=3.8in]{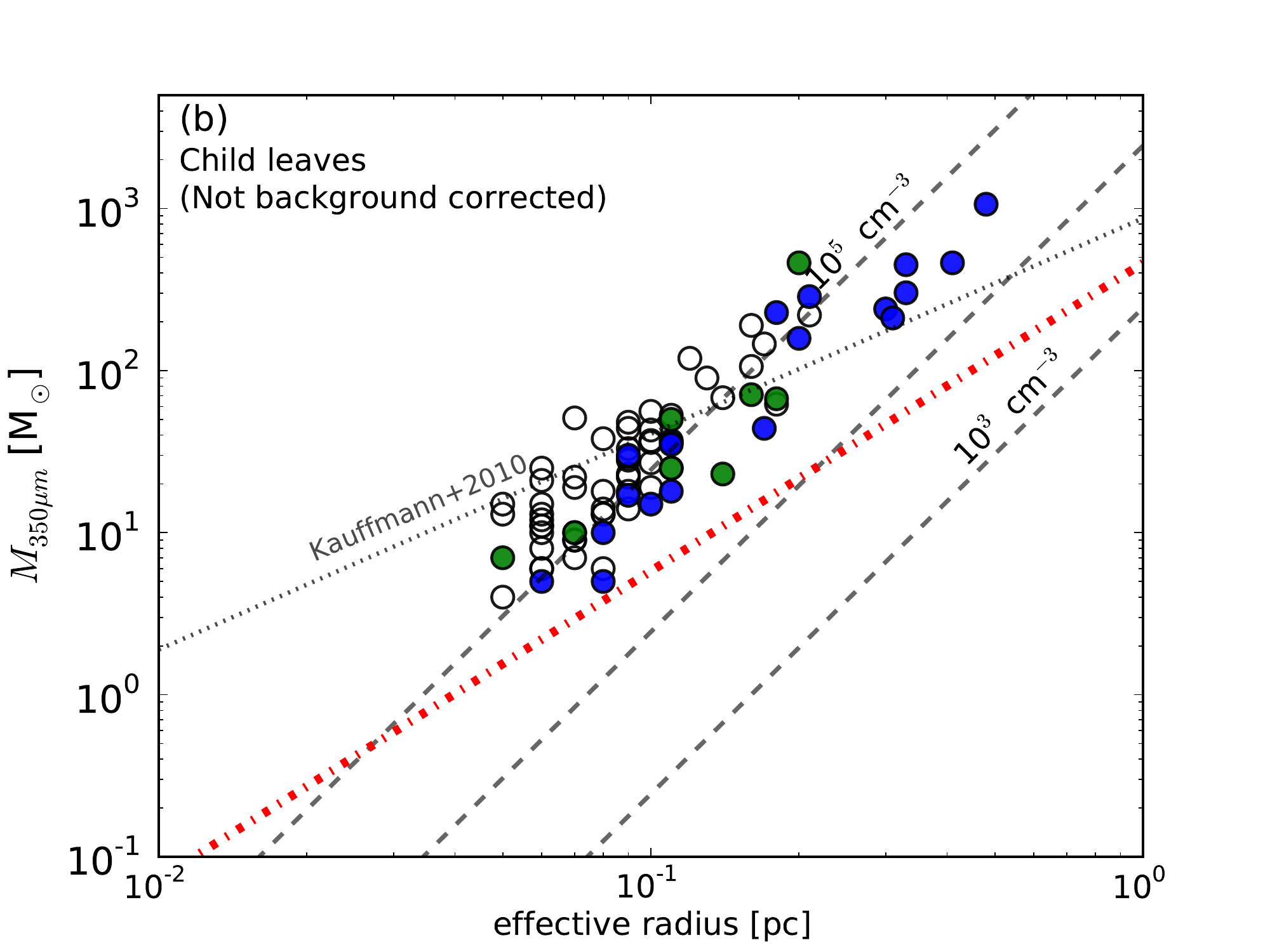}
\includegraphics[width=3.8in]{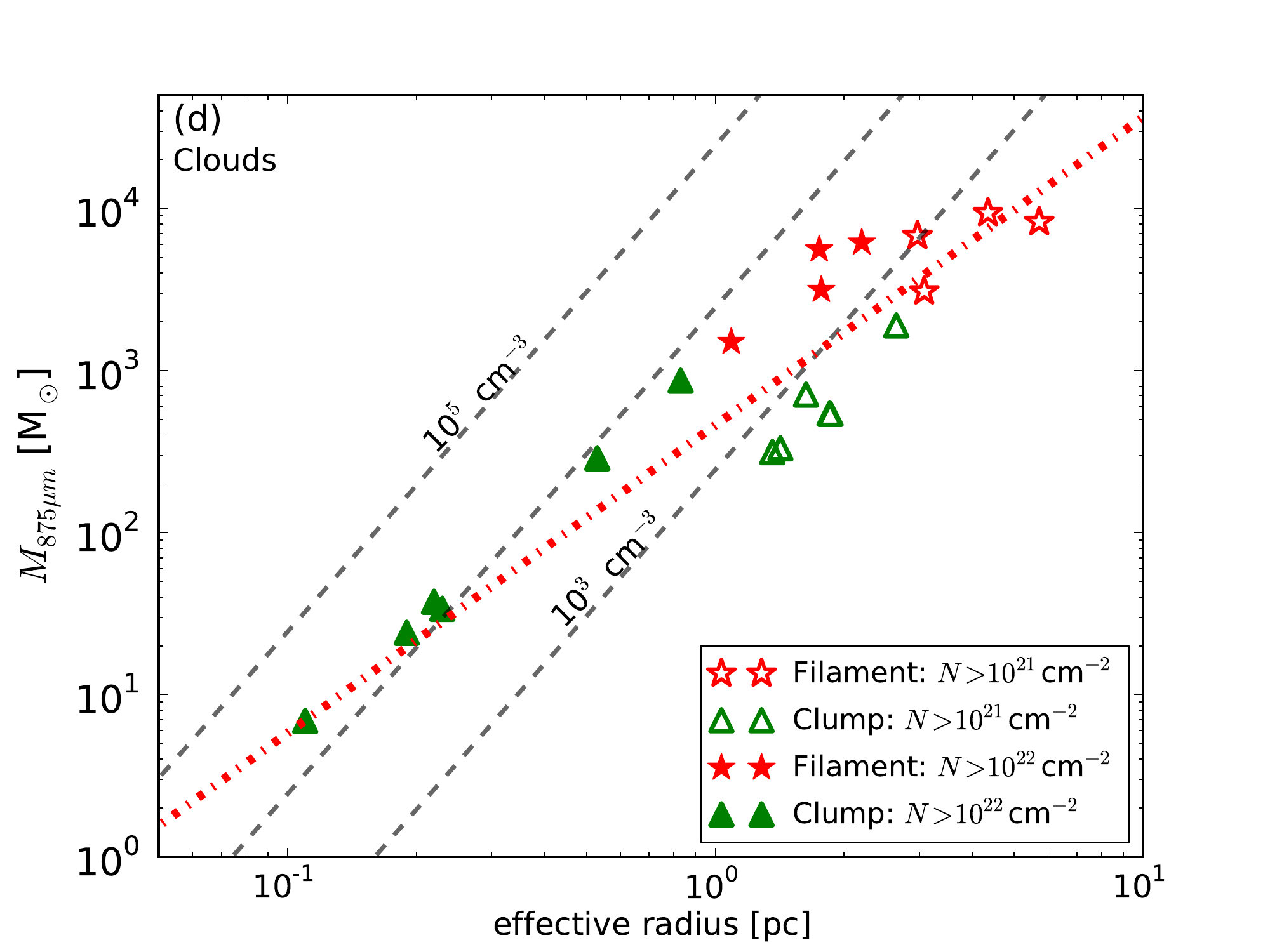}
}
\end{center}
\caption{\label{f:massradius} (a) Child leaf mass (corrected for the parent flux contribution) as a function of effective radius ($r_\mathrm{eff}$ = $\sqrt{A_{leaf} / \pi}$). The blue and green filled circles are the leaves recovered as PACS and cold cores, respectively, and open circles have no {\em Herschel} counterpart. The masses are calculated using the best fit temperatures, or the upper limit temperatures for leaves with no {\em Herschel} counterpart. The dashed grey lines are loci of constant number density, and the dashed-dotted red line is Larson's mass-size relation \citep{Larson1981}, which is approximately a line of constant column density, $N_{H_2} = 10^{22}$\,cm$^{-2}$. (b) Same as (a), but the leaf mass is not corrected for parent flux contribution. (c) Total mass in each trunk, no flux correction, as a function of effective radius. (d) Total cloud mass from the ATLASGAL survey maps as a function of effective radius, computed for the area above 10$^{21}$\,cm$^{-2}$ (empty symbols) and 10$^{22}$\,cm$^{-2}$ (filled symbols).  Filamentary clouds are plotted in red stars and clumpy clouds are plotted in green triangles. Note that the axis scales in (d) differ from (a) through (c).}
\end{figure*}

We examine the relationship between $M_{350{\mu}m}$ and effective radius, defined as $r_\mathrm{eff} = \sqrt{A_{leaf} / \pi}$ where $A_{leaf}$ is the area of a structure, in Figure~\ref{f:massradius}.  We plot several lines of reference in these diagrams: loci of constant number density (dashed grey), the \citet{Larson1981} relation in the red dash-dotted line, and the relation proposed by \citet{Kauffmann_masssize2}, which shows their empirical dividing line in concentration between local clouds devoid of massive stars (below line) and high-mass star-forming clouds (above line).  All but 14 leaves (10\%) fall below this criterion. Twelve of these 14 are found in filamentary IRDCs, and eight have {\em Herschel} counterparts. The leaves exceeding this criteria are noted in Table~\ref{tab:newcores}.  

In plot (a) we show the child leaves corrected for the parent structure flux. We indicate the leaves for which there are {\em Herschel} counterparts by filling in with blue (for PACS cores) or green (for cold cores). We find a trend of roughly constant volume density ($M \propto r_\mathrm{eff}^{2.9}$) for the ensemble of cores with a median value of $4.4 \times 10^4$\,cm$^{-3}$.  The trend is slightly steeper ($M \propto r_\mathrm{eff}^{3.2}$) when only leaves with {\em Herschel} counterparts are fit, though the median density is roughly the same. 

We show the same relationship in plot (b), but in this case $M_{350{\mu}m}$ is derived from the uncorrected flux value.  The trend is shallower in this case: $M \propto r_\mathrm{eff}^{2.3}$ for the full ensemble of leaves and $M \propto r_\mathrm{eff}^{2.6}$ for just those leaves with {\em Herschel} counterparts. The median densities in these populations are $1.2 \times 10^5$\,cm$^{-3}$ and $8.0 \times 10^4$\,cm$^{-3}$\footnote{We note that due to the chopping away of flux on large scales, the computed sizes of objects larger than the bolometer array size (FWHM $\sim$ 100$''$) should be treated with caution. This effect would tend to bias both the flux and size to smaller values.}, respectively, which compare favorably to recent interferometric measurements of IRDC cores \citep[e.g.][]{Beuther2013}. 

The roughly constant volume density trend for the compact objects shown in (a) is consistent with the \cite{Kainulainen2011a} extinction study of local clouds but in contrast to that which is observed in dust emission locally \citep[e.g.][]{Enoch2008}, which follows a trend of constant surface density, similar to the \cite{Larson1981} relationship for turbulent clouds using CO observations. This apparent discrepancy might easily be reconciled by examining the various scales identified as iso-surfaces in lower levels of the {\tt dendrogram}~tree. In plot (c) of Figure~\ref{f:massradius}, we show the relationship for all ``trunks'' of the {\tt dendrogram}, i.e. the sum of each independent flux structure. For example, in Figure~\ref{f:tree_diagram}, structures 3, 6, 7, and 8 are trunks, then so are structures 1+2 and 4+5, including their parent flux. This is analogous to applying a single column density contour to each cloud (see Table~\ref{tab:obstable}). Here the relation becomes shallower ($M \propto r_\mathrm{eff}^{2.3}$) with a median density of 4.3 $\times$ 10$^4$\,cm$^{-3}$. 

Taking this idea a step further, using ATLASGAL data \citep{ATLASGAL} we plot the same again in Figure~\ref{f:massradius}(d) for different column density boundary definitions of the cloud, and the result is consistent with the Larson relation. Here we also make a distinction between the clumpy and filamentary clouds in our sample, showing that while the filamentary sources tend to be more massive, they adhere to the same relationship as the clumpy clouds. We return to this discussion of morphology in Section~\ref{ss:morph}.

\begin{figure*}
\begin{center}
\includegraphics[scale=0.7,angle=90]{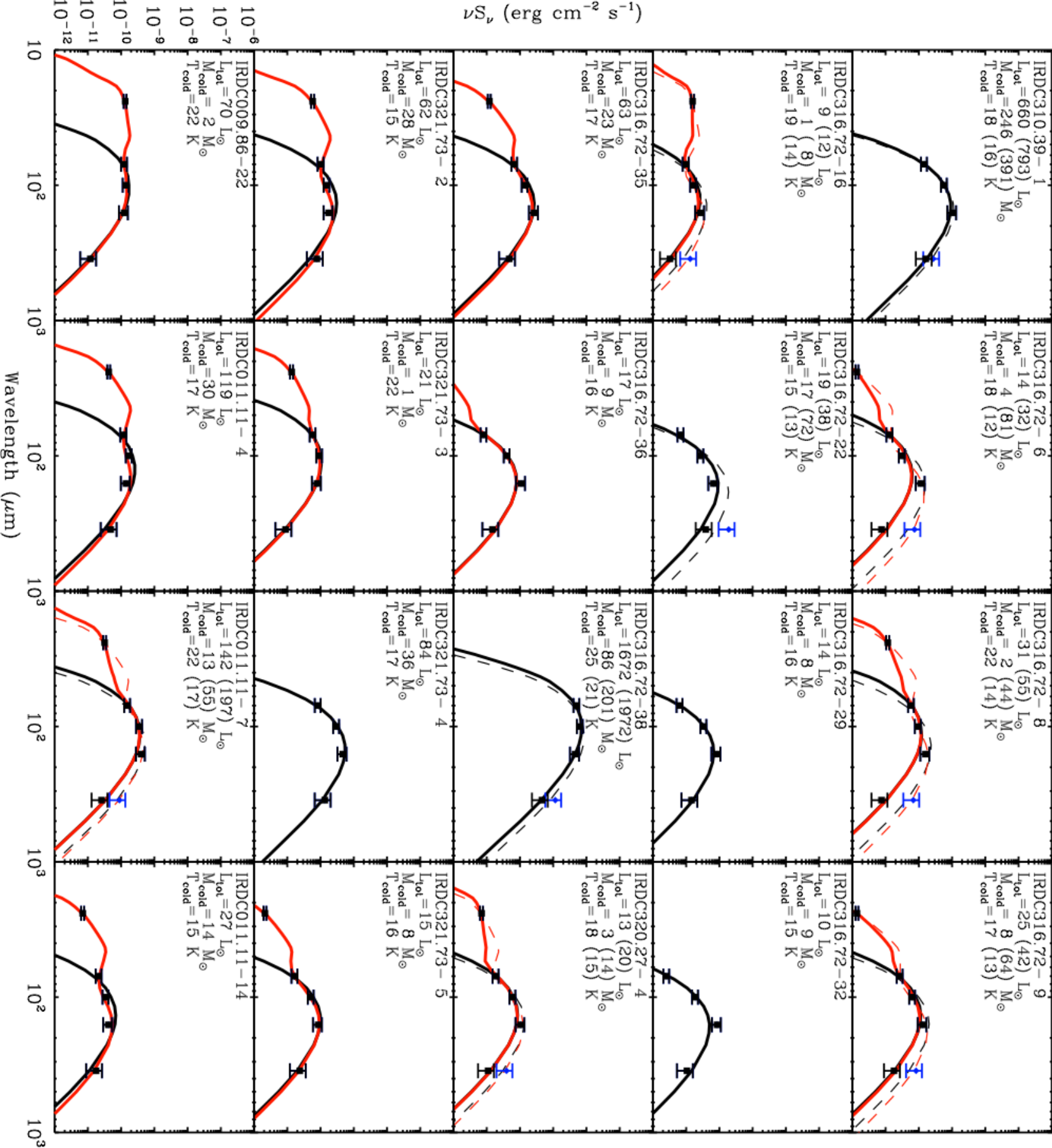}
\includegraphics[scale=0.7,angle=90]{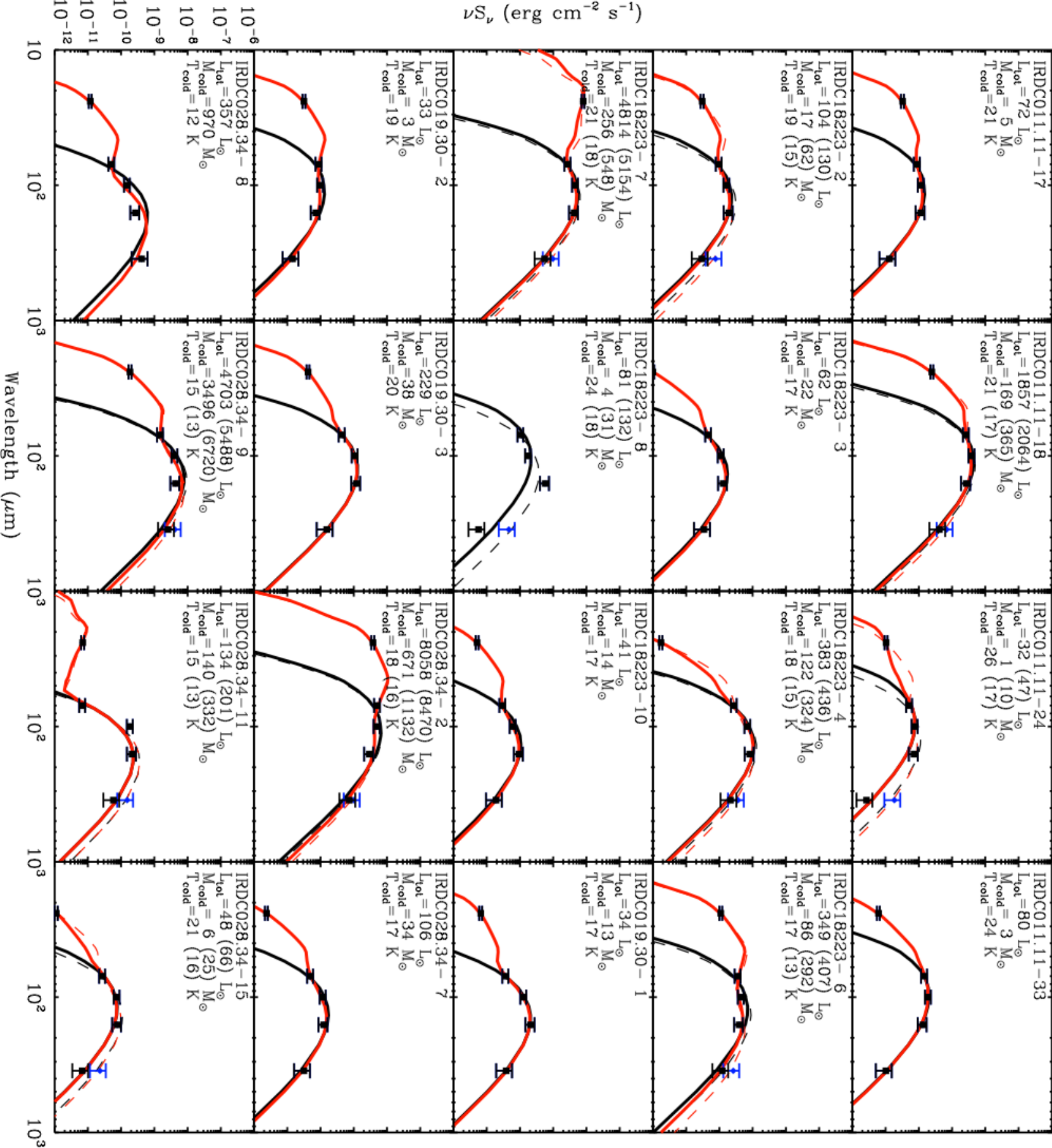}
\end{center}
\caption{\label{f:sed} SEDs of cores with 70\,$\mu$m counterparts. The IRDC name and ID number from Table~\ref{tab:newcores} are shown in the upper-left of each panel along with the best-fit properties to the cold component from the full-SED fit (the properties from the fit to the SED uncorrected for parent flux contribution are given in parentheses).  For all objects, the solid black line shows the single-temperature fit to the data excluding the 24\,$\mu$m counterpart.  In cases in which a 24\,$\mu$m counterpart was detected, the red line shows the summed spectrum of a two component fit.  The uncorrected 350\,$\mu$m flux is shown in blue, and the fits to the uncorrected SED are shown in dashed black (single component) and red (double component) lines. }
\end{figure*}

\begin{figure*}
\begin{center}
\includegraphics[scale=0.8,angle=90]{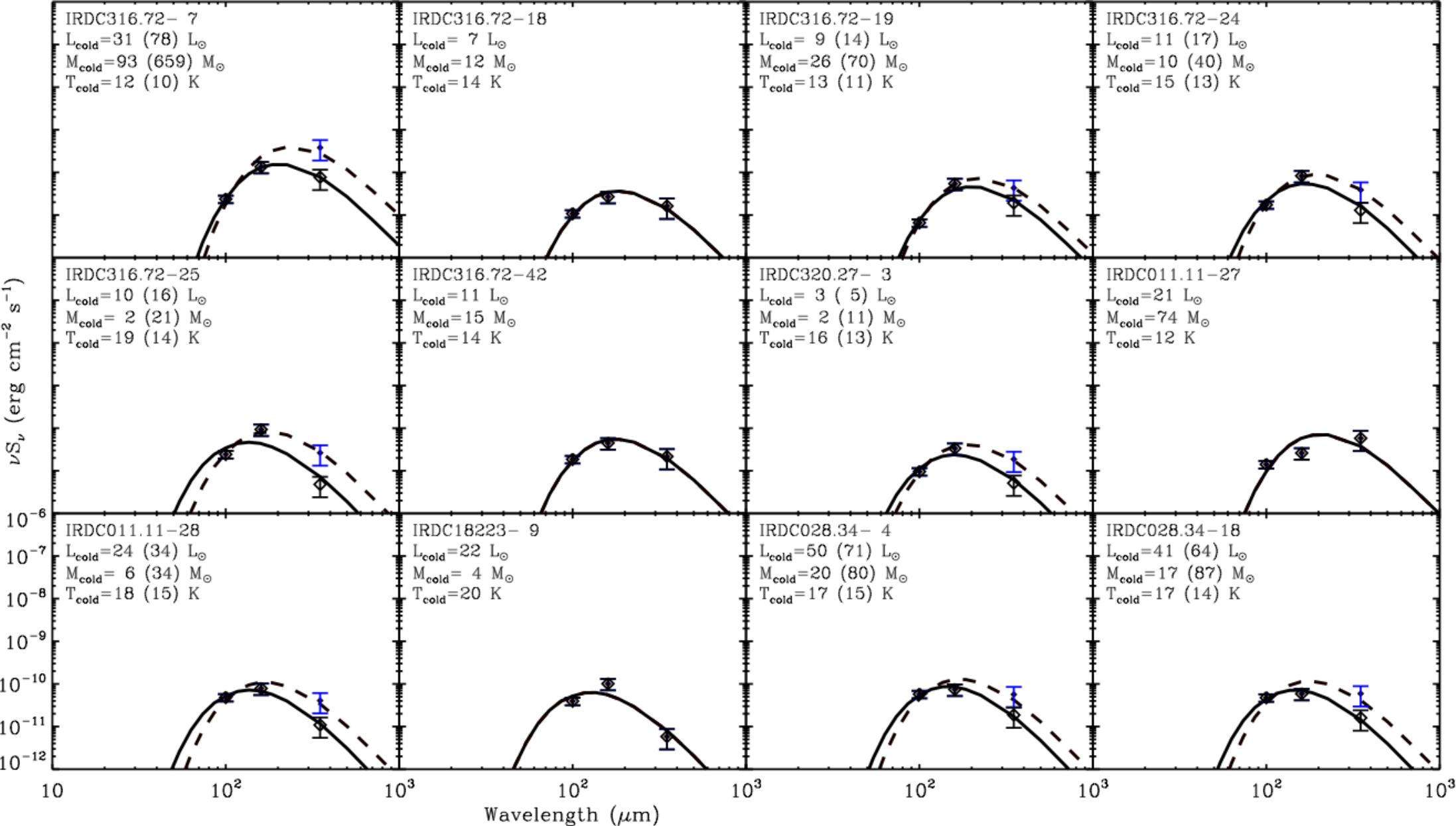}
\end{center}
\caption{\label{f:csed} SEDs of cores with no 70\,$\mu$m counterparts. The IRDC name and ID number from Table~\ref{tab:newcores} are shown in each panel. The best-fit single temperature modified blackbody fit to the SED with the with $S_{350{\mu}m}$ corrected for the parent flux is plotted in the solid line (the parameters are listed in each panel), and the fit to the SED including the uncorrected $S_{350{\mu}m}$ is shown with the dashed gray line.}
\end{figure*}

Is the steepening of the mass-size relation on the small scales a consequence of an observational bias or a real physical effect?  In an idealized case, a constant volume density trend is expected if IRDC cores are in pressure balance \citep[cf. the Pipe Nebula][]{Lada_pipe}.  However, there are several observational issues one must keep in mind.  First, we note that the spread in the relation is nearly 1 dex and, as mentioned above, bolometer mapping may bias us toward smaller radii \citep[see also][]{Enoch2006,Rosolowsky_BGPS}, which would steepen the relation.  Furthermore, the mass-radius relation has been shown to suffer from projection effects and thus poor mapping to physical structure \citep{BP2002,Shetty2010}, which also causes considerable scatter.  Finally, the complex dynamics play an important if not dominant role in the energy balance of IRDCs \citep[e.g.][]{Ragan2012}, and we lack this information on the core scale for this sample. Therefore, since the changing trends can be reproduced simply by considering different radii within the structure, the reader should be cautious in drawing conclusions on the core stability from these data alone.

\begin{table*}
\caption{Core properties, method comparison. \label{tab:coreprops}}
\begin{tabular}{llrrrcrrrl}
\hline \hline
IRDC & ID  &  $M_\mathrm{PACS}^b$ & $M_\mathrm{350{\mu}m}^c$ & $M^\mathrm{full-SED}_\mathrm{total}$ & $T_\mathrm{PACS}^b$ & $T^\mathrm{full-SED}_\mathrm{cold}$ & $L_\mathrm{PACS}^b$ & $L^\mathrm{full-SED}_\mathrm{cold}$ & Notes \\
 & no. $^a$ & ($\msun$) & ($\msun$)~~ & ($\msun$)~~ & (K) & (K)~~~~ & ($\lsun$) & ($\lsun$)~~~ & \\
\hline
IRDC\,310.39-0.30 & & & & & & & & & \\
 &1$^d$ &        303 &    229 (2054) &     246 (391) & 17 & 18 (16) &      716 &      660 (793) & \\ 
\hline
IRDC\,316.72+0.07 & & & & & & & & & \\
 &6 &          9 &      3 (30) &       4 (81) & 17 & 18 (12) &       19  &       14 (32) & \\ 
 &7 & 	      -- &     93 (462) &     93 (659) &  --  & 12 (10) &  --  &     31 (78) & \\ 
 &8 &          6 &      2 (17) &       2 (44) & 20 & 22 (14) &       34 &       31 (55) & \\ 
 &9 &          8 &      8 (36) &       9 (64) & 18 & 17 (13) &       26 &       25 (42) & \\ 
 &16 &         1 &      1 (5) &       1 (8) & 20 & 19 (14) &        5 &        9 (12) & \\ 
 &18 &        -- &       14 &     12 &  --  & 14 &  --  &      7 & \\ 
 &19 &        -- &     22 (50) &     26 (70) &  --  & 12 (11) &  --  &      9 (14) & \\ 
 &22$^d$ &     7 &     25 (177) &      18 (72) & 16 & 15 (13) &       12 &       19 (38) &  \\ 
 &24 &        -- &      8 (25) &     10 (40)  &  --  & 15 (13) &  --  &     11 (17) & \\ 
 &25 &        -- &      2 (10) &      2 (10) &  --  & 19 (14) &  --  &     10 (16) & \\ 
 &29 &        12 &      8 &       8  & 16 & 16 &       16 &       14 & \\ 
 &32 &        22 &      7 &       9  & 14 & 15  &       16 &       10 & \\ 
 &35 &        13 &     23 &      24  & 19 & 17  &       54 &       63 & \\ 
 &36 &        16 &      7  &       10  & 16 & 16  &       21 &       17 & \\ 
 &38 &        79 &     90 (228) &      86 (201) & 25 & 25 (21) &     1650 &     1672 (1972) & \\ 
 &42 &       --  &     17 &     15 &  --  & 14 &  --  &     11 & \\ 
\hline
IRDC\,320.27+0.29 & & & & & & & & & \\
 &3 &  		   --  &       2 (7) &      2 (11) &  --  & 16 (13) &  --  &      3 (5) & \\ 
 &4 &            5 &      3 (10) &       3 (14) & 18 & 18 (15) &       14 &       13 (20) & \\ 
\hline
IRDC\,321.73+0.05 & & & & & & & & & \\ 
 &2 &          3 &     34 &      28 & 22 & 15 &       32 &       62 & \\ 
 &3 &          1 &      2 &       1  & 23 & 22  &       16 &       21 & \\ 
 &4 &         24 &     43 &      36  & 18 & 17  &       71 &       84 & \\ 
 &5 &          4 &      9 &       8 & 18 & 16 &       12 &       15 & \\ 
\hline
IRDC\,009.86-0.04  & & & & & & & & & \\
 &2 &          2 &      3 &       2  & 24 & 22 &       37 &       70 & \\ 
\hline
IRDC\,011.11-0.12  & & & & & & & & & \\
 &4 &          5 &     36 &      30 & 23 & 17 &       68 &      119 & \\ 
 &7 &         25 &     10 (35) &      13 (55) & 20 & 22 (17) &      152 &      142 (197) & \\ 
 &14 &         2 &     17 &      14 & 21 & 15 &       16 &       27 & \\ 
 &17 &         4 &      6 &       5  & 23 & 21  &       51 &       72 & \\ 
 &18 &        80 &    183 (302) &     169 (365) & 24 & 21 (17) &     1444 &     1857 (2064) & \\ 
 &24 &         3 &      1 (5) &       1 (10) & 22 & 26 (17) &       32 &       32 (47) & \\ 
 &27 &        -- &    116 &      75 & -- & 12 &      --  &       21  & \\ 
 &28 &        -- &      6 (23) &       7 (34) & -- & 18 (15) &      --  &       24 (34) & \\ 
 &33 &         4 &      3 &       3 & 24 & 24  &       73 &       80  & \\ 
\hline
IRDC\,18223            & & & & & & & & & \\
 &2 &        10 &     17 (44) &      17 (62) & 21 & 19 (15) &       79 &      104 (130) & \\ 
 &3 &         9 &     25 &            23  & 20 & 17  &       51 &       62 & \\ 
 &4 &        61 &    143 (240) &     122 (324) & 20 & 18 (15) &      323 &      383 (436) & \\ 
 &6 &        12 &     99 (211) &      86 (292) & 24 & 17 (13) &      200 &      349 (407) & \\ 
 &7 &       185 &    254 (450) &     256 (548) & 22 & 21 (18) &     1979 &     4814 (5154) & IRAS\,18223-1243  \\ 
 &8 &        51 &      2 (15) &       4 (31) & 18 & 24 (18) &      156 &       81 (132) & \\ 
 &9 & 	     -- &      3 &     			4      & -- & 20      &      --  &       22 & \\ 
 &10 &        7 &     14 &      14  & 20 & 17 &       33 &       41 & \\ 
\hline
IRDC\,019.30+0.07 & & & & & & & & & \\
 &1 &         12 &     13 &      13 & 18 & 17 &       33 &       34  & \\ 
 &2 &        0.8 &      4 &       3 & 25 & 19 &       19 &       33  & \\ 
 &3 &         37 &     36 &      38 & 20 & 20 &      219 &      229  & \\ 
\hline
IRDC\,028.34+0.06 & & & & & & & & & \\
 &2 &         93 &    782 (1062) &     671 (1132) & 28 & 18 (16) &     3577 &     8058 (8470) & MSX G028.3373+00.1189 \\ 
 &4 &         -- &     22 (67) &     20 (80) &  --  & 17 (15) &  --  &     50 (71) & \\ 
 &7 &         14 &     37 &      34 & 20 & 17 &       85 &      106 & \\ 
 &8 &         49 &   1369 &     971 & 18 & 12 &      151 &      357 & \\ 
 &9$^d$ &    534 &   4490 (9249) &    3496 (6720) & 20 & 15 (14) &     2950 &     4703 (5488) & MSX G028.3937+00.0757 \\ 
 &11 &       265 &    110 (286) &     140 (332) & 14 & 15 (13) &      177 &      134 (201) & \\ 
 &15 &         9 &      5 (18) &       6 (25) & 20 & 21 (16) &       52 &       48 (66) & \\ 
 &18 &       --  &     19 (71) &     17 (87) &  --  & 17  (14) &  --  &     41 (64) & \\ 

\hline
\end{tabular}

\tablefoottext{a}{from Table~\ref{tab:newcores}}

\tablefoottext{b}{from \citet{Ragan2012b}}

\tablefoottext{c}{Derived with the best fit dust temperature (column 6 of this table).}

\tablefoottext{d}{Leaf overlaps with multiple {\em Herschel} cores. SABOCA flux is scaled according to luminosity ratio of cores (see Section~\ref{sec:corerecovery}).}

\end{table*}

\subsection{SEDs and core properties}
\label{s:sed}

\begin{figure*}
\centerline{
\includegraphics[width=2.35in]{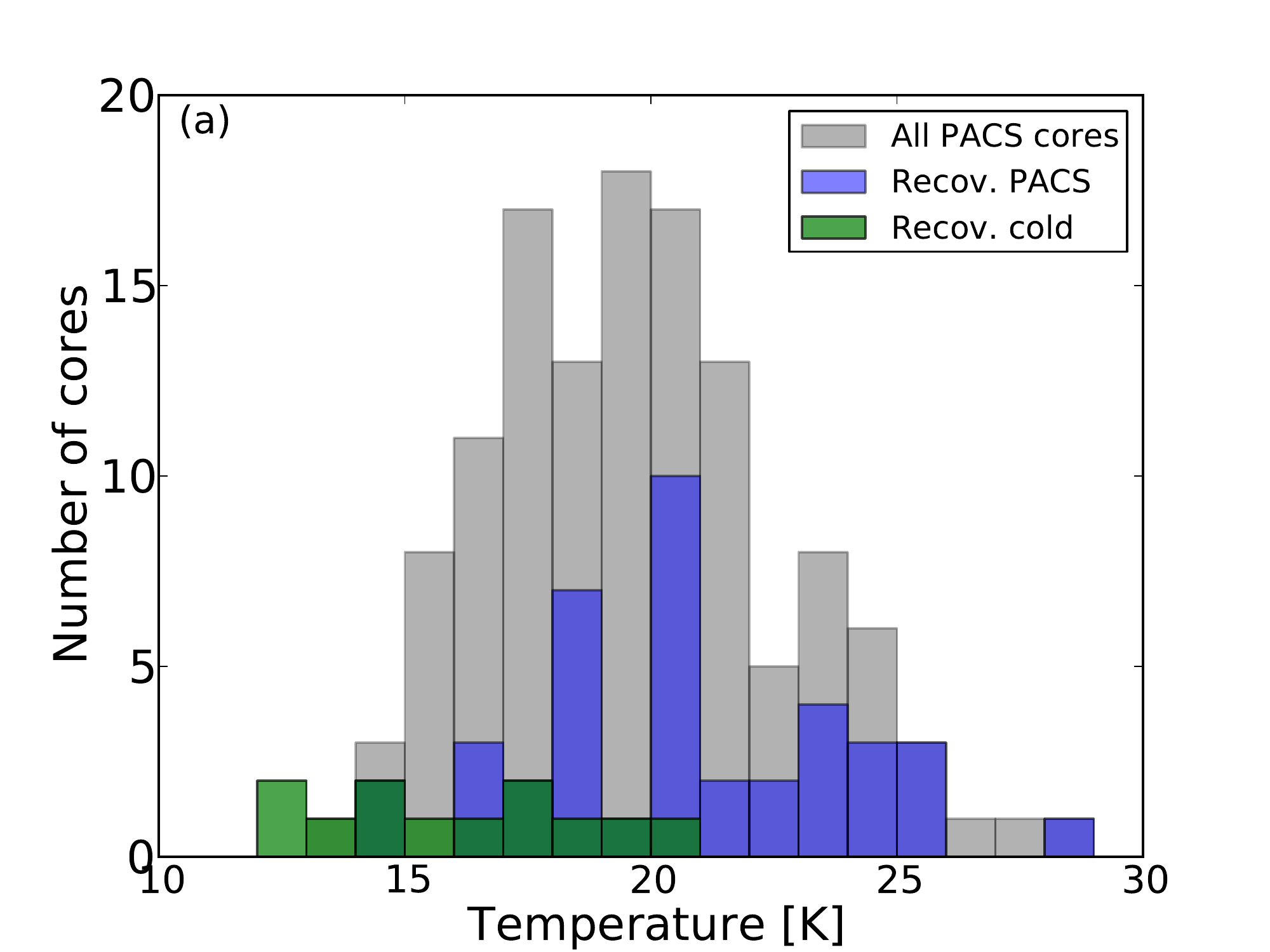}
\includegraphics[width=2.35in]{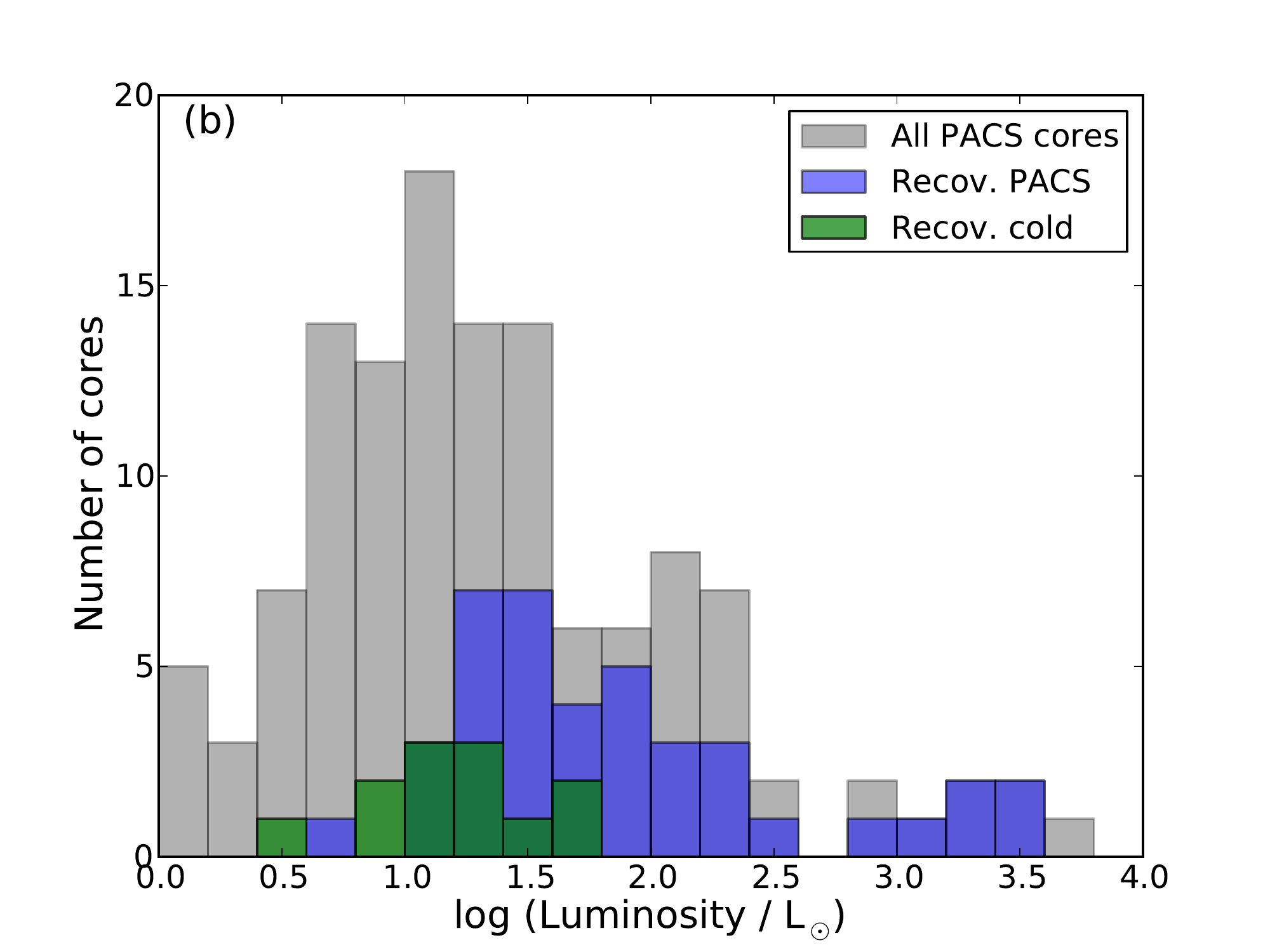}
\includegraphics[width=2.35in]{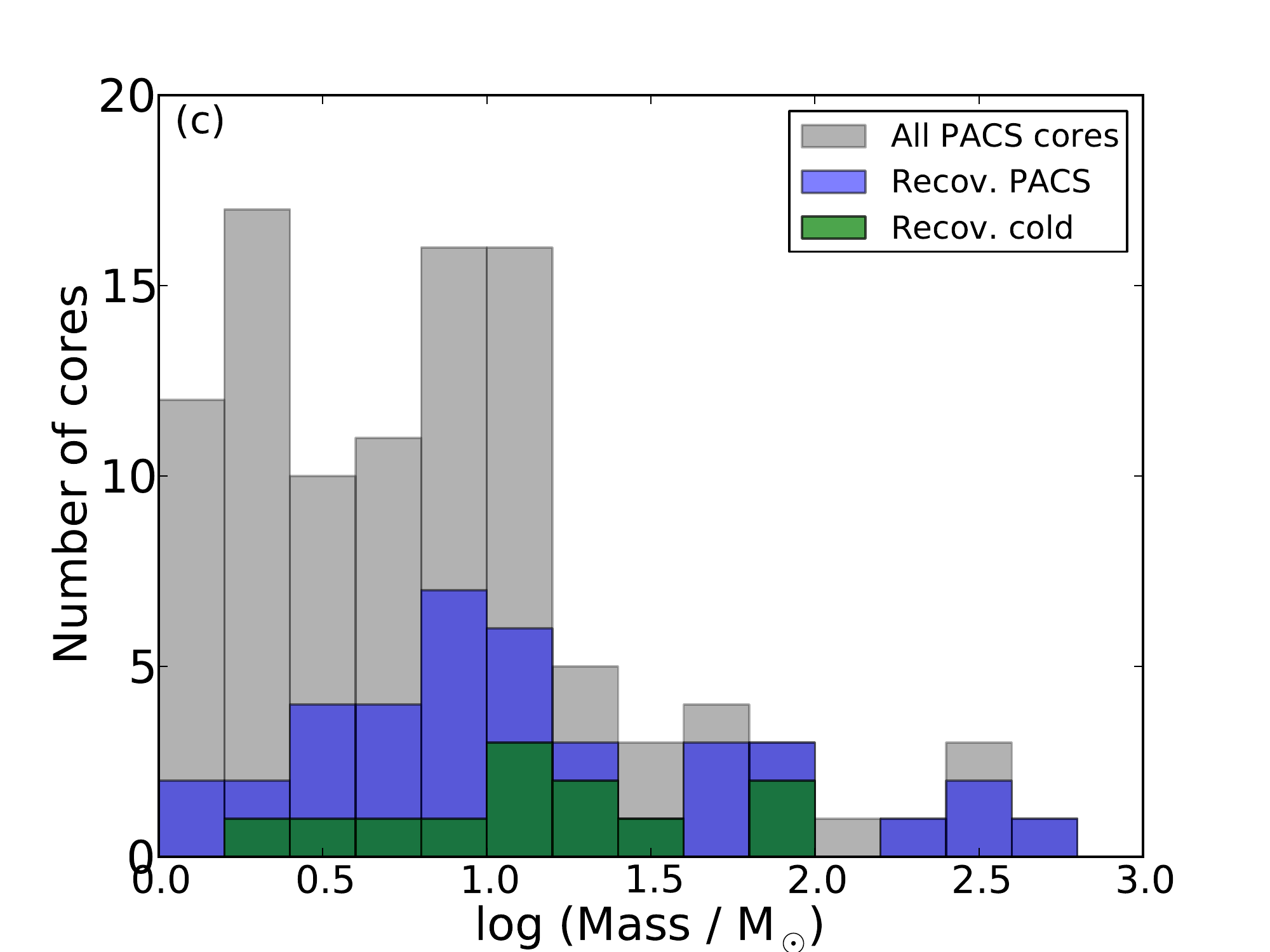}
}
\caption{\label{f:recover} The gray histogram shows the (a) temperature, (b) luminosity, and (c) mass distribution of all PACS cores found in this sample by R12. The blue histogram shows the subset of these cores which matched SABOCA leaves. The green histogram shows distributions of  cold cores (see Figure~\ref{f:csed}).}
\end{figure*}

In R12, we fit blackbodies to the SEDs of all cores detected in all three PACS bands, but left out {\em Herschel}/SPIRE data for longer wavelengths because the angular resolution was not sufficient to isolate the 0.1\,pc scale. Because SABOCA boasts similar angular resolution as PACS at 350\,$\mu$m, we are now able to overcome that limitation. Table~\ref{tab:coreprops} lists all cores for which a SABOCA leaf matches a {\em Herschel}/PACS core identified in R12.  For a complete comparison, we also list in Table~\ref{tab:newcores} the $M_{350{\mu}m}$ (see Equation~\ref{eq:m350} and Section~\ref{ssec:structure}) using the best-fit temperature (in cases of a recovered core, noted in the last column) or the upper-limit temperature (in case the leaf has no {\em Herschel} counterpart).  For the latter cases, using an upper-limit temperature will result in a lower-limit mass.  

Figure~\ref{f:sed} shows the full SEDs for the 40 overlapping objects.  First, we fit a single temperature modified blackbody function to the SED from 70 to 350\,$\mu$m SABOCA data, shown with the solid black line, representing the ``cold'' component.  We follow the method exactly as we did for fitting the PACS SEDs in R12. The SED-fitting algorithm assumes the dust is optically thin and takes into account the frequency-dependence of the dust opacity, $\kappa_{\nu}$, for which we adopt OH5. We compare parameters derived from the new fits ($M_\mathrm{total}^\mathrm{full-SED}$, $T_\mathrm{cold}^\mathrm{full-SED}$, and $L_\mathrm{cold}^\mathrm{full-SED}$) to the PACS+SABOCA SED to those found in R12, which use just the PACS data ($M_\mathrm{PACS}$, $T_\mathrm{PACS}$, and $L_\mathrm{PACS}$), in Table~\ref{tab:coreprops}. 

Next, for cores with 24\,$\mu$m counterparts, we fit the full SED (from 24\,$\mu$m to 350\,$\mu$m, a total of five data points) with two temperature components henceforth referred to as the ``warm'' and ``cold'' components, since a single temperature modified blackbody does not capture the shape of the full spectral range well.  We again assumed OH5 dust for both the warm and cold components\footnote{We note that OH5 dust may not be the optimal choice for the relatively warm inner component of the SED, where the protostar has presumably heated the region. If we instead adopt a dust opacity model with no ice mantles \citep[e.g.][column 2]{ossenkopf_henning}, or OH2, for the warm inner component, the mean deviation from our reported mass in Table~\ref{tab:coreprops} is about 7\%. Therefore, for simplicity, we adopt uniform dust properties (OH5) throughout the analysis.} of our SED.  We list the best fit parameters to the cold component in each panel of Figure~\ref{f:sed} and in Table~\ref{tab:coreprops}. In parentheses, we show the results from a fit using the 350\,$\mu$m flux value which is not corrected for parent flux (blue data point at 350\,$\mu$m). Those fits, always with lower temperatures and higher masses and luminosities, are shown in the dashed curves.

The summed (warm + cold) SED fit is shown with the red line in Figure~\ref{f:sed}. For those cases, the ``full-SED'' parameters given in Table~\ref{tab:coreprops} are that of the cold, outer component.  In most cases, the inclusion of a second blackbody component to account for the 24\,$\mu$m flux causes $T_\mathrm{cold}^\mathrm{full-SED}$ to be lower than was fit with the PACS data alone. This supports the picture presented in R12 where the warm temperatures in 24\,$\mu$m-bright cores are attributed to their marginally more advanced evolutionary stage. The warm dust component contributes slightly to the 70\,$\mu$m emission, and with the second fit component invoked here, the temperature of the cold component decreases. The summed SED for the fit including the uncorrected value of $S_{350{\mu}m}$ is shown in the red dashed curve.

Figure~\ref{f:csed} shows the SEDs for objects which were only detected in PACS 100 and 160\,$\mu$m bands and were recovered by the SABOCA observations. We fit a single temperature modified blackbody to the three data points on the SED, for which we list the derived properties from the best fit in Table~\ref{tab:coreprops} and in the individual panels of Figure~\ref{f:csed}. We also show the 350\,$\mu$m flux uncorrected for the contribution from the parent structure (blue data point) and the corresponding derived properties in parentheses.  Again, the uncorrected value for $S_{350{\mu}m}$ results in lower temperatures and higher masses, but given the uncertainties, it is less clear whether the correction, which is on the order of a factor of 3.6$\pm$1.0, improves the blackbody fits or not. 

Figure~\ref{f:recover} shows the temperature distribution for all PACS cores, which had the mean and median of 20\,K. Based on the SED fits, the 70\,$\mu$m-dark ``cold'' cores are colder on average (mean and median 16\,K) than the 70\,$\mu$m-bright cores (mean and median 19\,K and 18\,K).  The cold core temperatures compare well with the temperatures found in the starless globules modeled in \citet{Launhardt2013}, whereas the PACS cores more closely match the peak temperatures found for the protostellar cores.  The center and right panels of Figure~\ref{f:recover} show that while the cold cores occupy the same range in mass as the 70\,$\mu$m-bright cores, the luminosities of 70\,$\mu$m-bright cores tend to be higher (median 52\,$\lsun$) than 70\,$\mu$m-dark cores (median 17\,$\lsun$).

\begin{table*}
\begin{center}
\caption{Core population \label{tab:corepop}}
\begin{tabular}{lrrrrc}
\hline \hline
IRDC & $N_{PACS}$ & $N_{70dark}$ & $N_{SABOCA}$ & $N^{overlap}_{PACS}$ &  $N^{overlap}_{70dark}$\\
name &           (1)   &           (2)        &             (3)        &              (4)                 &                   (5)  \\
\hline
IRDC\,310.39-0.30   &   2  &  0  & 2   & 1 (2)  	& 0 (0) \\
IRDC\,316.72+0.07 &  14 & 9  & 42  & 10 (12) & 6 (6) \\
IRDC\,320.27+0.29  &   3   &  6 & 4   & 1 (1) 		& 1 (1) \\
IRDC\,321.73+0.05  &   10 &  1 & 5   & 4 (4) 		& 0 (0) \\ 
IRDC\,004.36-0.06   &   1  &  3 & 4    & 0 (0) 		& 0 (0) \\
IRDC\,009.86-0.04   &   3  &  1 & 2    &  1 (1) 	& 0 (0) \\
IRDC\,011.11-0.12 &  20 & 28 & 35  & 7 (11) 	& 2 (4) \\
IRDC\,015.05+0.09  &   3  &  0  &  8   & 0 (1)		& 0 (0) \\
IRDC\,18223           &  13 & 11 & 10  & 7 (9)		& 1 (1) \\
IRDC\,019.30+0.07  &   6  &  4  &  3   & 3 (3) 	& 0 (0) \\
IRDC\,028.34+0.06 &  14 & 10 & 20 &  6 (8) 	& 2 (2) \\
\hline
{\bf Total:}      &  89 & 73 & 135 & 40 (52) & 12 (14) \\
\hline
\end{tabular}
\end{center}

\tablefoottext{1}{Detected at 70, 100, and 160\,$\mu$m, as reported in \citet{Ragan2012b} \\
overlapping with mapped area.}

\tablefoottext{2}{Detected at 100 and 160\,$\mu$m, but not 70\,$\mu$m.}

\tablefoottext{3}{Number of ``leaves'' detected in SABOCA 350\,$\mu$m map with {\tt dendrograms}.}

\tablefoottext{4}{Number of overlapping cores between column (1) and (3). The number in parentheses is the number of PACS cores coincident with any SABOCA emission (not just leaves).}

\tablefoottext{5}{Number of overlapping cores between column (2) and (3). The number in parentheses is the number of ``70dark'' cores coincident with any SABOCA emission (not just leaves).}

\end{table*}

\subsection{Core mass estimates}

\begin{figure}
\includegraphics[width=\linewidth]{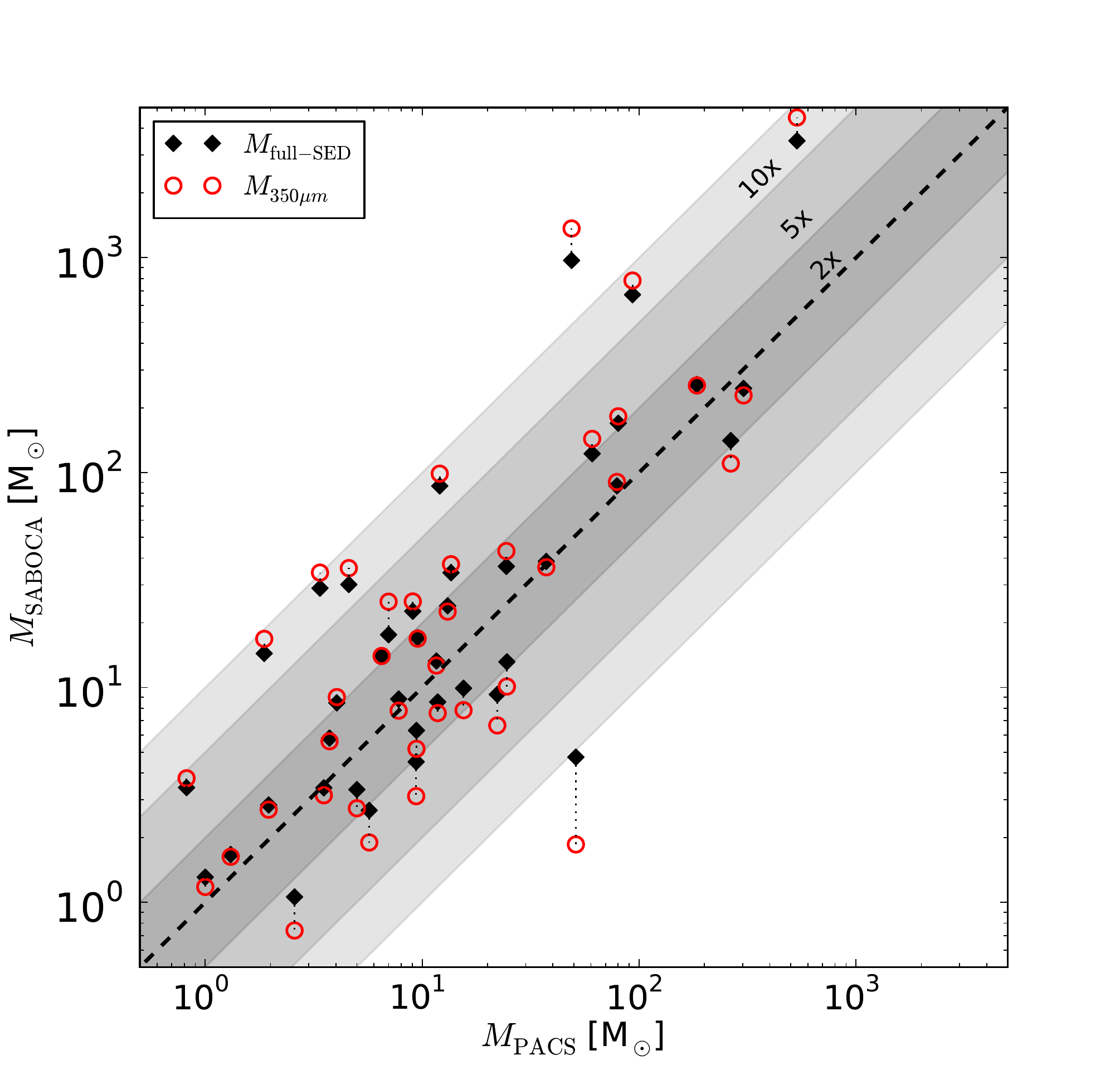}
\caption{\label{f:mcompare} Comparison between the originally derived mass ($M_\mathrm{PACS}$) to the newly derived mass including the SABOCA 350\,$\mu$m data point ($M_\mathrm{full-SED}$, black diamonds) or $M_{350{\mu}m}$ derived from Equation~\ref{eq:m350} in red open circles. Ranges of mass differences between the two methods are shown. Of the 40 PACS cores, 63\% change by less than a factor of 2, and 83\% change by less than a factor of 5. All but one core mass changes by less than a factor of 10.}
\end{figure}

Calculating accurate core masses is fundamental in constraining the initial conditions for star formation in IRDCs. With the present dataset, we calculate mass in two ways; one is using just the 350\,$\mu$m flux ($M_{350{\mu}m}$: Equation~\ref{eq:m350}) and the other is using SED fits described in Section~\ref{s:sed} ($M_\mathrm{full-SED}$).  A weakness inherent in the former method is that one must assume a dust temperature (in Table~\ref{tab:newcores} we assume a uniform temperature of 20\,K), where in the latter method the temperature is fit simultaneously with the other parameters. We compare the previous estimate of core mass, $M_\mathrm{PACS}$, from R12 to these two methods using SABOCA data in Figure~\ref{f:mcompare}. Below 100\,$\msun$, the use of the full-SED or $M_{350{\mu}m}$ results in a higher mass twice as often as it results in a smaller mass. The mass increases are more extreme for the most massive cores, though this is a result of the large angular extent of these objects in SABOCA.

We can compare the three methods for the 40 cores that appeared in R12 that were recovered by SABOCA. In 25 of the 40 cases (63\%), the inclusion of the SABOCA data point altered the mass estimate by less than a factor of 2, and in 33 cases (83\%), the mass adjustment was under a factor of 5. From Figure~\ref{f:mcompare} we see that in most (68\%) of the cases the previous PACS-only mass estimate was lower than the full-SED mass. We also plot $M_{350{\mu}m}$ evaluated at the full-SED fit temperature result values typically within 20\% of $M_\mathrm{full-SED}$.  In the following, we adopt the $M_\mathrm{full-SED}$ and the corresponding cold component temperature (see Section~\ref{s:sed}).

A similar comparison can not be made for cold cores, as it was impossible to model their SEDs with PACS data only. However, there is good agreement between their $M_\mathrm{full-SED}$ and $M_{350{\mu}m}$, within 50\% for all twelve cores.  The maximum mass of a cold core is 93\,$\msun$. Since these cores have lower temperature than the PACS cores, any inaccuracy in the temperature results in larger errors in the mass than for warmer cores. 

Since $M_{350{\mu}m}$ agrees fairly well with $M_\mathrm{full-SED}$, we can confidently estimate masses for the 83 SABOCA leaves that have no {\em Herschel} counterpart using Equation~\ref{eq:m350}. We stress that this entirely new population of cores are of particular interest because with no {\em Herschel} counterpart, they could be cold, dense, starless structures, representing the pristine earliest phase of the high-mass star formation process.

\subsection{Evolutionary diagnostics}
\label{s:evol}

\begin{figure}
\includegraphics[width=\linewidth]{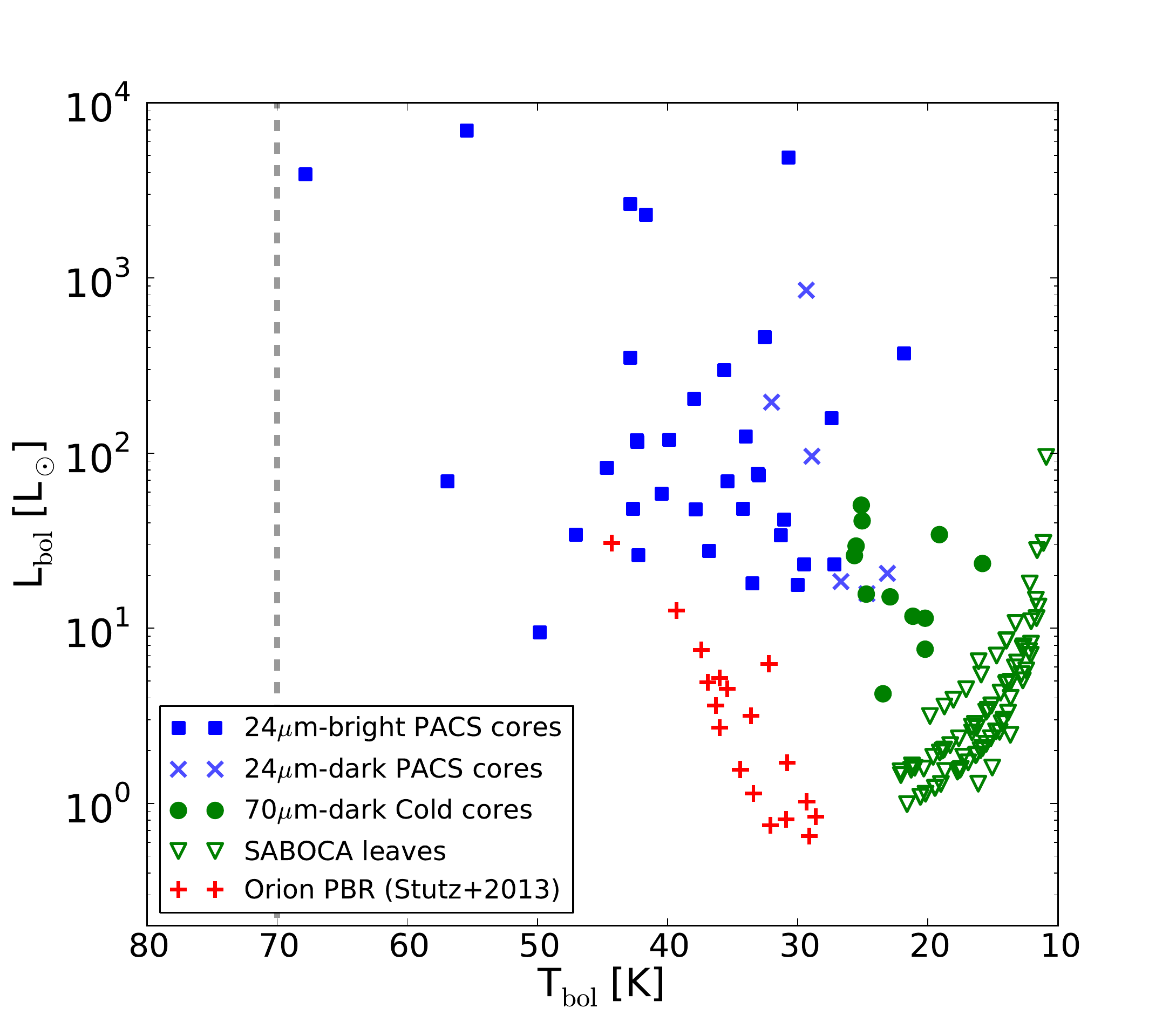}
\caption{\label{f:blt} Bolometric luminosity ($L_\mathrm{bol}$) plotted as a function of bolometric temperature ($T_\mathrm{bol}$). The cores detected in all PACS bands are plotted in blue (squares for 24\,$\mu$m-bright, $\times$ for 24\,$\mu$m-dark), and the 70\,$\mu$m-dark ``cold cores'' are plotted in filled green circles.  The green triangles represent upper limits for objects which were detected only at 350\,$\mu$m with SABOCA. For comparison, the PACS bright red objects (PBRs) in Orion \citep{Stutz2013} are shown in red crosses. The dividing line between Class I and Class 0 defined for low-mass protostars at 70\,K \citep{Andre1993} is shown for reference.}
\end{figure}

The bolometric temperature ($T_\mathrm{bol}$) and luminosity ($L_\mathrm{bol}$) are important evolutionary indicators of a protostar, and the bolometric luminosity-temperature (BLT) diagram \citep[e.g.][]{MyersLadd1993} can be used to compare protostars. Since {\em Herschel} samples the peak of the SEDs, we can reliably estimate both quantities \citep{Dunham2013} and place them into context.  

We calculate the bolometric temperature following \citet{MyersLadd1993}:

\begin{equation}
T_\mathrm{bol} = 1.25 \times 10^{-11}~\frac{\int~\nu S_{\nu}d\nu}{\int~S_{\nu}d\nu}~~\mathrm{K}.
\end{equation}

\noindent $T_\mathrm{bol}$ is integrated over the full SED and will reflect the contribution shortward of 70\,$\mu$m, thus the values are higher than the $T_\mathrm{cold}^\mathrm{full-SED}$ reported in Table~\ref{tab:coreprops}. Similarly, $L_\mathrm{bol}$ is integrated over all frequencies and will be higher than $L_\mathrm{cold}^\mathrm{full-SED}$, which includes the blackbody fit from 70\,$\mu$m \footnote{IRDC316.72-L38 is 24\,$\mu$m bright, but saturated in the MIPS band. For these calculations, we use the flux-luminosity relation presented in R12 (their Fig. 8) to estimate $S_{24{\mu}m}$ = 3.4\,Jy. }.  We plot the results in Figure~\ref{f:blt}, and we include upper limits for the SABOCA leaves with no {\em Herschel} counterpart. The entire population has $T_\mathrm{bol}$ below the 70\,K boundary for Class 0 objects in low-mass regions \citep{Andre1993}.  We plot 24\,$\mu$m-dark PACS cores, which have $T_\mathrm{bol} <$ 35\,K, but $L_\mathrm{bol}$ in the same range as 24\,$\mu$m-bright cores. This is consistent with our previous finding that both populations are in fact protostars, only 24\,$\mu$m-dark cores are either slightly less evolved (redder) or the warm dust nearest to the protostar is obscured due to geometry (e.g. a disk or region of excess extinction, see R12). 

Cores dark at 70\,$\mu$m (filled green circles) all have $T_\mathrm{bol} < 30$\,K, in the range of prestellar cores defined for the low-mass regime \citep[e.g.][]{YoungEvans2005}. We plot the upper limits of both $L_\mathrm{bol}$ and $T_\mathrm{bol}$ for SABOCA leaves with no {\em Herschel} counterpart at all (green triangles). There is a strong upper limit on  $T_\mathrm{bol}$ of 25\,K, though the $L_\mathrm{bol}$ of SABOCA leaves can span the same range as cold cores.

We plot the values for the \citet{Stutz2013} sample of PACS Bright Red Objects (PBRs) in Orion and find they are consistent with the protostellar cores but at lower luminosities. The IRDC sources presented here are significantly more luminous, as one must bear in mind that they are also much more distant than Orion.  Consequently, the probability that what we see as a ``core'' is in fact multiple embedded sources is higher.  Therefore, we emphasize that the $T_\mathrm{bol}$ divisions between classes of protostars that has often been used for low-mass regions (as is shown for reference in Figure~\ref{f:blt}) can not be straightforwardly applied here. The modeling of evolutionary tracks of high-mass sources on such a diagram, which requires a broadened consideration of accretion models, enhanced feedback, and the clustered environment of high-mass sources, is a growing area of research \citep[cf.][M.D. Smith 2013, submitted]{Hosokawa2010,Hosokawa2011}.

\begin{figure}
 \includegraphics[width=\linewidth]{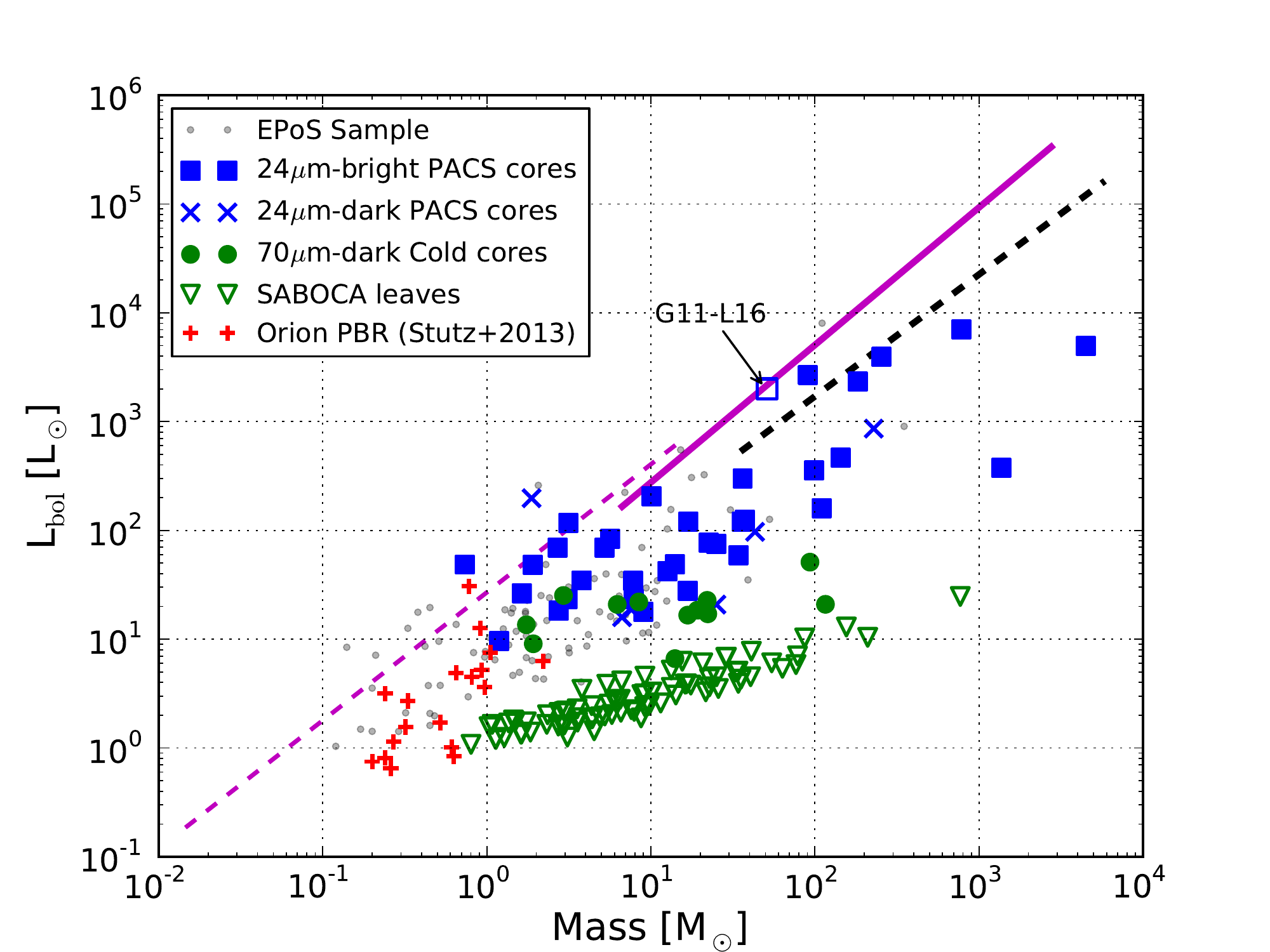}
 \caption{\label{f:lummass} $L_\mathrm{bol}$-Mass diagram for all cores in the EPoS sample (R12, gray dots), those protostellar cores (70\,$\mu$m-bright) recovered by SABOCA (blue squares), candidate starless cores (70\,$\mu$m-dark cold cores, green circles).  The $M_{350{\mu}m}$ (at the upper limit temperature) and upper limit to $L_\mathrm{bol}$ are plotted in the empty green triangles, also candidate starless objects. The UCHII region in IRDC\,011.11 (Leaf 16) is indicated in the empty blue square. We plot the Orion PBR sample \citep{Stutz2013} for comparison in red crosses.  The magenta lines are the empirical boundaries between ``Class 0''-like (below line) and ``Class II''-like (above line) found for high-mass \citep[solid magenta line][]{Molinari2008} and low-mass \citep[thin dashed magenta line][]{Saraceno1996}. The black dashed line is the empirical fit to the \citet{Molinari2008} ``MM'' sources, analogous to Class 0 in the low-mass regime. See text for details.}
\end{figure}

Figure~\ref{f:lummass} shows the mass-luminosity relation for the two core populations, plotted with the full distribution of PACS cores from R12. The mass used here is from the ``full-SED'' method of calculation for recovered cores and $M_{350{\mu}m}$ evaluated at their upper-limit temperature for the SABOCA leaves. We also plot the upper limits in $L_\mathrm{bol}$ for the SABOCA leaves, which lie already distinctly below the recovered core population. Cold cores have similar $L_\mathrm{bol}$ to PACS cores in their common mass range. 

For comparison, we plot the luminosities and masses derived in R12 for all cores in the eleven IRDCs of the present sample, and the sample of the youngest Orion protostars (PBRs) from \citet{Stutz2013}.  We show the empirical loci derived by \citet{Molinari2008} and \citet{Saraceno1996}. To mirror the familiar ``Class 0/I/II'' distinction in low-mass star formation, \citet{Molinari2008} put forth a high-mass analogy of ``MM/IR-P/IR-S'' sources based on their infrared and millimeter properties. The dashed magenta lines are the fits to Class I objects from \citet{Saraceno1996} and Molinari's high-mass equivalent ``IR-P'' source locus is plotted in the solid magenta line.  Sources above the loci are more evolved ``Class II'' (``IR-S'') sources.  Below the locus, earlier ``Class 0'' (``MM'') stages are found.  Based on these divisions, our PACS cores agree with the Class 0 (``MM'') regime, and Class I (``IR-P'') in a few cases, similar to the Orion PBRs but with higher masses and luminosities.  We note that the relations for high-mass objects were formulated based on IRAS, MSX, and sub-mm data of inhomogeneous resolution, but always with larger beams, thus the bolometric quantities are integrated over larger areas on average. 

The 70\,$\mu$m-dark cold cores (filled green circles) and the upper limits for SABOCA leaves (green triangles) occupy the region below $T_\mathrm{bol}$ = 30\,K. Due to their similarity in $T_{bol}$, masses, and densities discussed above, it is plausible that the {\em Herschel} cold cores are simply more luminous versions of the unrecovered SABOCA leaves at a similar evolutionary stage. We note that the bolometric luminosity not only scales with the mass but also can be enhanced by external irradiation \citep[e.g. ][R12]{Beuther_18454,Pavlyuchenkov2012}.

\begin{figure}
\includegraphics[width=\linewidth]{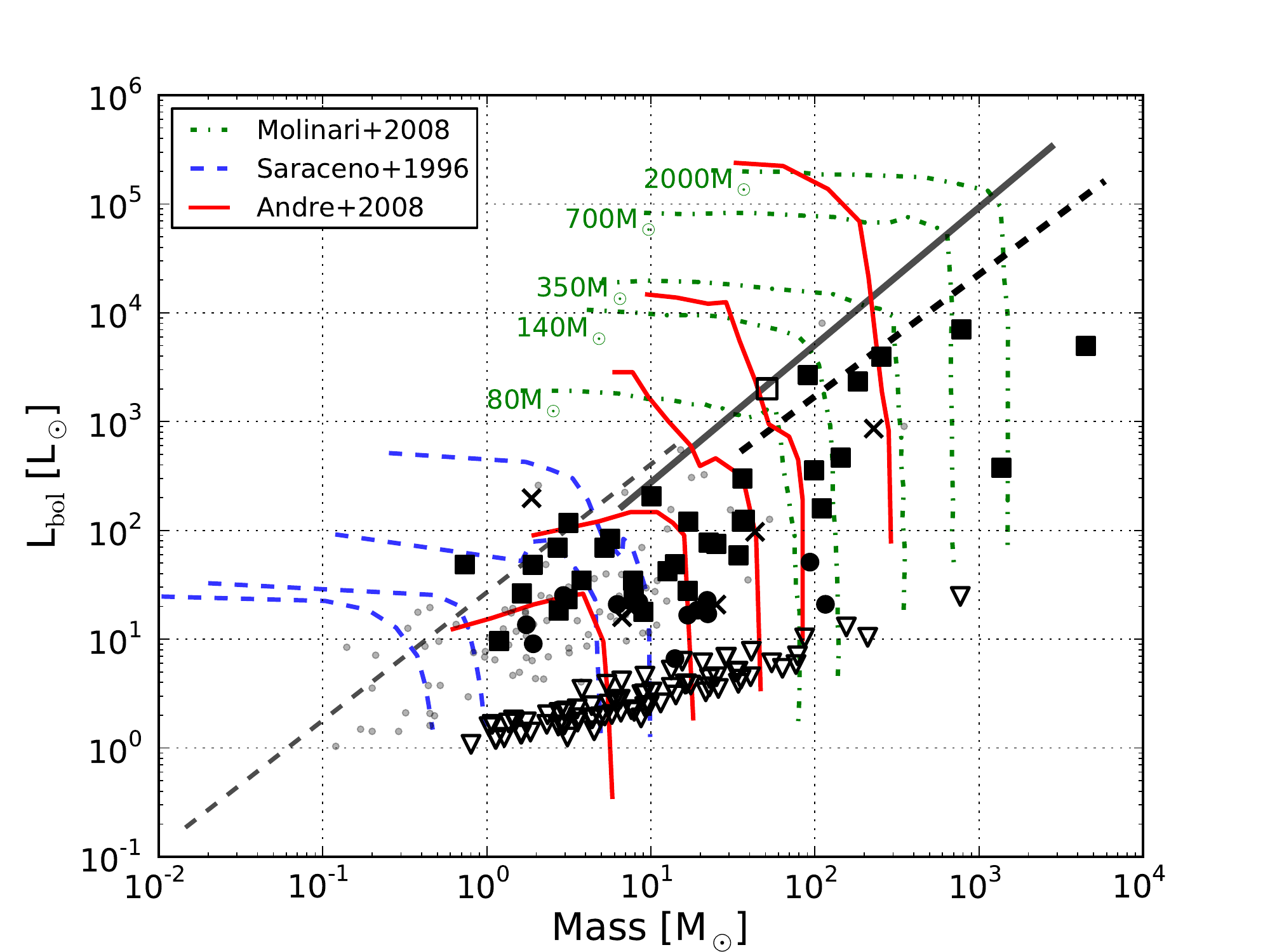}
\caption{\label{f:lummass_tracks} Same as Figure~\ref{f:lummass} (our data are shown in the corresponding black symbols for clarity), but with various evolutionary models overplotted. In red solid lines are the \citet{Andre2008} tracks for final stellar masses of 1,3,8,15, and 50$\msun$ (left to right). The \citet{Saraceno1996} tracks for initial masses of 0.5, 1, 2, and 4\,$\msun$ are shown in blue dashed lines, and the \citet{Molinari2008} tracks for initial envelope masses 80, 140, 350, 700, and 2000\,$\msun$ are shown in green dashed-dotted lines. The diagonal lines are the same loci fitting the Class I-like sources (grey dashed and solid) and ``MM'', Class 0-like soures (dashed black) as in Figure~\ref{f:lummass}.}
\end{figure}

Figure~\ref{f:lummass_tracks} shows the same $L_\mathrm{bol}$-Mass diagram but with various evolutionary tracks overplotted. The \citet{Molinari2008} tracks for the indicated initial envelope masses are shown in green, and the \citet{Saraceno1996} tracks for initial masses of 0.5, 1, 2, and 4\,$\msun$. These two models assume accelerating accretion scenarios.  For comparison, the \citet{Andre2008} decelerating accretion evolutionary tracks are shown for final stellar masses 1,3,8,15, and 50$\msun$. Our data all fall in the earliest phases of the evolution (from bottom right to upper left along the tracks), but indicate that the cores clearly represent the early stages of a large mass range of stars and clusters, depending on which accretion scenario is favored.

\subsection{Individual sources}
\label{ss:individual}

Our sample boasts great variety in structure. Below, we note the general properties of each region and highlight the most massive objects and other regions of interest.

\smallskip

\noindent {\it IRDC310.39-0.30:}~~ Two leaves are detected in this IRDC, the most massive one is coincident with a {\em Herschel} core, which has a mass of 320\,$\msun$ at $T$ = 17\,K. Leaf 1 reported in Table~\ref{tab:newcores} overlaps with two PACS cores (cores 1 and 2 from R12). For the purposes of SED fitting and bolometric quantity calculations, we use the luminosity ratio of those cores from R12 and scale the 350\,$\mu$m flux by 0.74 which is the fraction contributed by the main source.

\smallskip

\noindent {\it IRDC316.72+0.07:}~~ This complex filament hosts the greatest number of leaves and also the most overlapping with {\em Herschel} sources.  A small, bright, embedded cluster is known to be forming at the southeast end of the filament (main component leaf 38), and the infrared-dark filament to the northwest hosts several new cores. The leaves in the filament are of modest mass, though because of the several hierarchical levels to the filament, the corrections for parent flux are also the largest (see Table~\ref{tab:coreprops}).

A bright high-mass protostellar object (HMPO) resides at the southeast tip of the filament. Due the the prevalent diffuse infrared emission in the area, the three {\em Herschel} cores there may underrepresent the true population. SABOCA detects 15 leaves in the region, three of which can be associated with the known PACS cores and two are cold cores. We note all leaves in this region in Table~\ref{tab:newcores} and note that the non-detection of a corresponding {\em Herschel} core may be a consequence of confusion. Leaf 22 overlaps with the position of two PACS cores (cores 15 and 16 from R12), thus for SED-fitting, the 350\,$\mu$m flux is scaled by 0.42 which is the fraction of the total luminosity attributable to the closest matching core to the leaf position. 

This IRDC hosts six 70\,$\mu$m-dark {\em Herschel} cores, most of which have temperatures below 15\,K. For leaves 19, 24, and 25, the 100\,$\mu$m data point fits better with the uncorrected value of the 350\,$\mu$m flux, indicating that these cores may be closer to their upper-limit masses and lower-limit temperatures.

\smallskip

\noindent {\it IRDC320.27+0.29:}~~ Four leaves are detected in this IRDC, two of which correspond to {\em Herschel} cores in the main cloud.  While the {\em Herschel}/SPIRE images show a strong stream-like component to the west, it is very weak in the SABOCA images, indicating that it is a relatively flat or diffuse structure, whose structure was chopped away in the SABOCA observations.

\smallskip

\noindent {\it IRDC321.73+0.05:}~~ This IRDC hosts five leaves, 4 of which overlap with {\em Herschel} cores. The mass in this cloud is organized in three clumps, all centered on at least one {\em Herschel} core. Each of these main cores are recovered with SABOCA (the two northern clumps host leaves with masses of 28 and 36\,$\msun$), though the nearby, low-luminosity sources are not. 

\smallskip

\noindent {\it IRDC004.36-0.06:}~~ We recover none of the {\em Herschel} cores in this IRDC. All but the western-most leaf (id \# 4) appear as absorption features at 100\,$\mu$m, but only range between 1 and 6\,$\msun$. 

\smallskip

\noindent {\it IRDC009.86-0.04:}~~ The central {\em Herschel} source is recovered (id \# 2), and a new neighboring source to the north, though it appears faint in PACS bands, has similar properties.  Interestingly, the prominent absorption feature to the northeast, which strongly emits in NH$_3$\,(1,1) and (2,2) \citep[known tracers of cold, dense gas][]{Ragan2011a} is not recovered in SABOCA emission.

\smallskip

\noindent {\it IRDC011.11-0.12:}~~ A total of 35 leaves are found in this filamentary IRDC, 10 of which have {\em Herschel} counterparts. The central, dominant star forming region \citep[P1 in ][]{Johnstone_G11} corresponds to leaf 18 with a mass of 169\,$\msun$. Leaf 27 is a cold core with 74\,$\msun$ and $T$ = 12\,K. Leaf 16 has no {\em Herschel} counterparts, but has a high column density of 5.32 $\times$ 10$^{22}$ cm$^{-2}$ and 131\,$\msun$ if one takes the upper limit $T$ = 14\,K. 

Leaf 16 is a confirmed HII region \citep{Urquhart2009b}, and its associated $v_{lsr}$ of 31.3 km s$^{-1}$ is consistent with that measured in the IRDC. In R12, this diffuse source did not meet our PSF-fitting criteria, thus its SED was not fit. However, using literature values and bolometric flux measurement \citep{Mottram2011}, the bolometric luminosity is 2 $\times$ 10$^3 \lsun$ assuming the 3.4\,kpc distance. The SABOCA emission totals 12.6\,Jy, which equates to 51\,$\msun$ assuming 20\,K. Together, this places the source among the most evolved sources in the sample (see Figure~\ref{f:lummass}).

\smallskip

\noindent {\it IRDC015.05+0.09:}~~ There are 8 leaves in this IRDC, though none correspond to {\em Herschel} cores.  A 46\,$\lsun$ core sits in the center of the region, but does not overlap with the SABOCA emission.

\smallskip

\noindent {\it IRDC18223:}~~ This well-studied IRDC filament stems from IRAS 18223-1243 to the north, extends south, and is actually part of a much larger-scale bubble edge \citep{Tackenberg2013a}. Ten leaves are found in this IRDC, all but the faintest two correspond to {\em Herschel} cores.

\smallskip

\noindent {\it IRDC019.30+0.07:}~~ All three leaves are the main {\em Herschel} cores in the two main clumps in this IRDC. The main absorption feature visible in the 100\,$\mu$m image is not recovered in SABOCA emission.

\smallskip

\noindent {\it IRDC028.34+0.06:}~~ There are 20 leaves in this IRDC, eight of which correspond to {\em Herschel} cores. The two brightest leaves (2 and 9) correspond to MSX sources, which also have large mass corrections from the PACS mass derived in R12. This is most likely due to source confusion: in the PACS images, what appeared as an extended emission region in MSX was resolved into point sources. As is evident in Figure~\ref{fig:g2834}, the northern bright source (leaf 9) appears extended such that one leaf encompasses three PACS cores (cores 18, 19, and 20 from R12). For SED-fitting, the 350\,$\mu$m flux was scaled by 0.78, which is the fraction of the luminosity contributed by the closest matching core to the leaf position.  Leaf 6 is infrared-dark with a mass of 473\,$\msun$ adopting the upper-limit temperature of 14\,K.

\section{Discussion}

The objective of this study was to isolate the flux originating from the core such that we could better constrain the spectral energy distributions of pre- and protostellar cores that we originally characterized with {\em Herschel}. These cores are closely associated with large reservoirs of (relatively colder) dense gas and dust.  While {\em Herschel} allowed us to identify internally-heated cores, the high uncertainty in the mass measurement made it difficult to determine the total amount of mass in cores compared to the natal cloud. These observations enable us to quantify this relationship in IRDCs.  

We also characterize the earliest (possibly prestellar) stages of cores in IRDCs by combining the {\em Herschel}/PACS  and SABOCA data.  When a leaf corresponds to a PACS core (i.e. has a counterpart at 70, 100, and 160\,$\mu$m), we will refer to it as ``star-forming'' as it requires an internal heating source to produce the observed SED \citep[cf. R12,][]{Stutz2013}.  A leaf lacking a 70\,$\mu$m counterpart, either a ``cold core'' or normal ``leaf'', appear not to be star-forming. We discuss the caveats and considerations regarding this divide below.

\subsection{The nature of structure seen by SABOCA}
\label{ss:nature}

The hierarchical nature of molecular clouds is one of their defining characteristics, and our observations show that IRDCs are no exception. Even above our relatively high column density threshold ($\sim$10$^{22}$\,cm$^{-2}$), we find that most (61\%) leaf structures branch from a larger parent structure (see Figure~\ref{f:tree_diagram}). As is clear from Figures~\ref{fig:g31039} through \ref{fig:g2834}, there exists relatively diffuse extended structure that has been chopped away.  The full hierarchical structure of these clouds will be explored in a forthcoming publications (Ragan et al. in preparation).  Our \dendro~deconstruction enables us to quantify the contribution of the parent, large-scale structure to which a given leaf belongs (e.g. a clump or cloud). The fraction of mass in the core relative to the parent structure can vary from roughly equal levels to a core comprising less than 10\% of the flux.  This result challenges the assumptions underpinning the linear scaling of fluxes, which have been implemented to extract the SED from unresolved sources at these wavelengths \citep[e.g.][]{NguyenLuong2011,getsources}. 

In the following, we examine the physical properties of the population as a whole, then dividing the leaves between ``protostellar'' (coincident with a PACS core) and ``starless'' (everything else).  In Section~\ref{ssec:structure} we compute the average column density of each leaf (integrated flux divided by the area). Taking the leaf population as a whole, the mean column density is 2.9 $\times$ 10$^{22}$\,cm$^{-2}$. The starless leaves have lower column densities, averaging 2.5 $\times$ 10$^{22}$\,cm$^{-2}$ compared to those leaves with PACS core, the mean $N_{H_2}$ is 3.7$\times$ 10$^{22}$\,cm$^{-2}$.  

In comparing different methods for mass estimation, we find that the masses derived from SED-fits to the PACS data only (70 to 160\,$\mu$m) in R12 are within a factor of two of the redone SED-fit including the 350\,$\mu$m data point 63\% of the time, and is within a factor of five 83\% of the time. Encouragingly, the mass computed from the 350\,$\mu$m flux alone ($M_{350{\mu}m}$, Equation~\ref{eq:m350}) is always less than 20\% different than the full-SED method. The advantage of using the full-SED method is that one simultaneously fits for the temperature, which for these low temperature cores makes a big difference in the final mass estimates.  

Figure~\ref{f:massradius} shows the relation between leaf mass and effective radius on different scales. The median child leaf volume density is $4.4 \times 10^4$\,cm$^{-3}$ when the background parent structure emission is subtracted (plot a), $1.2 \times 10^5$\,cm$^{-3}$ if such a correction is not made (plot b). The background-corrected leaves follow a constant volume density locus in this diagram, while the uncorrected leaves, trunks, and whole cloud structures follow progressively shallower trends, reaching Larson's relation at the last step. Because we do not have kinematic data on the appropriate angular scales for these data, we reserve our discussion of cloud and core stability for a later paper. However, we note that the mass we derive is that above a high column density threshold ($\sim 10^{22}$\,cm$^{-2}$), so the factor that changes from plot to plot is the area by which the mass is divided (i.e. very little additional mass is included above that threshold as the area increases). This serves as a demonstration of some of the problems that have been found with the mass-size relation in mapping simulations to the observational plane \citep{BP2002, Shetty2010}.

We compare our sample of leaves to the concentration relation proposed by \citet{Kauffmann_masssize2}. Using the corrected mass values, we find fourteen leaves that their definition have the potential to form high-mass stars. Interestingly, not all of these leaves harbor signs of star formation. Seven of the 14 leaves indeed exhibit (at least one) associated PACS core, while the remaining 7 are either a cold core or a leaf with no {\em Herschel} counterpart at all.  These latter objects would be of special interest in pursuit of objects in the earliest ``quiescent'' phases of high-mass star formations, and follow-up kinematic studies are underway to explore their conditions. For now our observations only disclose good candidates for this elusive phase, and we will discuss what can be inferred about their lifetimes in Section~\ref{s:lifetimes}.

\subsection{Filamentary versus Clumpy IRDCs}
\label{ss:morph}

\begin{figure}
\includegraphics[width=\linewidth]{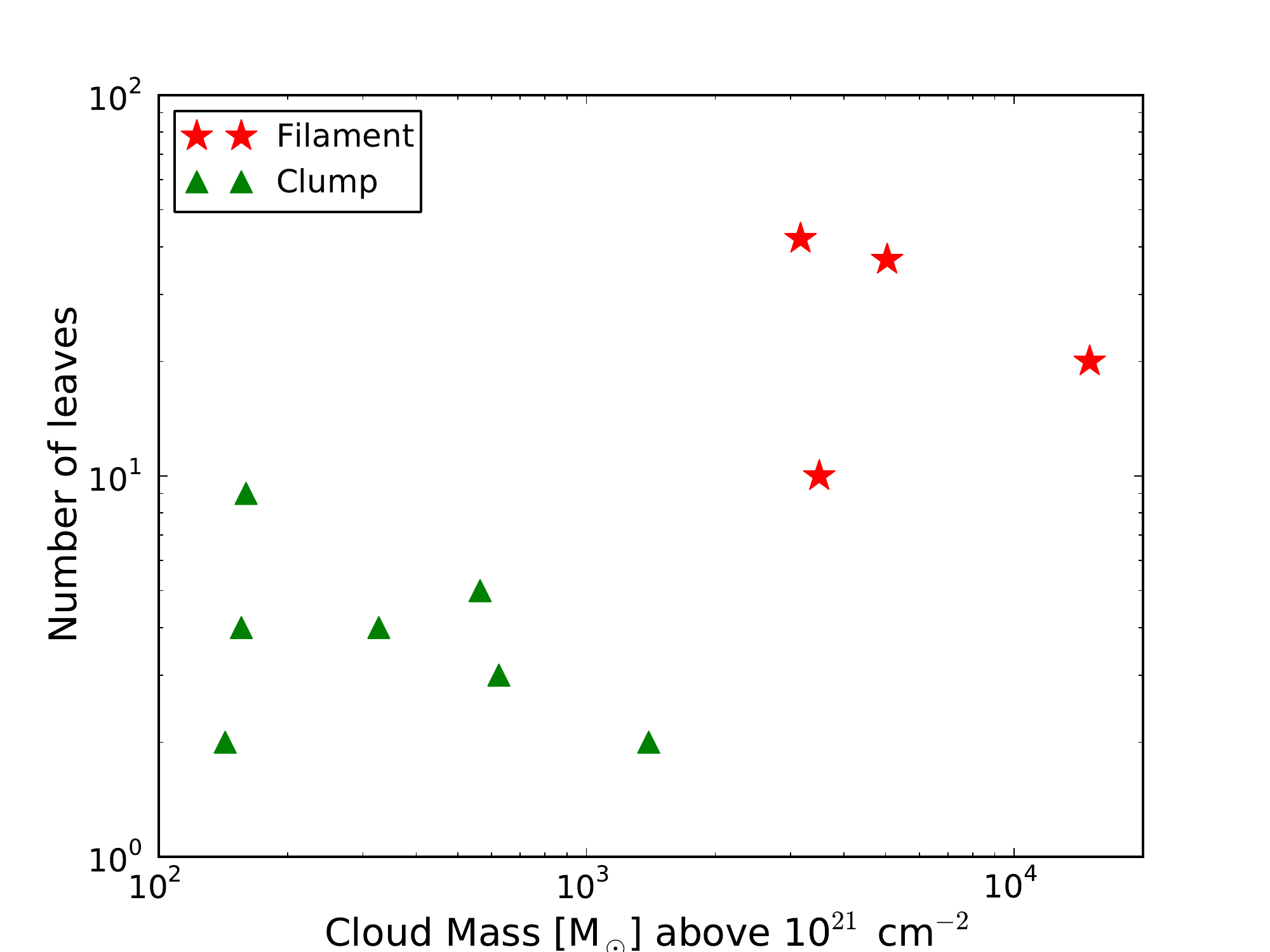}
\caption{\label{f:nsab_mass} The total number of leaves extracted in a cloud as a function of the total cloud mass (see Table~\ref{tab:obstable}). Filamentary IRDCs are plotted in red stars, and clumpy IRDCs are plotted in green triangles.}
\end{figure}

\begin{figure}
\includegraphics[width=\linewidth]{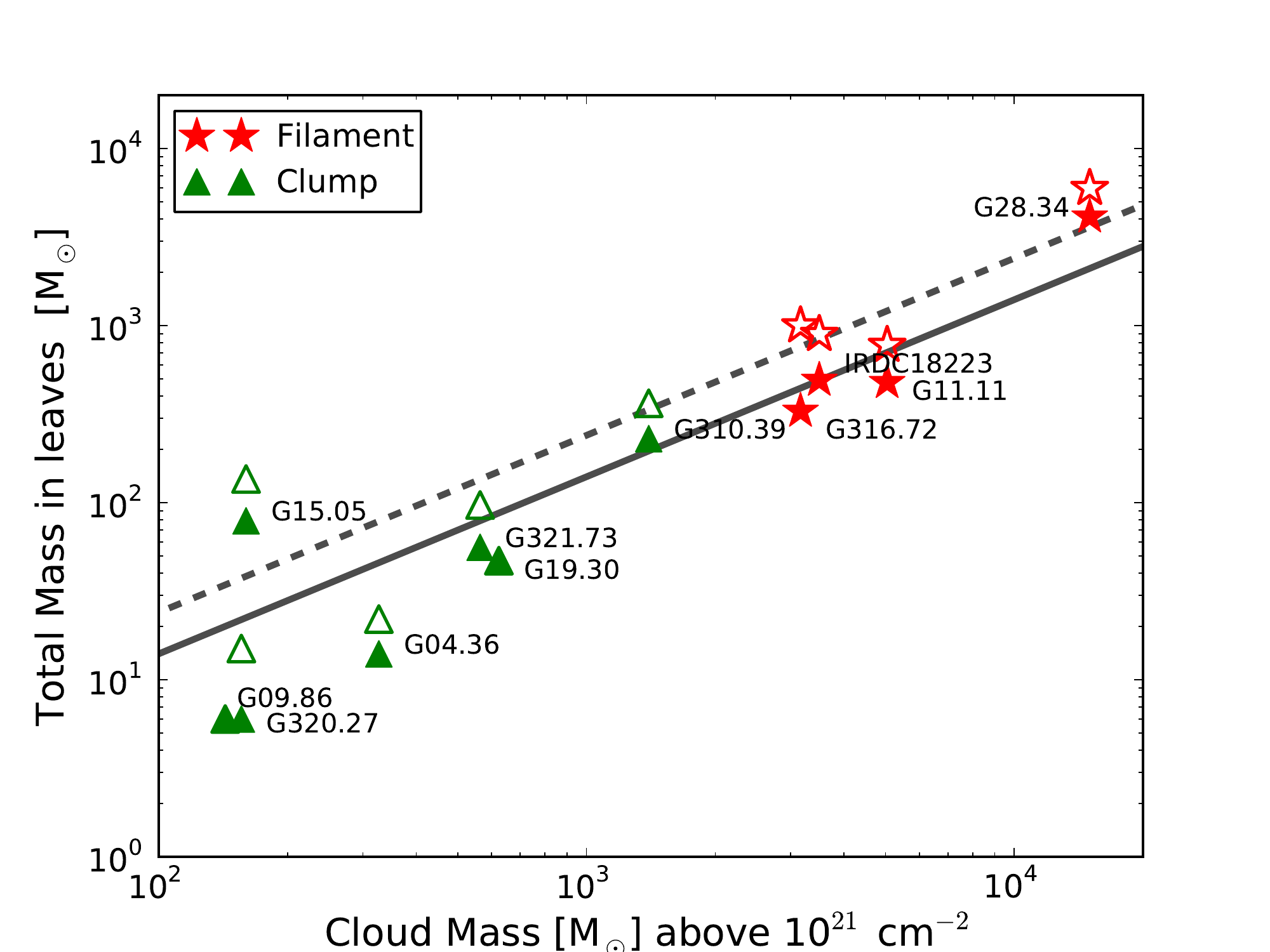}
\caption{\label{f:massinleaves} Total mass in dendrogram leaves (corrected for parent structure flux assuming 20\,K) as a function of the total mass of the cloud above a column density threshold $N_{H_2}\sim\,$10$^{21}$\,cm$^{-2}$ (see Table~\ref{tab:obstable}). Filamentary clouds are marked with filled red stars, and clumpy clouds are plotted in filled green triangles. The totals uncorrected for parent flux contributions are shown in the corresponding empty symbols. The solid line shows the average fraction (14\%) for mass values corrected for parent flux, and the dashed line shows the average for uncorrected mass values (24\%). }
\end{figure}

{\em Herschel} has ushered the return of filaments back into focus as an instrumental feature of star formation regions. Due to a confluence of effects -- from formation via the convergence of turbulent flows \citep[e.g.][]{Heitsch2008c} and enhancement through ionizing feedback \citep[e.g.][]{DaleBonnell2011} -- star formation is confined to the dense ridges and intersections of filaments \citep[cf.][]{Schneider2012, Hennemann2012}. Our sample features a mix of filamentary IRDCs and clumpy IRDCs, so we can qualitatively examine any differences between morphologies. 

The four prominent filaments in our sample are not only the most massive but are also the most complex in terms of their substructure, which is shown in Figure~\ref{f:nsab_mass}.  Here we plot the number of leaves detected in a cloud versus its mass, computed from ATLASGAL data, above the column density threshold of 10$^{21}$\,cm$^{-2}$.  These four IRDCs contain 107 of the 135 total leaves (79\%) in the sample. We note that the two nearest filaments (IRDC\,316.72 and IRDC\,011.11) have about twice as many leaves as the two farthest, although the difference in distance is only $\sim$1\,kpc.  We see no such difference in the clumpy IRDCs. Thus we do not appear biased to detect more substructure in the nearest clouds.

Based on this small sample, we can speculate on the importance of filaments in core formation in IRDCs. In Figure~\ref{f:massinleaves}, we show the amount of mass in leaves as a function of the total cloud mass. Here we assume a uniform temperature for all leaves (20\,K) to be consistent with the assumptions going into the total cloud mass calculation. Overall, an average of 14\% of the cloud mass is found in leaves. Looking at the cloud types separately, we find hardly any difference: filaments have 15\% in leaves, clumps have 13\%. The trend is similar if one takes the leaf masses uncorrected for parent flux contributions instead: 24\% of the total mass in leaves, 28\% in filaments and 21\% in clumpy clouds. We conclude that leaves form as efficiently in filaments as in clumpy IRDCs. It may be the case that filamentary structure is the best way to accumulate large masses, and subsequently more cores (leaves), coherently before massive stars form and destroy the cloud.

\subsection{Starless and protostellar cores in IRDCs}

Our SED-fitting allows us to compute useful quantities to place our sources in an evolutionary context. Figures~\ref{f:blt} and \ref{f:lummass} show how the cores and leaves compare to other known populations of objects in their masses, bolometric temperatures and luminosities. Our sample falls entirely within the ``Class 0''-like regime. Compared to the youngest protostars in Orion \citep{Stutz2013}, the nearest region of high-mass star formation, the PACS cores appear to be at a similar evolutionary stage.

Cold cores and SABOCA leaves lacking a {\em Herschel} counterpart are alike in their $T_\mathrm{bol}$ and mass range, but differ in $L_\mathrm{bol}$.  We hypothesize that SABOCA leaves are simply the low-luminosity population of objects in roughly the same or slightly earlier evolutionary stage as cold cores.  The infrared properties, $L_\mathrm{bol}$ and $T_\mathrm{bol}$, of both populations indicate that they are candidate starless cores which require further observations to confirm their status.  

While (due to sensitivity limits) we do not recover all cores reported in R12 with our SABOCA maps, nor do all structures found in the SABOCA images correspond to {\em Herschel} cores, we can consider these populations together as an ensemble of substructures of IRDCs at different evolutionary stages.  For example, various evolutionary tracks have been proposed \citep[e.g.][]{Saraceno1996, Molinari2008, Andre2008} which directly connect an initial phase among the SABOCA leaves to the PACS cores a few 10$^5$\,yr later (see Figure~\ref{f:lummass_tracks}).  Furthermore, based on the BLT evolutionary diagram (see Figure~\ref{f:blt}), SABOCA leaves and cold cores have $T_\mathrm{bol} < 30$\,K, which is the regime for ``prestellar'' cores \citep[e.g. ][]{YoungEvans2005}, which supports our assessment that these objects could represent the initial phase.

\subsection{Core lifetimes}
\label{s:lifetimes}

One elusive quantity needed to constrain theoretical models is the lifetime of the earliest ``starless'' phase ($\tau_{SL}$) for high-mass star forming cores/clumps \citep{Motte_cygX, Chambers2009, Miettinen2012, Tackenberg2012}. For such a calculation, we first assume that objects that are mid-infrared (MIR) dark ($\lambda < 100\,\mu$m) are starless, and cores bright at 70\,$\mu$m (our PACS cores) are actively forming protostars, which has been shown as a good diagnostic for embedded protostars \citep[][R12]{Dunham2008, Stutz2013}. This differs slightly from commonly adopted criteria of the previous studies that lacked supporting {\em Herschel} data. For instance, by our definition, an object is only considered ``starless'' if it lacks counterparts at 70\,$\mu$m and shortward, whereas the previous studies use 8 or 24\,$\mu$m as the cutoff.  Consequently, a 70\,$\mu$m bright object with no shortward counterpart, which constitutes 34\% of the protostars in R12, would be counted as starless in the previous studies. 

On the other hand, some previous studies probe for star formation activity in different ways which do not depend on the critical 70\,$\mu$m data. While {\em Spitzer} may have missed the deeply embedded protostars that {\em Herschel} now captures, indirect signposts, such as molecular outflows traced by SiO \citep[e.g.][]{Motte_cygX} and excess 4.5\,$\mu$m emission \citep[e.g.][]{Cyganowski2008, Chambers2009}, help to provide a more complete census of star formation. Currently, such data for our sample are not available, though this would be a useful follow-up for the starless candidates. 

While our measure of embedded star formation is imperfect, and the sensitivity of our SABOCA leaves is limited, we can compute the relative populations of protostellar to starless objects and thus an upper limit to the lifetime of the starless phase for the most massive objects. Because we lack spectral information at shorter wavelengths (in part due to high extinction and poor sensitivity in distant clouds such as IRDCs) which in nearby clouds provide empirical divisions between Class 0/I/II protostellar stages in nearby clouds, we do not further refine the protostellar classification scheme here. However, based on our evolutionary diagnostics, we assume that the PACS cores are in the high-mass protostellar object (HMPO) stage \citep{sridharan_irdc, Beuther2002, Motte_cygX}, previous to the formation of an HII region.

In the simplest case, if we assume that all the candidate starless objects will ultimately form protostars, and the rate of this progession is independent of the core mass, then the ratio of the number of starless to protostellar cores is the same as the ratio of the lifetimes \citep[see][]{Enoch2008}.  If we assume that our protostars are predominantly in the ``Class 0''-like phase as our evolutionary diagnostics indicate (see Figures~\ref{f:blt}, \ref{f:lummass}, and \ref{f:lummass_tracks}), we can adopt the 0.10\,Myr lifetime of Class 0 source found in local clouds \citep{Evans2009} and derive a lifetime of starless phase of 0.17\,Myr, consistent with other estimates of ``IR-dark'' lifetimes with equivalent assumptions \citep[e.g.][]{Chambers2009, Wilcock2012a, Miettinen2012}.  Considering only the massive objects ($M_{350{\mu}m} >$ 40$\msun$), the starless lifetime is only 0.07\,Myr. 

These assumptions carry several caveats when applied to regions of higher mass. For example, it has been shown that cores/clumps of higher masses have shorter lifetimes \citep[e.g.][]{Motte_cygX, Hatchell2008}.  The HMPO lifetime in Cygnus X is 3.2 $\times$ 10$^4$\,yr \citep{Motte_cygX}. Adopting this timescale for the massive cores, then the upper limit to the starless lifetime is 2.1 $\times$ 10$^4$\,yr. This is consistent with the findings of \citet{Tackenberg2012} study which focused on the high-mass population of IRDC clumps, though not as short as that derived for the 1.7\,kpc distant Cygnus X, which is found to be below 10$^3$\,yr.

Whether all of our candidate ``starless'' leaves are (a) actually starless and (b) will ultimately collapse and form stars is another over-simplification of the situation. If either of these assertions is wrong, the starless lifetime could decrease significantly. Furthermore, the lifetimes are shorter than the estimated free-fall times ($\tau_{ff} \sim$ 0.1 - 0.2\,Myr) for these objects, which argues against a quasi-static paradigm of core evolution. A detailed investigation of the cloud dynamics on small scales is needed to understand these early phases.

\section{Summary and Conclusions}

We present APEX/SABOCA 350\,$\mu$m maps of eleven IRDCs. These 7.8$''$ resolution maps probe the Rayleigh-Jeans tail of the spectral energy distribution of the cold dust comprising IRDCs on similar scales as our previous PACS observations of these clouds.  We deconstruct the SABOCA emission using the \dendro~algorithm, which enables us to differentiate flux on large and small scales.  We isolate the smallest structures, known as  ``leaves'', atop the \dendro~tree and connect them to the structure seen in the {\em Herschel} observations. These structures correspond to the upper boundary to the ``core'' size scale \citep{BerginTafalla_ARAA2007}, but due to the large distances, we can not say these are structures on their way to forming a single star, but rather more likely multiple stars or clusters.

We recover the cores that are found both in emission and absorption in the original {\em Herschel} study.  We construct SEDs of the recovered cores including the PACS data at 70, 100, and 160\,$\mu$m plus the new data at 350\,$\mu$m.  Cores with components from 70-350\,$\mu$m are termed ``PACS cores'' and those with counterparts at 100-350\,$\mu$m (and not 70\,$\mu$m) are termed ``cold cores.''  We find that the core SEDs are well-fit by modified blackbody functions under the optically thin assumption. With the SED properties, including bolometric luminosities and temperatures, we place the cores in a broader evolutionary context. We summarize our main conclusions below.

\begin{enumerate}

\item The overall dense structure of the clouds known from {\em Herschel} and previous measurements is nicely recovered by SABOCA at 350\,$\mu$m.  Our \dendro~analysis reveals hierarchical structure, beginning with ``branches'' and ``trunks'' on large scales and leaves on the smallest scales. The structure of our four filamentary clouds is considerably more complex than in the clumpy IRDCs, which exhibit a more monolithic nature. 

\item  In the eleven clouds, at total 135 \dendro~``leaf'' structures are found.  Forty leaves are directly associated with a {\em Herschel} core from the R12 catalog, and twelve leaves overlap with 70\,$\mu$m-dark {\em Herschel} cores and have been modelled here for the first time. The remaining 83 leaves show no indication of a {\em Herschel} counterpart. The latter two populations are considered as candidate starless core populations. 

\item Using \dendro~provides information about the larger structures in which cores reside. We correct the fluxes used in our measurements for this large-scale ``parent'' flux, which ranges from a factor of 0.5 to 15, even within a given cloud. The median value of the correction is 4.8, and the factor is independent of the leaf's surface brightness. Taking these corrections into account, the core masses computed from the SABOCA flux alone ($M_{350{\mu}m}$) are always within 20\% of the mass found from SED-fitting ($M_\mathrm{full-SED}$).

\item While filamentary IRDCs are the most massive of our small sample and exhibit the most substructure, the fraction of the total cloud mass above $N_{H_2} \sim 10^{21}$\,cm$^{-2}$ residing in ``leaf'' structures is roughly the same as that found in clumpy IRDCs. The average fraction of the total cloud mass found in leaves is 14\%.

\item To study the prospect of {\em high-mass} star formation in these cores, we examine their concentration in the mass-radius relation. Fourteen leaves (10\% of the sample) meet the criteria concentration criteria set by \citet{Kauffmann_masssize2}, eight of which correspond to {\em Herschel} (PACS and cold) cores. Ten of these 12 leaves occur in the filamentary IRDCs in the sample. These are promising candidates from our sample for the pristine initial phases of high-mass stars. The remaining 121 leaves are at various early phases of intermediate or low-mass star formation.


\item On the smallest size scales, cores exhibit a rough trend of constant volume density ($M_{350{\mu}m} \propto r_\mathrm{eff}^2.9$), where on larger ``clump'' and ``cloud'' scales, the relation flattens to match a locus of constant column density ($M_{350{\mu}m} \propto r_\mathrm{eff}^2.3$).  We assert that these relations rely strongly on the boundaries drawn around the structures of interest.

\item For the 52 leaves overlapping with a {\em Herschel} point source, we fit the SEDs with blackbodies. We derive bolometric temperatures and luminosities and find that all cores fall below the $T_\mathrm{bol} < $70\,K definition for Class 0 sources, though the luminosities are significantly higher than Class 0 sources in nearby Gould Belt clouds. 

\item Our evolutionary indicators show that SABOCA leaves (with no {\em Herschel} counterpart) may simply be low-luminosity cold cores. Both populations show no indication in the infrared of embedded star formation, similar mass ranges, (upper-limit) bolometric temperatures below 30\,K, which is the fiducial mark below which cores are catergorized ``starless '' or ``prestellar.''  Based on evolutionary models in the literature, the leaves are in the earliest phases of what might become solar mass stars up to stars of 50\,$\msun$ or greater.

\item We calculate an upper limit to the lifetime of massive ($M > 40\,\msun$) starless cores of $2.1 \times 10^4$\,yr, based on statistical lifetimes of high-mass protostars, significantly shorter than starless timescales in low-mass clouds. 

\end{enumerate}

\begin{acknowledgements}
The authors thank the anonymous referee whose comments led to significant improvements to this paper.
SR is grateful to Frederic Schuller for his assistance in SABOCA data reduction, Thomas Robitaille for help with the Python {\tt astrodendro} package, Anika Schmiedeke for {\em Herschel} data reduction, and Amelia Stutz for useful discussions. SR is supported by the Deutsche Forschungsgemeinschaft priority program 1573 (``Physics of the Interstellar Medium''). This research has made use of NASA Astrophysics Data System. This research made use of APLpy, an open-source plotting package for Python hosted at http://aplpy.github.com
\end{acknowledgements}



\appendix

\section{Image gallery}
\label{s:imgal}

\begin{figure*}
\begin{center}
\begin{tabular}{lr}
\includegraphics[width=0.46\linewidth]{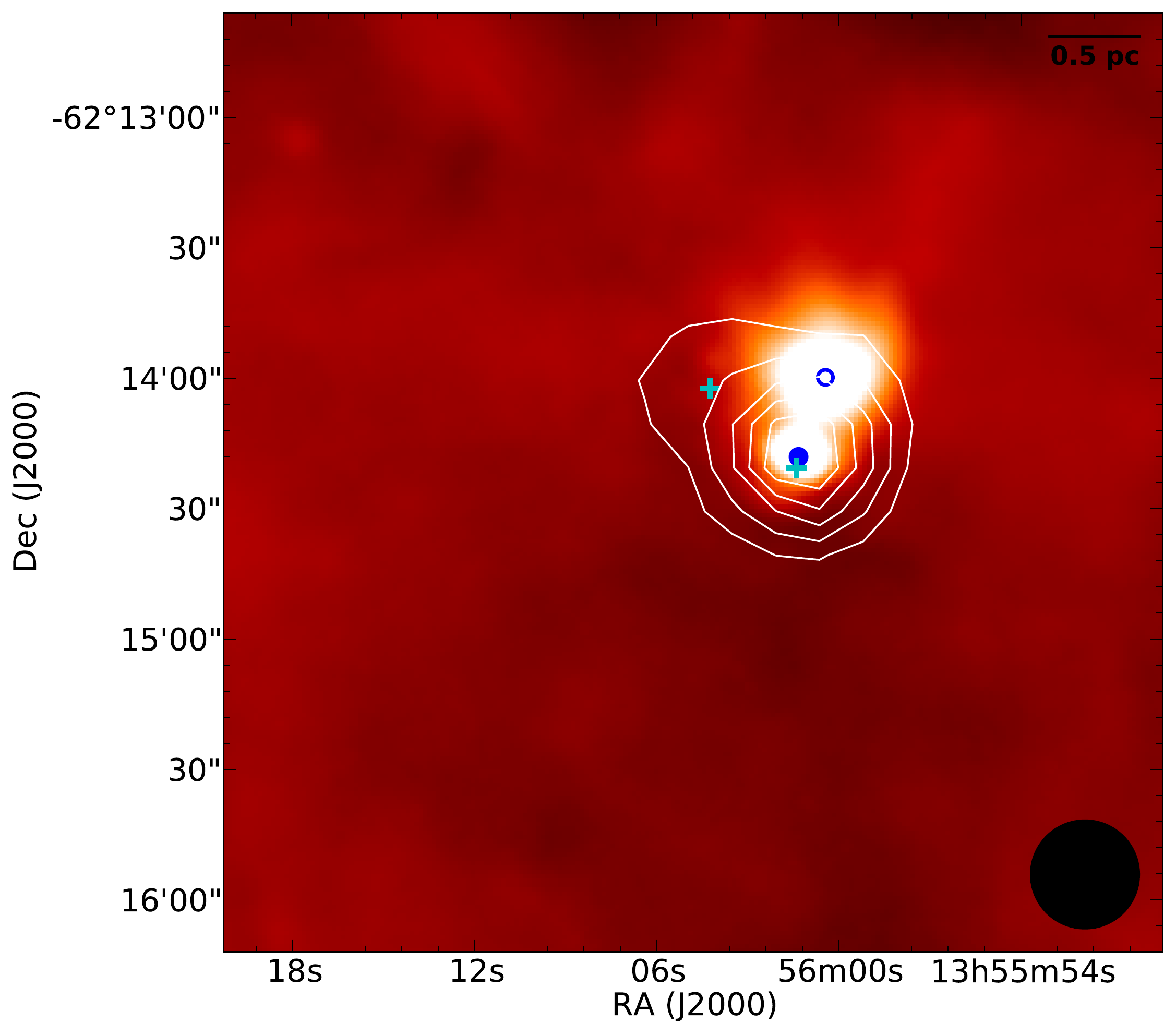} & \includegraphics[width=0.47\linewidth]{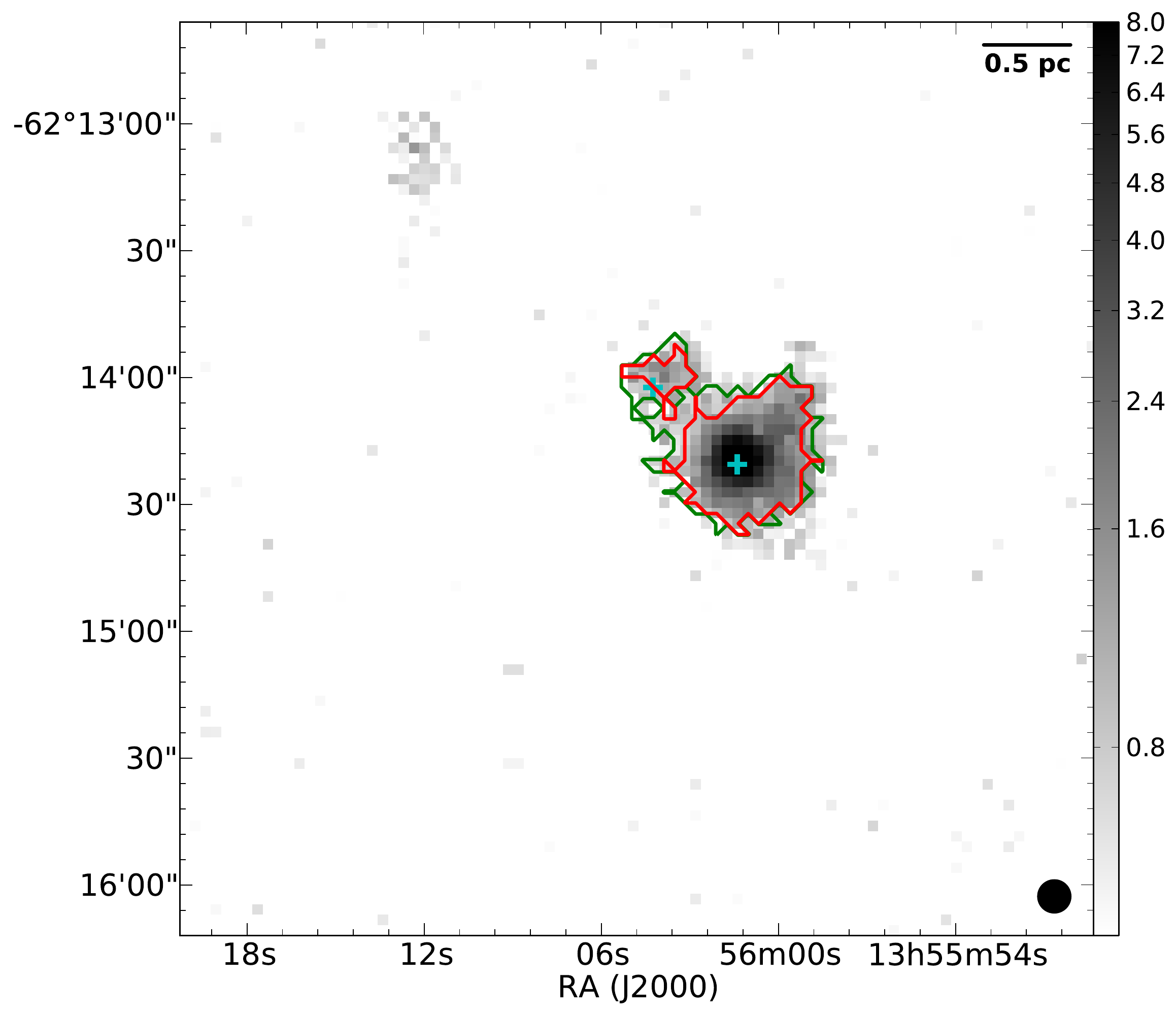}
\end{tabular}
\end{center}
\caption{IRDC310.39-0.30: {\it Left:} The image is the PACS 100\,$\mu$m with SPIRE 350\,$\mu$m contours overplotted. 
The white contour levels begin at 14\, Jy beam$^{-1}$ and increase in steps of 5\, Jy beam$^{-1}$. 
The blue circles are PACS cores identified in \citet{Ragan2012b}, and green circles are candidate 70\,$\mu$m-dark cores. A circle is filled if it is recovered as a ``leaf'' with SABOCA, and it is left open if it is not. Cyan crosses are the positions of {\it all} \dendro~ leaves.
{\it Right:} SABOCA 350\,$\mu$m image is plotted in greyscale with red contours showing the ``leaf''  structures identified with \dendro~ and green contours show lower-level structures (``branches'') in emission, if any.  The scalebar is in units of Jy beam$^{-1}$.  See Section~\ref{sec:dendro} for details.
\label{fig:g31039}}
\end{figure*}


\begin{figure*}
\begin{center}
\begin{tabular}{lr}
\includegraphics[width=0.46\linewidth]{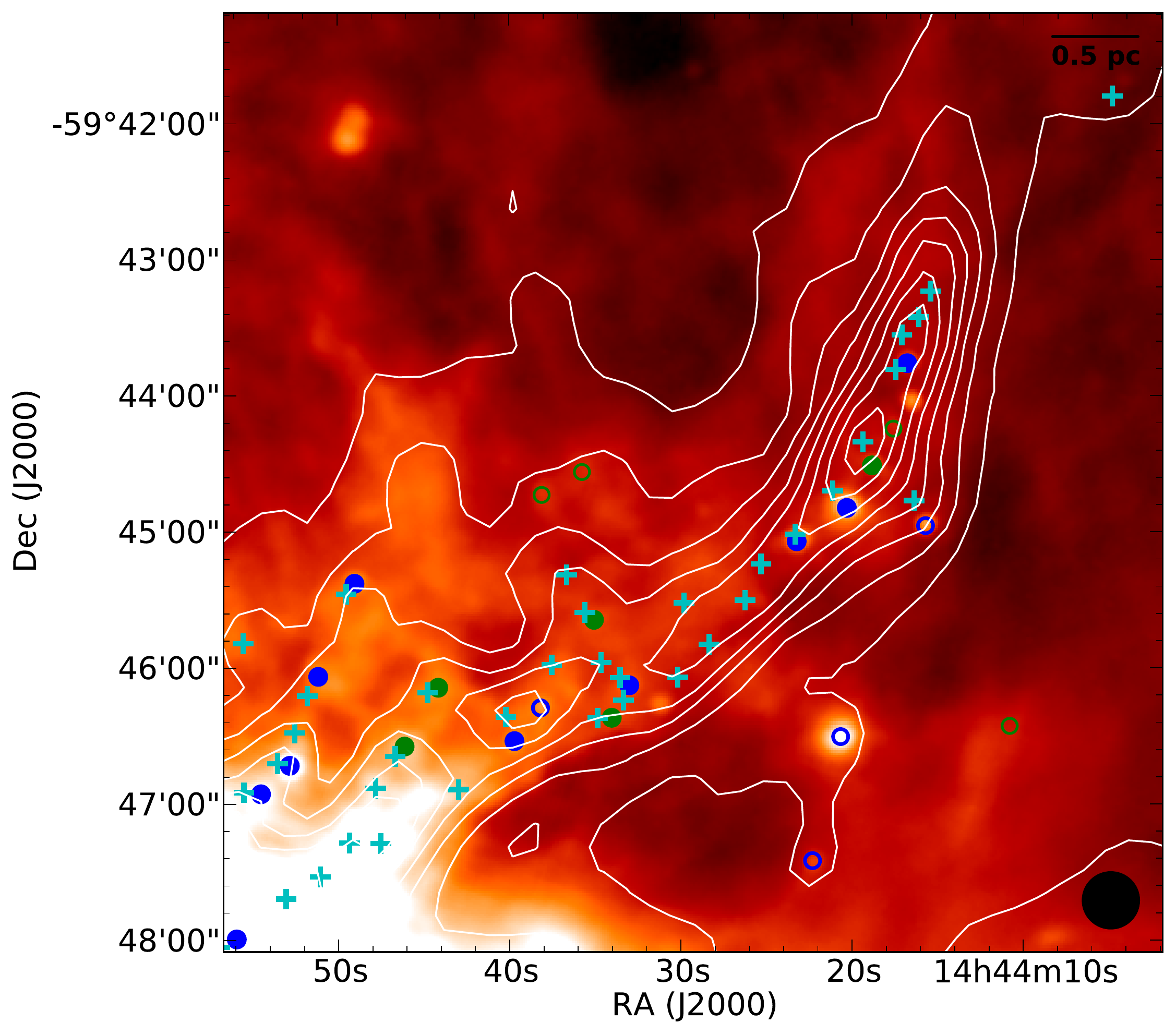} &
\includegraphics[width=0.47\linewidth]{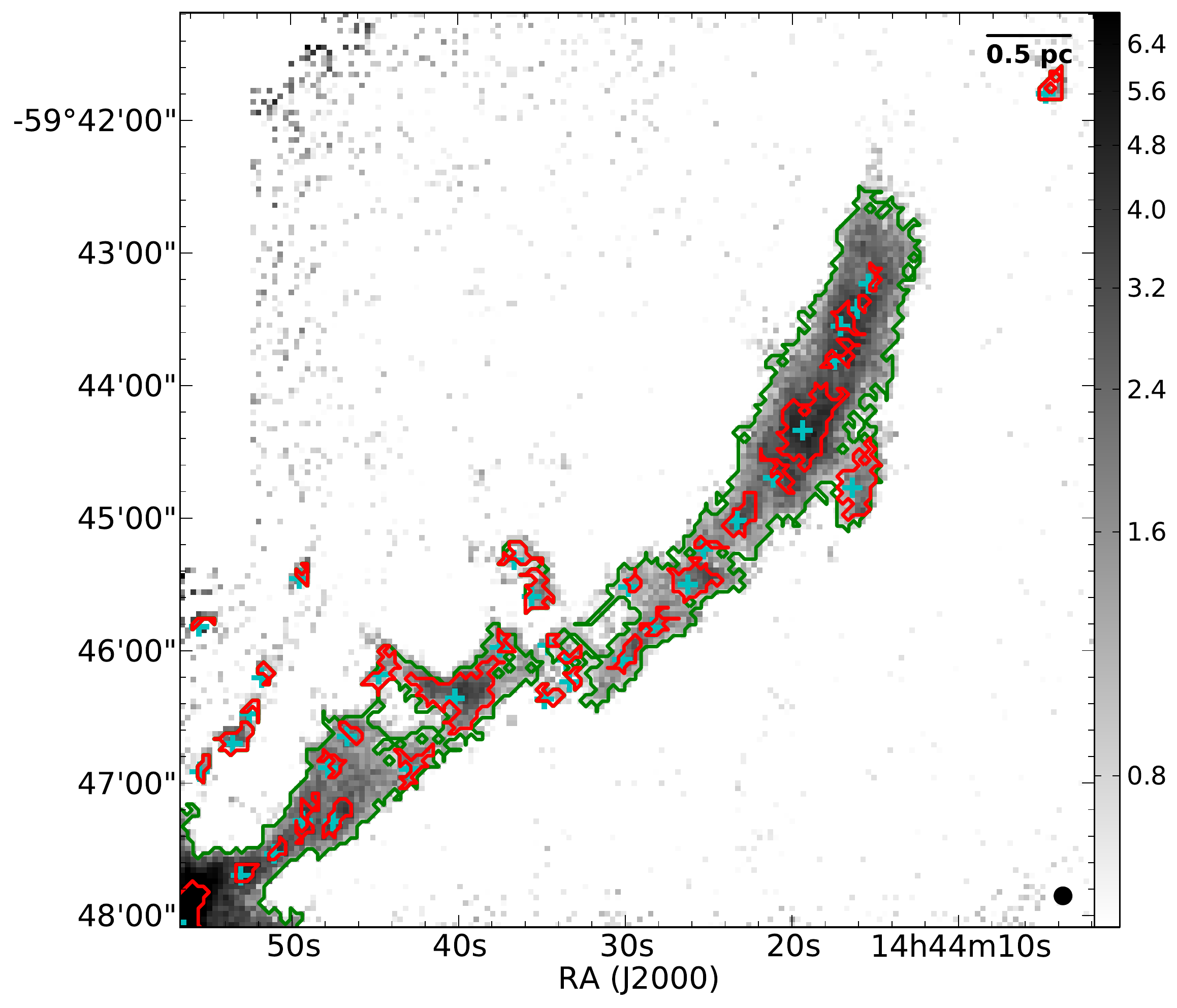}
\end{tabular}
\end{center}
\caption{IRDC316.72+0.07: {\it Left:} The image is the PACS 100\,$\mu$m with SPIRE 350\,$\mu$m contours overplotted. 
The white contour levels begin at 2\, Jy beam$^{-1}$ and increase in steps of 4\, Jy beam$^{-1}$.
The blue circles are PACS cores identified in \citet{Ragan2012b}, and green circles are candidate 70\,$\mu$m-dark cores. A circle is filled if it is recovered as a ``leaf'' with SABOCA, and it is left open if it is not. Cyan crosses are the positions of {\it all} \dendro~ leaves.
{\it Right:} SABOCA 350\,$\mu$m image is plotted in greyscale with red contours showing the ``leaf''  structures identified with \dendro~ and green contours show lower-level structures (``branches'') in emission, if any.  See Section~\ref{sec:dendro} for details.
\label{fig:g31672}}
\end{figure*}

\clearpage

\begin{figure*}
\begin{center}
\begin{tabular}{lr}
\includegraphics[width=0.46\linewidth]{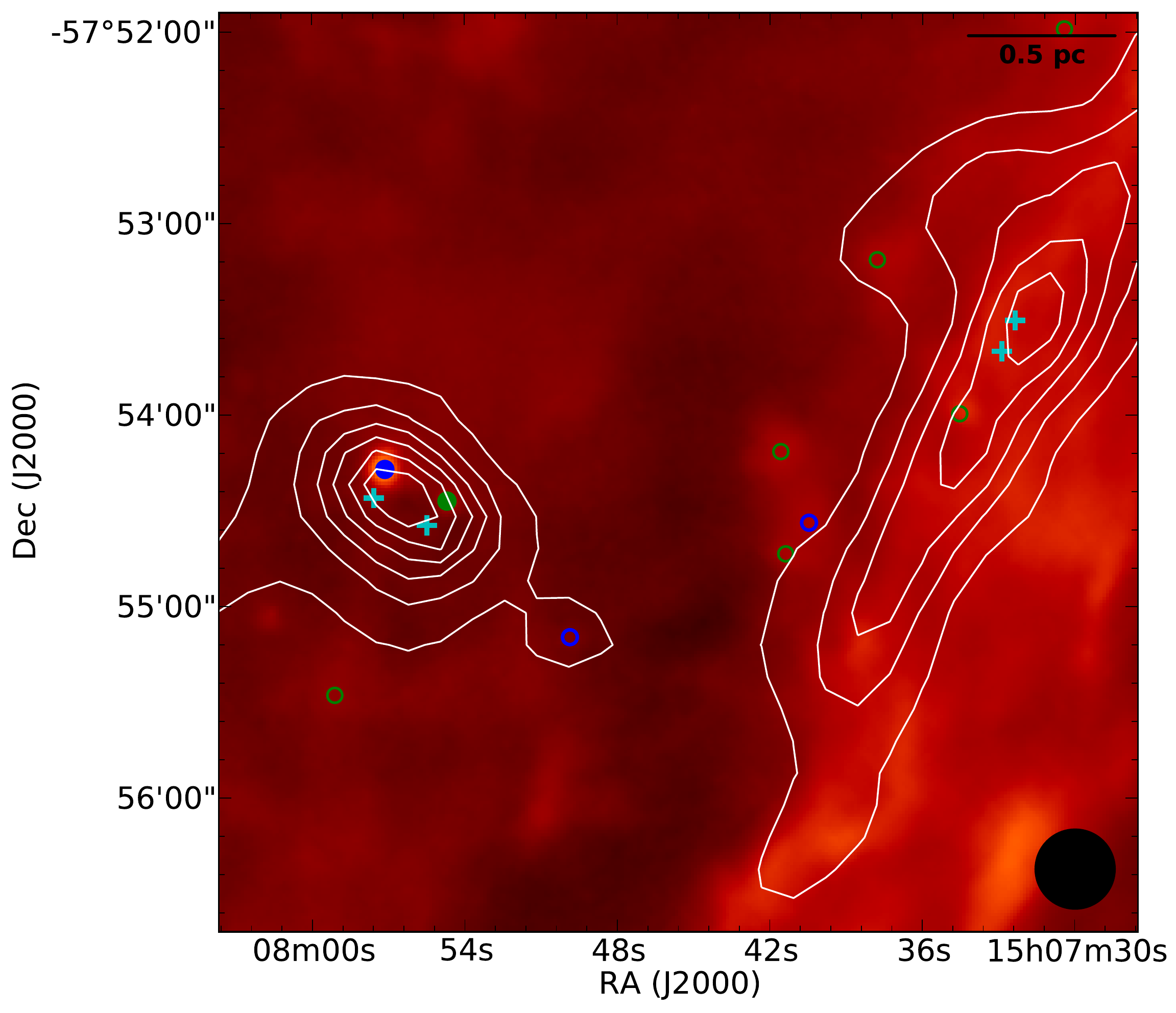} & \includegraphics[width=0.47\linewidth]{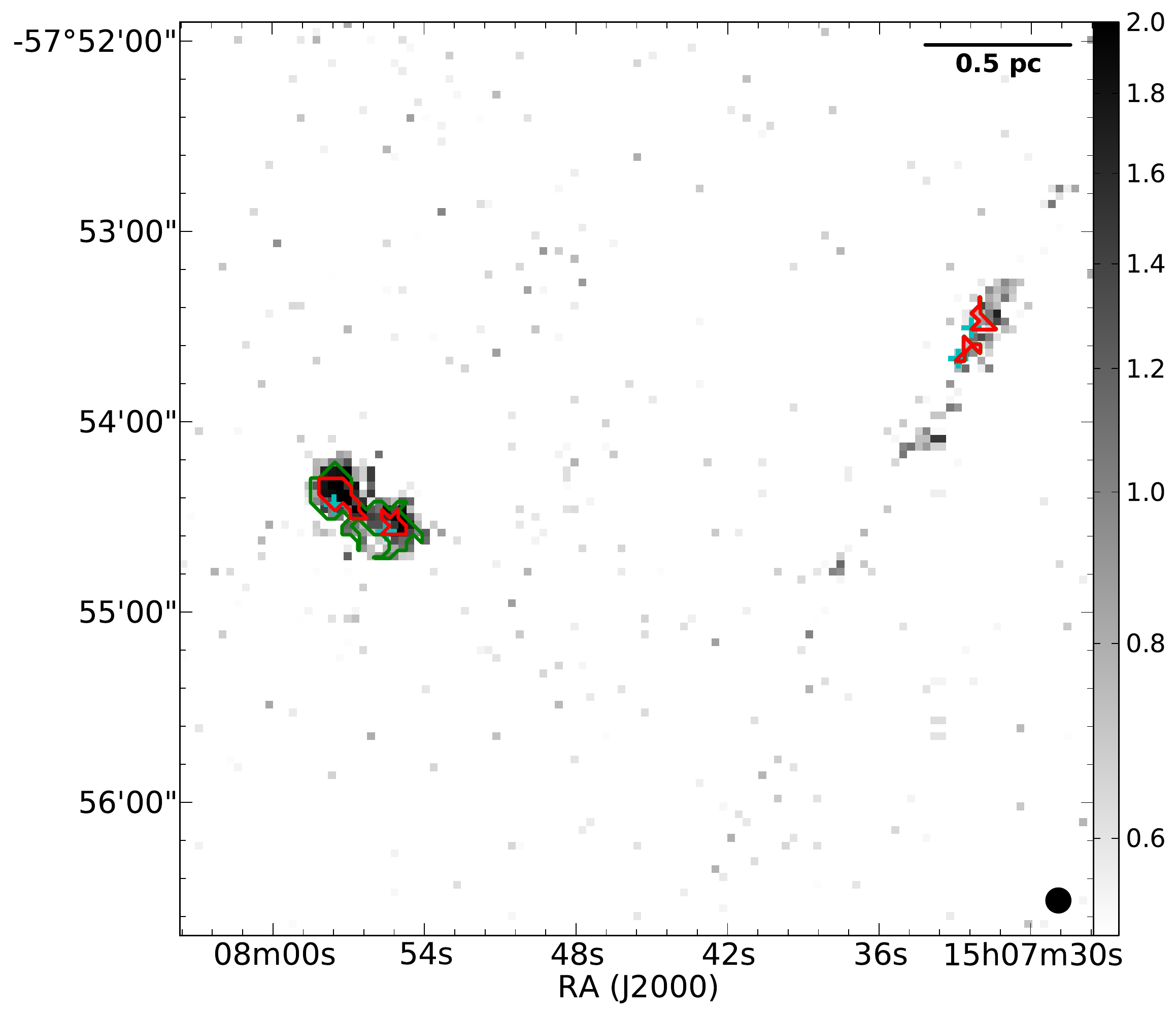}
\end{tabular}
\end{center}
\caption{IRDC320.27+0.29: {\it Left:} The image is the PACS 100\,$\mu$m with SPIRE 350\,$\mu$m contours overplotted. 
The white contour levels begin at 3\, Jy beam$^{-1}$ and increase in steps of 1\, Jy beam$^{-1}$.
The blue circles are PACS cores identified in \citet{Ragan2012b}, and green circles are candidate 70\,$\mu$m-dark cores. A circle is filled if it is recovered as a ``leaf'' with SABOCA, and it is left open if it is not. Cyan crosses are the positions of {\it all} \dendro~ leaves.
{\it Right:} SABOCA 350\,$\mu$m image is plotted in greyscale with red contours showing the ``leaf''  structures identified with \dendro~ and green contours show lower-level structures (``branches'') in emission, if any.  See Section~\ref{sec:dendro} for details.
\label{fig:g32027}}
\end{figure*}


\begin{figure*}
\begin{center}
\begin{tabular}{lr}
\includegraphics[width=0.46\linewidth]{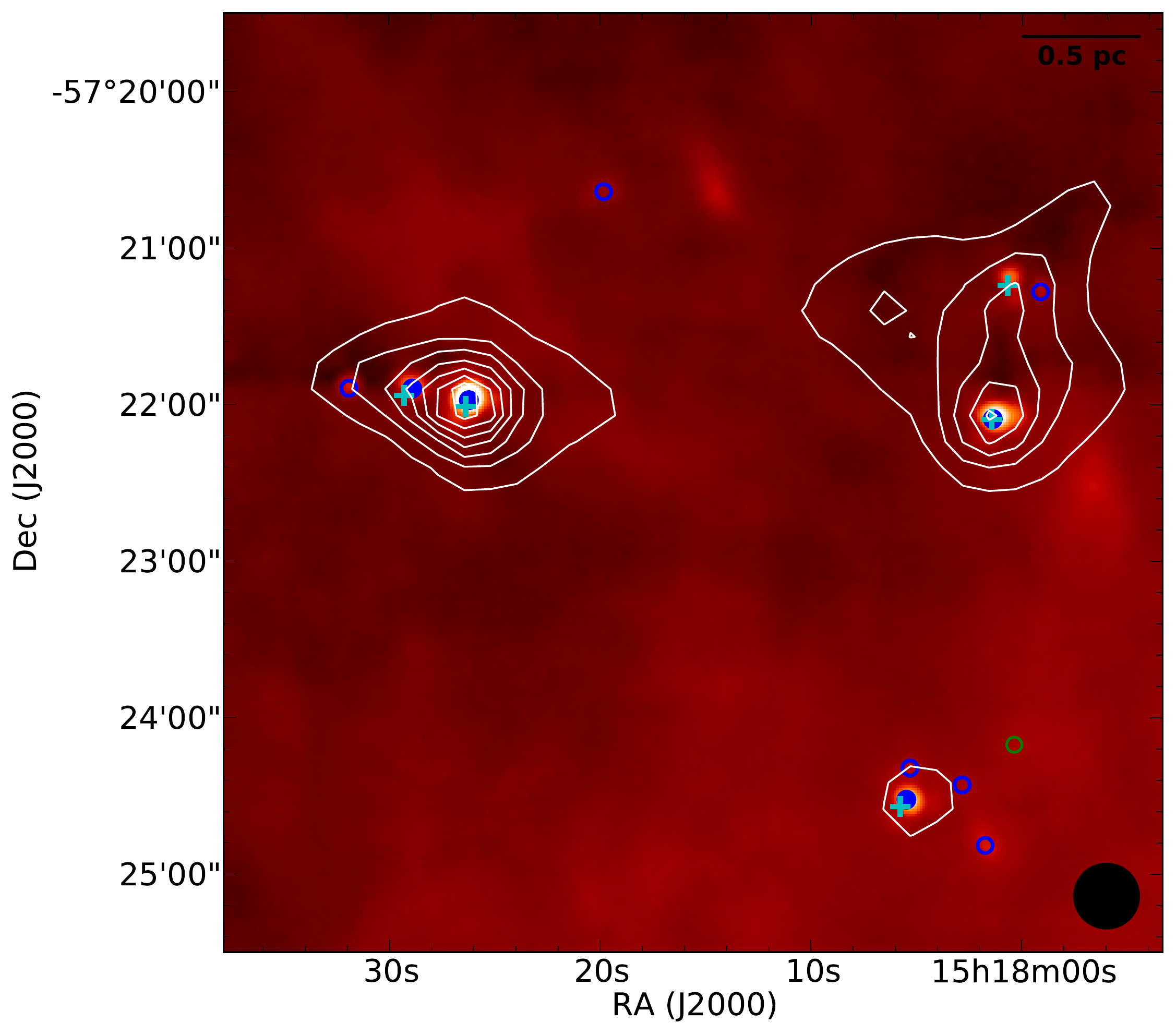} & \includegraphics[width=0.47\linewidth]{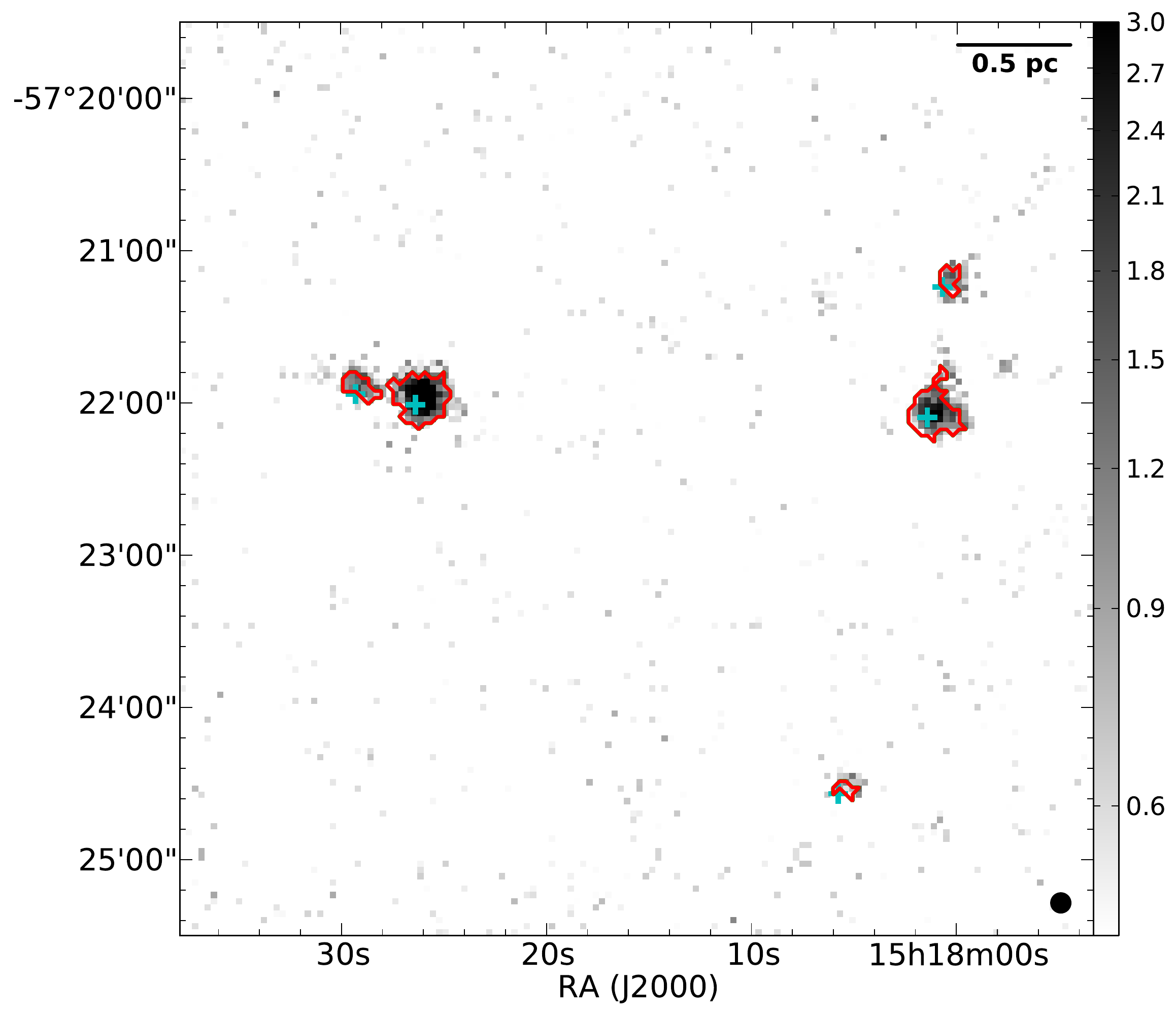}
\end{tabular}
\end{center}
\caption{IRDC321.73+0.05: {\it Left:} The image is the PACS 100\,$\mu$m with SPIRE 350\,$\mu$m contours overplotted. 
The white contour levels begin at 5\, Jy beam$^{-1}$ and increase in steps of 2\, Jy beam$^{-1}$.
The blue circles are PACS cores identified in \citet{Ragan2012b}, and green circles are candidate 70\,$\mu$m-dark cores. A circle is filled if it is recovered as a ``leaf'' with SABOCA, and it is left open if it is not. Cyan crosses are the positions of {\it all} \dendro~ leaves.
{\it Right:} SABOCA 350\,$\mu$m image is plotted in greyscale with red contours showing the ``leaf''  structures identified with \dendro~ and green contours show lower-level structures (``branches'') in emission, if any.  See Section~\ref{sec:dendro} for details.
\label{fig:g32173}}
\end{figure*}

\begin{figure*}
\begin{center}
\begin{tabular}{lr}
\includegraphics[width=0.46\linewidth]{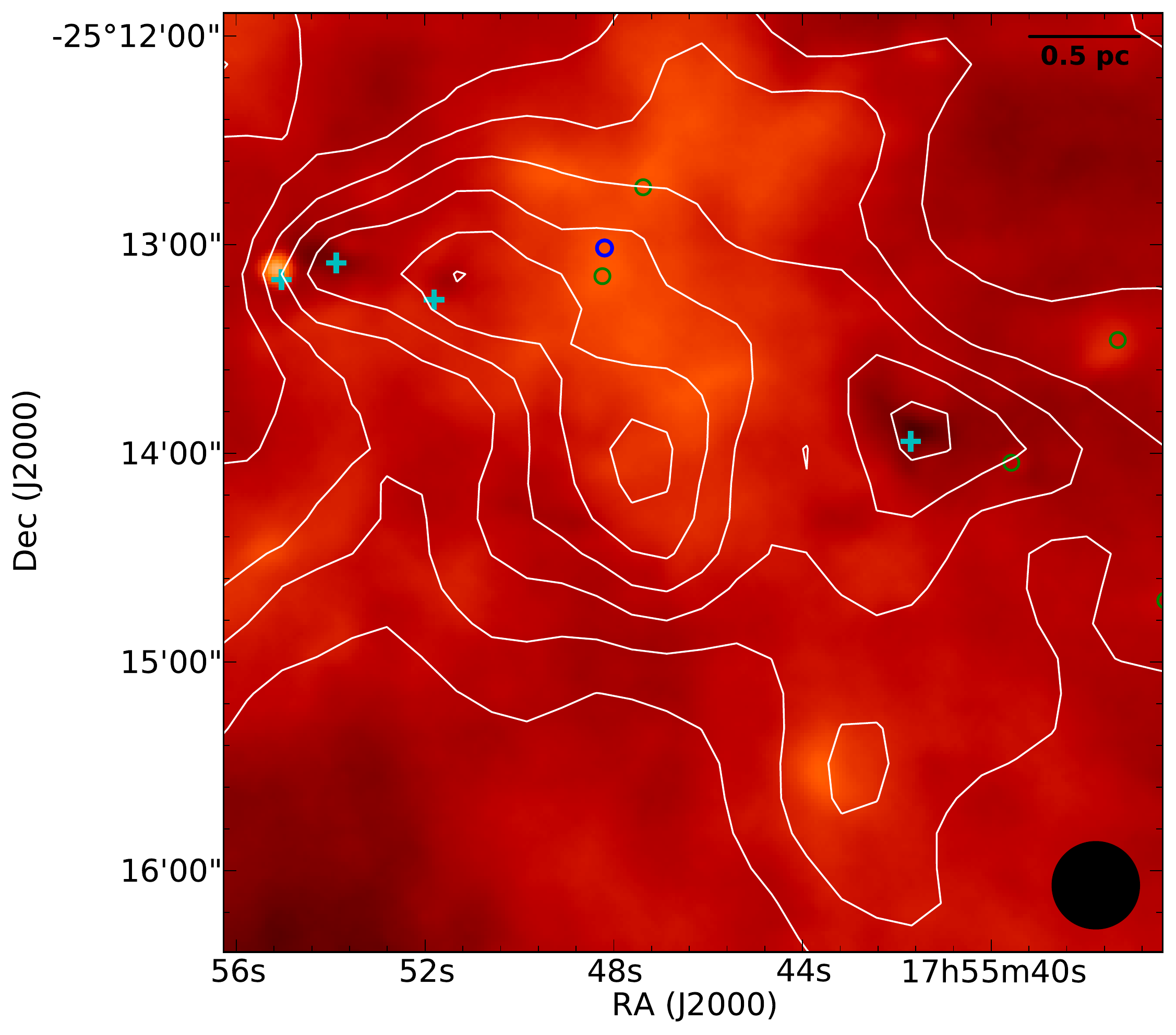} & \includegraphics[width=0.47\linewidth]{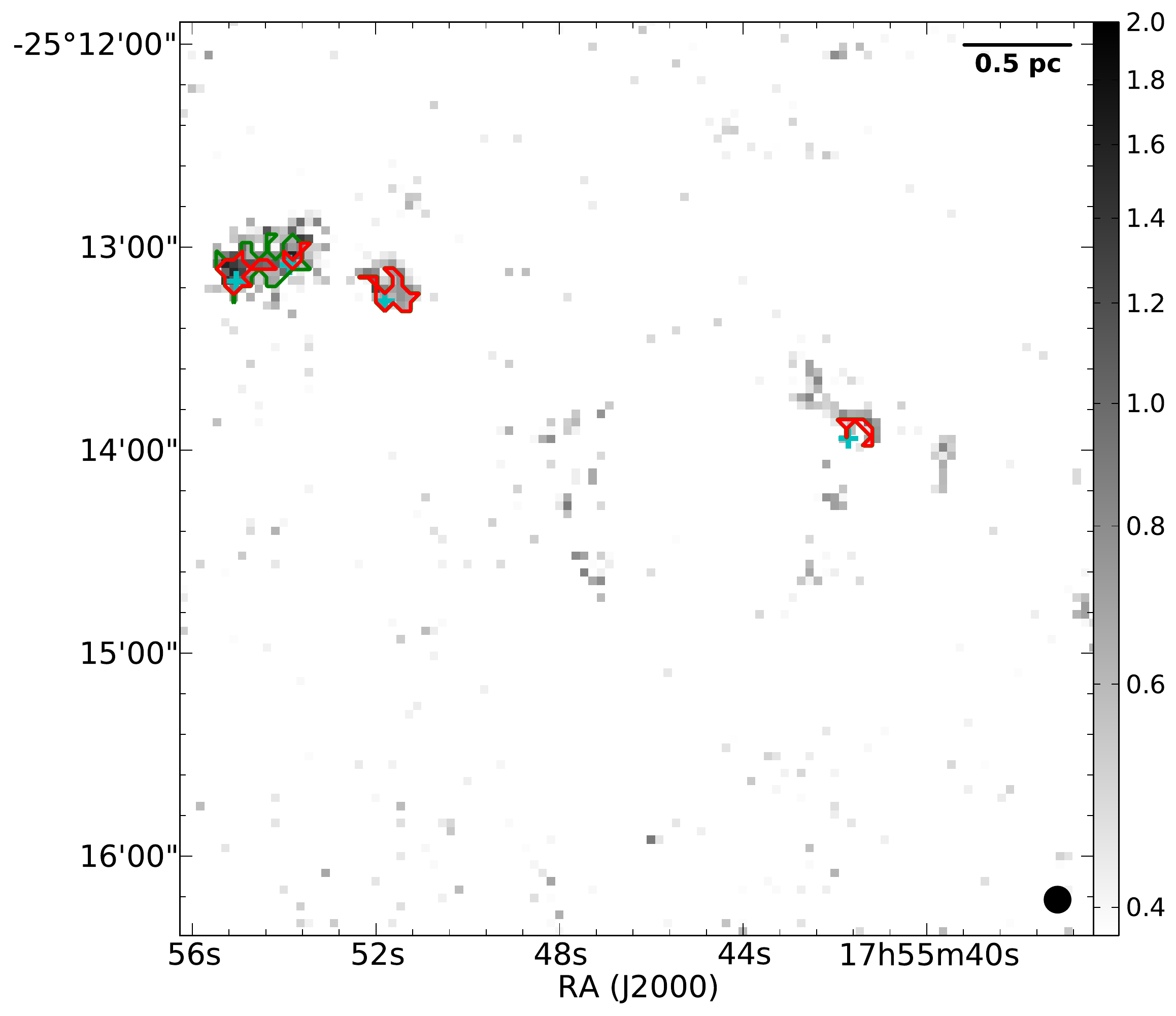}
\end{tabular}
\end{center}
\caption{IRDC004.36-0.06: {\it Left:} The image is the PACS 100\,$\mu$m with SPIRE 350\,$\mu$m contours overplotted. 
The white contour levels begin at 4\, Jy beam$^{-1}$ and increase in steps of 1\, Jy beam$^{-1}$.
The blue circles are PACS cores identified in \citet{Ragan2012b}, and green circles are candidate 70\,$\mu$m-dark cores. A circle is filled if it is recovered as a ``leaf'' with SABOCA, and it is left open if it is not. Cyan crosses are the positions of {\it all} \dendro~ leaves.
{\it Right:} SABOCA 350\,$\mu$m image is plotted in greyscale with red contours showing the ``leaf''  structures identified with \dendro~ and green contours show lower-level structures (``branches'') in emission, if any.  The scalebar is in units of Jy beam$^{-1}$. See Section~\ref{sec:dendro} for details. 
\label{fig:g0436}}
\end{figure*}


\begin{figure*}
\begin{center}
\begin{tabular}{lr}
\includegraphics[width=0.46\linewidth]{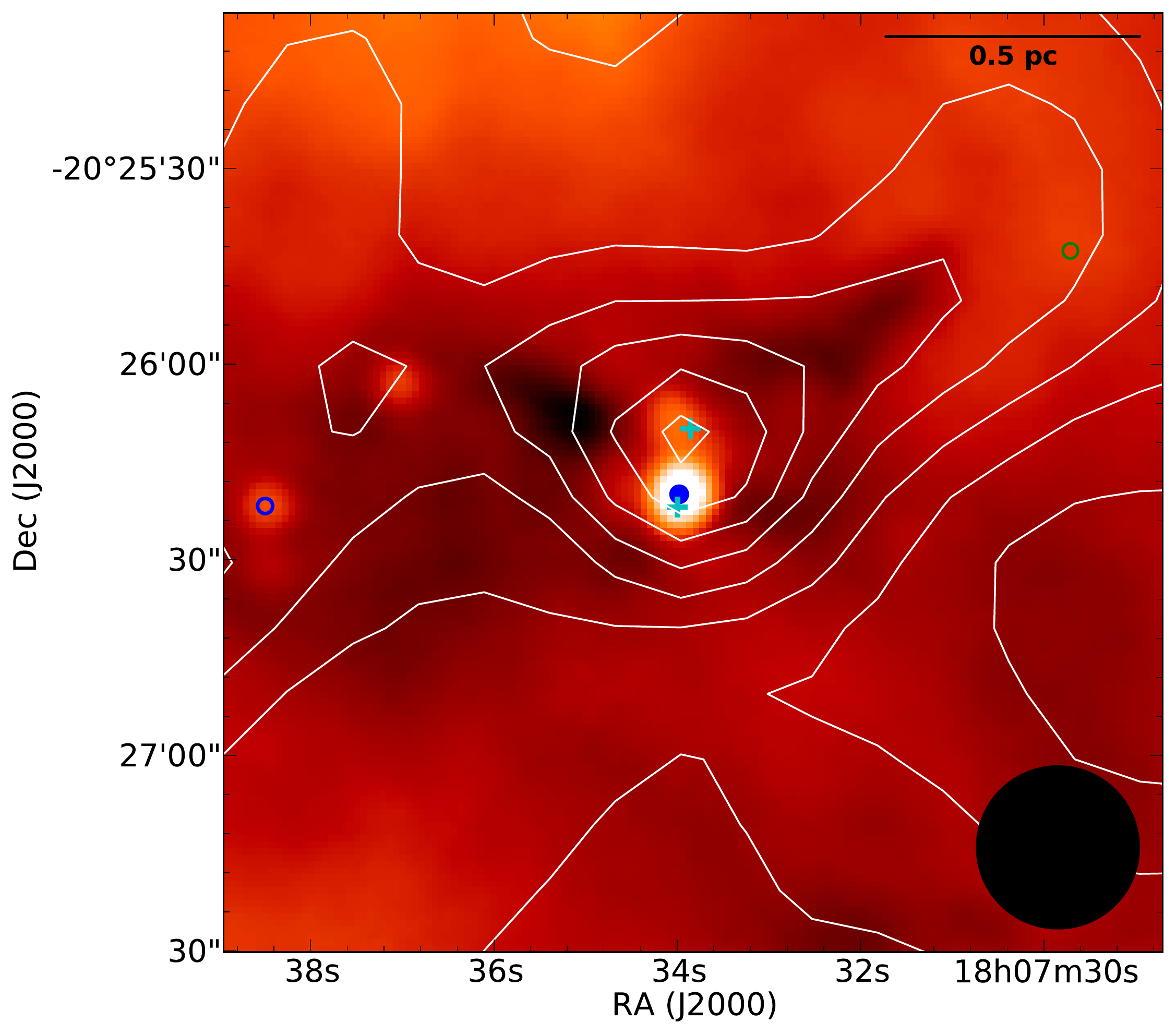} & \includegraphics[width=0.475\linewidth]{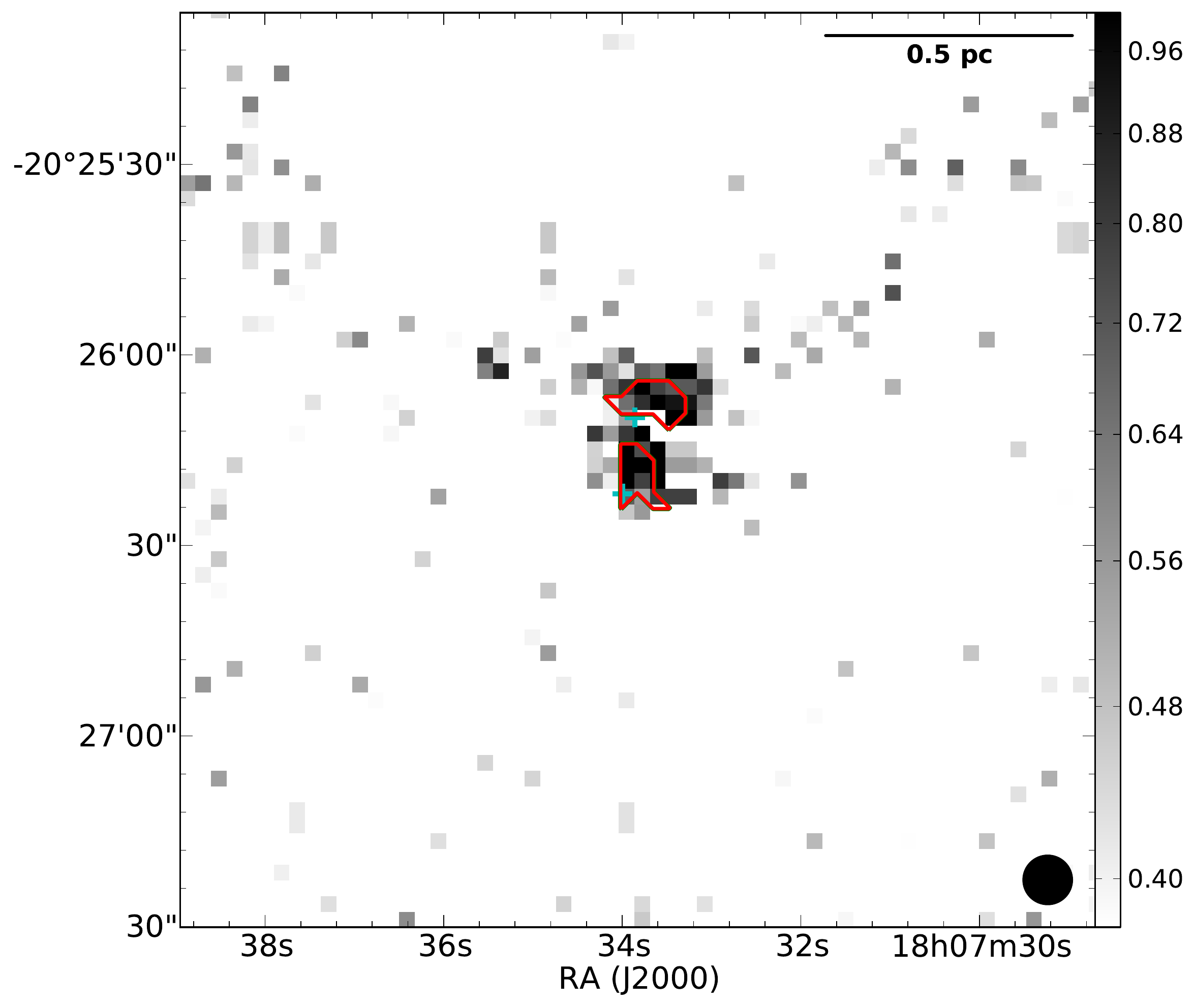}
\end{tabular}
\end{center}
\caption{IRDC009.86-0.04: {\it Left:} The image is the PACS 100\,$\mu$m with SPIRE 350\,$\mu$m contours overplotted. 
The white contour levels begin at 4\, Jy beam$^{-1}$ and increase in steps of 1\, Jy beam$^{-1}$.
The blue circles are PACS cores identified in \citet{Ragan2012b}, and green circles are candidate 70\,$\mu$m-dark cores. A circle is filled if it is recovered as a ``leaf'' with SABOCA, and it is left open if it is not. Cyan crosses are the positions of {\it all} \dendro~ leaves.
{\it Right:} SABOCA 350\,$\mu$m image is plotted in greyscale with red contours showing the ``leaf''  structures identified with \dendro~ and green contours show lower-level structures (``branches'') in emission, if any.  The scalebar is in units of Jy beam$^{-1}$.  See Section~\ref{sec:dendro} for details.
\label{fig:g0986}}
\end{figure*}

\clearpage

\begin{figure*}
\begin{center}
\begin{tabular}{lr}
\includegraphics[width=0.46\linewidth]{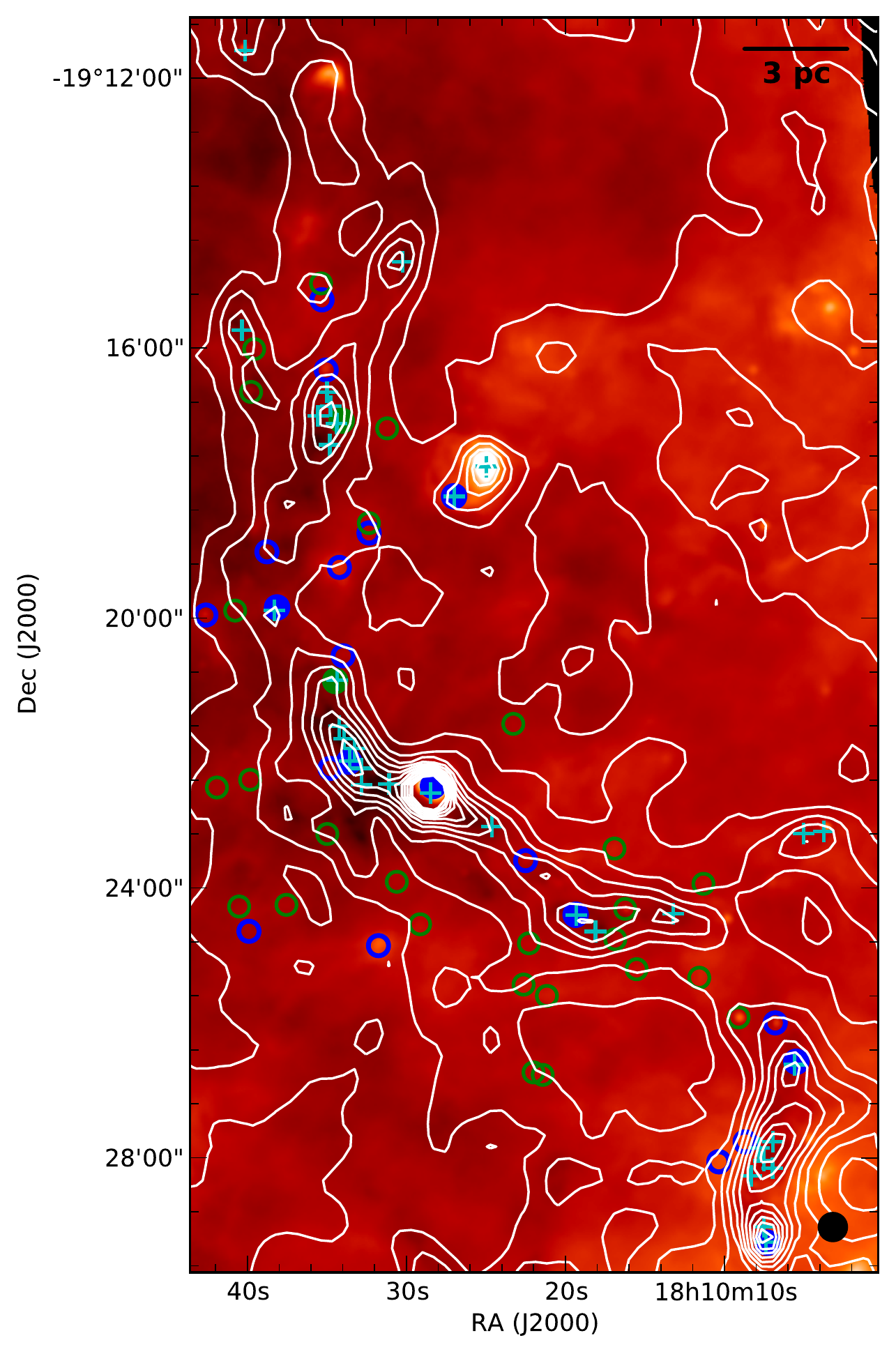} & \includegraphics[width=0.48\linewidth]{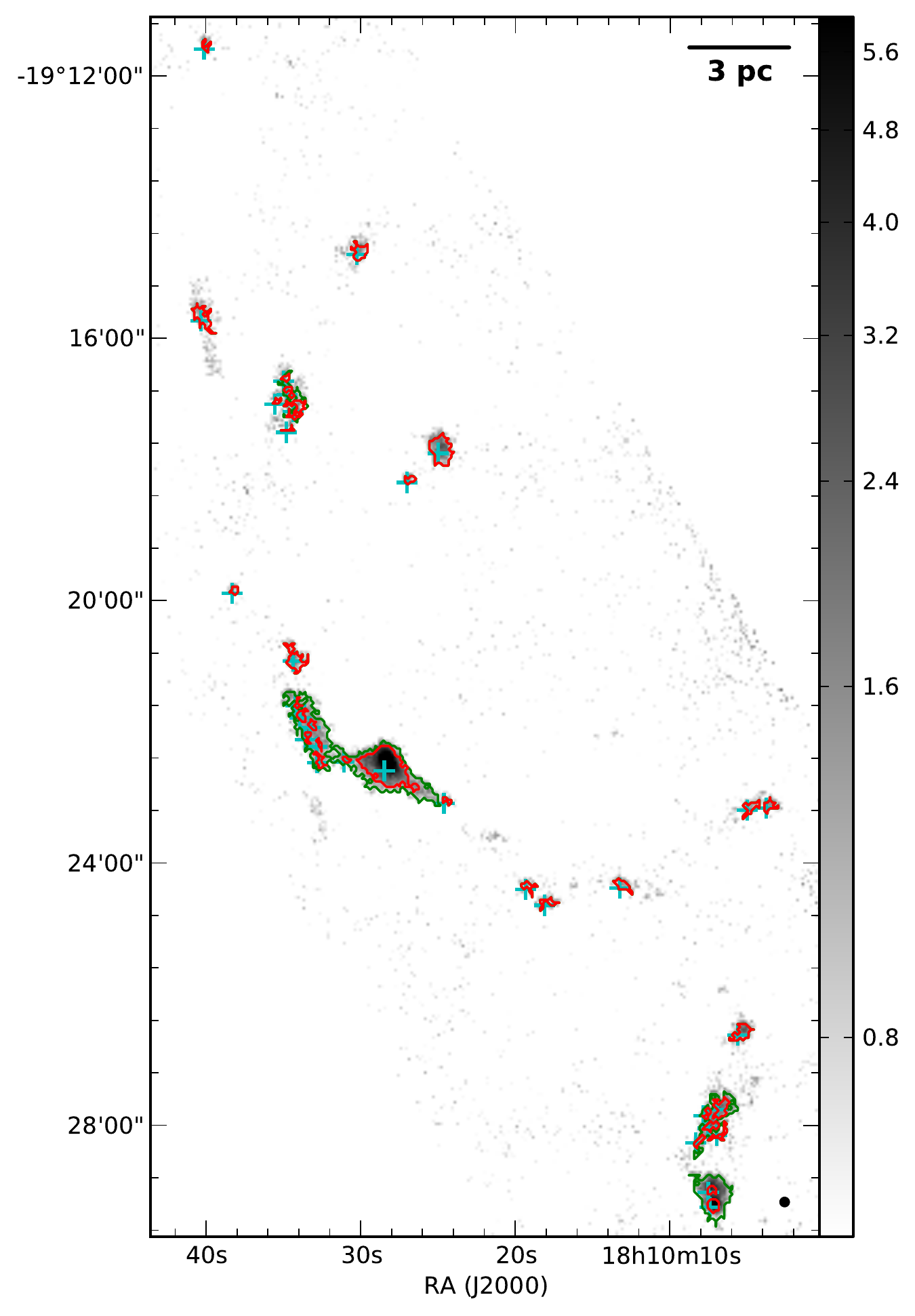}
\end{tabular}
\end{center}
\caption{IRDC011.11-0.12: {\it Left:} The image is the PACS 100\,$\mu$m with SPIRE 350\,$\mu$m contours overplotted. 
The white contour levels begin at 4\, Jy beam$^{-1}$ and increase in steps of 2\, Jy beam$^{-1}$. 
The blue circles are PACS cores identified in \citet{Ragan2012b}, and green circles are candidate 70\,$\mu$m-dark cores. A circle is filled if it is recovered as a ``leaf'' with SABOCA, and it is left open if it is not. Cyan crosses are the positions of {\it all} \dendro~ leaves.
{\it Right:} SABOCA 350\,$\mu$m image is plotted in greyscale with red contours showing the ``leaf''  structures identified with \dendro~ and green contours show lower-level structures (``branches'') in emission, if any.  The scalebar is in units of Jy beam$^{-1}$.  See Section~\ref{sec:dendro} for details.
\label{fig:g1111}}
\end{figure*}


\begin{figure*}
\begin{center}
\begin{tabular}{lr}
\includegraphics[width=0.46\linewidth]{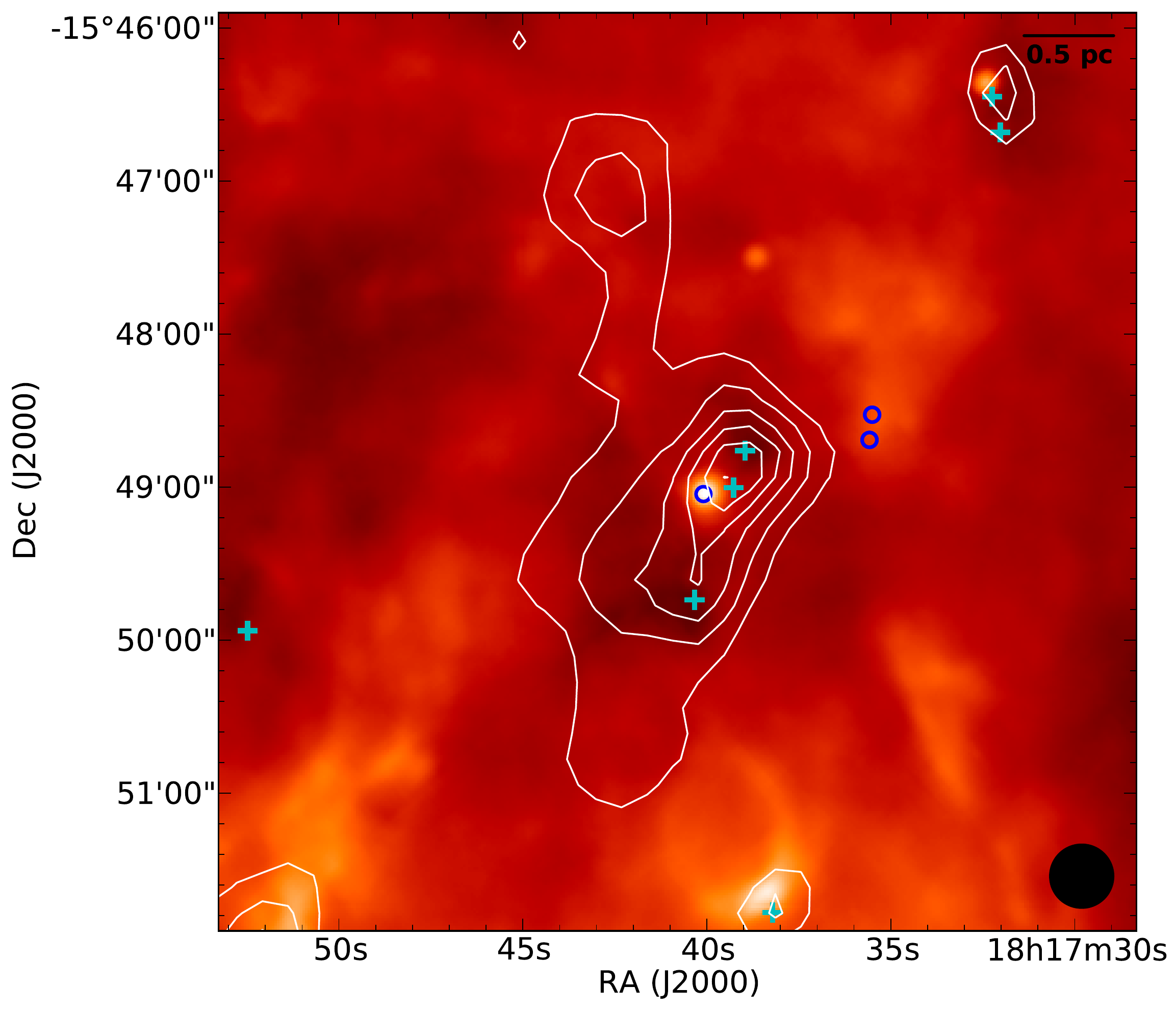} & \includegraphics[width=0.47\linewidth]{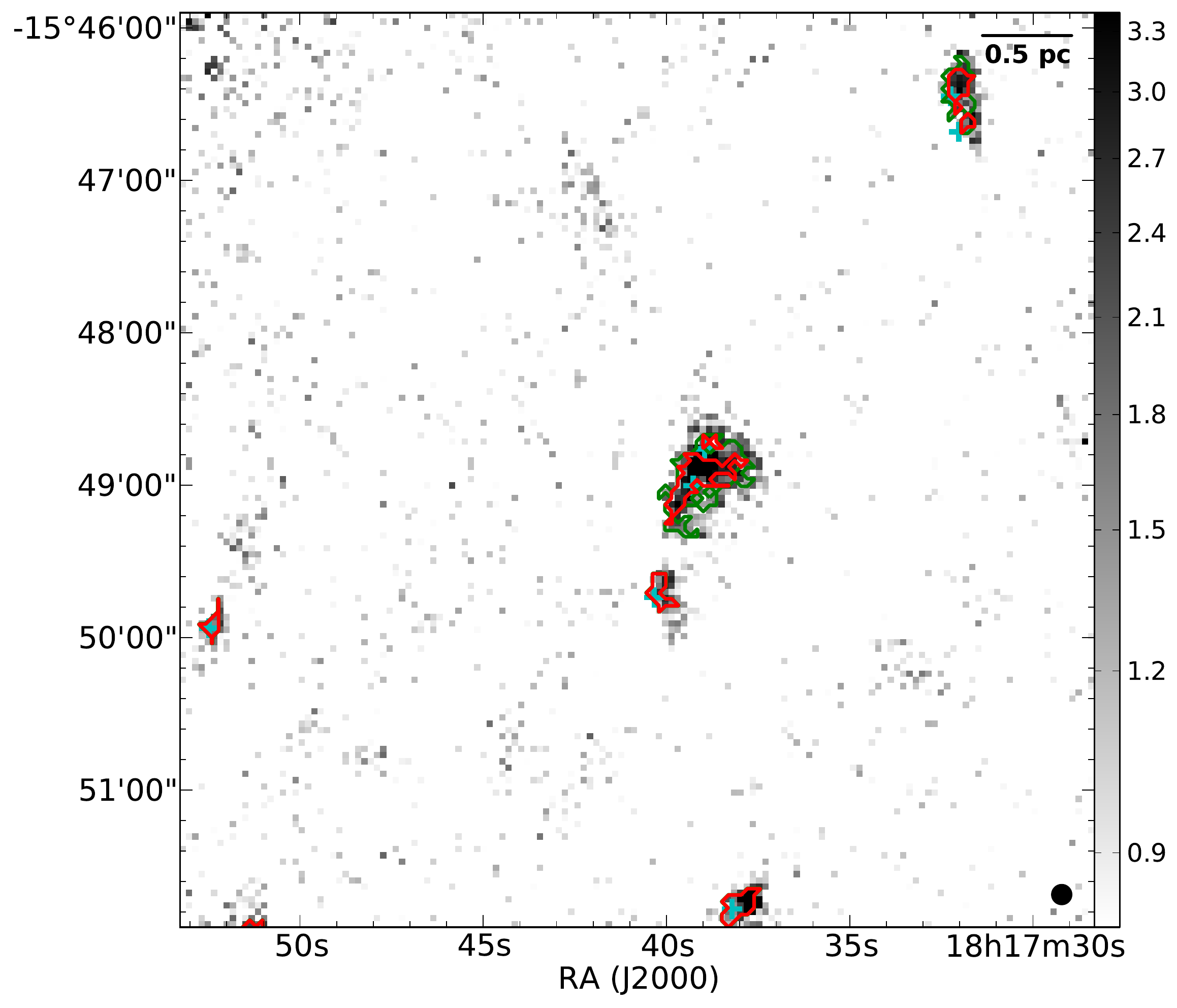}
\end{tabular}
\end{center}
\caption{IRDC015.05+0.09: {\it Left:} The image is the PACS 100\,$\mu$m with SPIRE 350\,$\mu$m contours overplotted. 
The white contour levels begin at 8\, Jy beam$^{-1}$ and increase in steps of 1\, Jy beam$^{-1}$.
The blue circles are PACS cores identified in \citet{Ragan2012b}, and green circles are candidate 70\,$\mu$m-dark cores. A circle is filled if it is recovered as a ``leaf'' with SABOCA, and it is left open if it is not. Cyan crosses are the positions of {\it all} \dendro~ leaves.
{\it Right:} SABOCA 350\,$\mu$m image is plotted in greyscale with red contours showing the ``leaf''  structures identified with \dendro~ and green contours show lower-level structures (``branches'') in emission, if any.  The scalebar is in units of Jy beam$^{-1}$.  See Section~\ref{sec:dendro} for details.
\label{fig:g1505}}
\end{figure*}

\clearpage

\begin{figure*}
\begin{center}
\begin{tabular}{lr}
\includegraphics[width=0.4\linewidth]{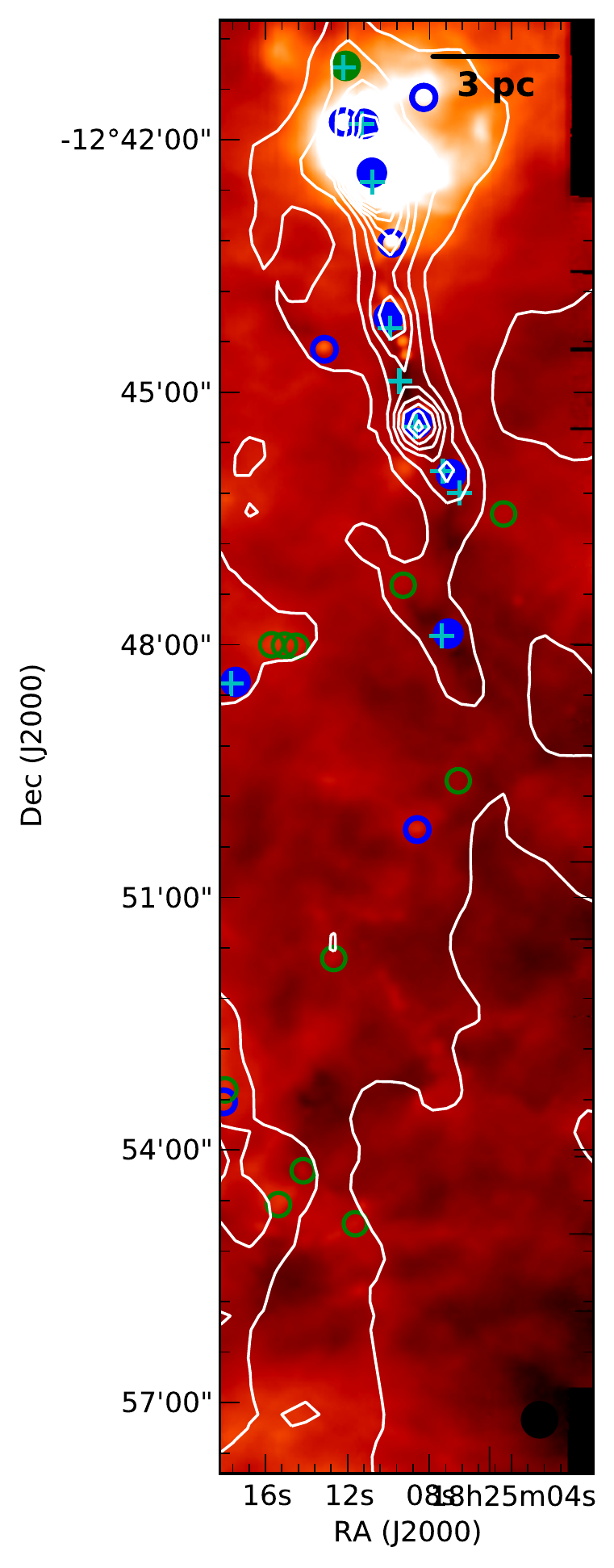} & \includegraphics[width=0.42\linewidth]{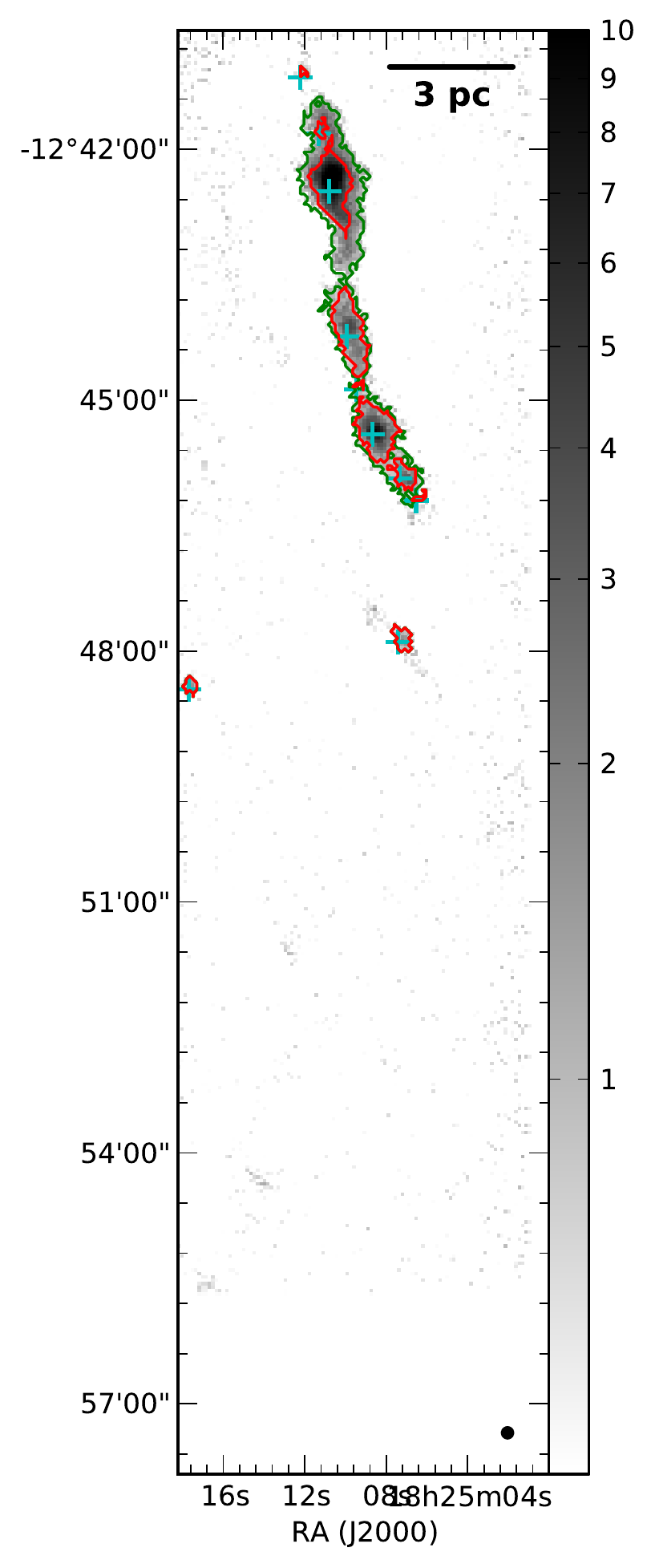}
\end{tabular}
\end{center}
\caption{IRDC\, 18223: {\it Left:} The image is the PACS 100\,$\mu$m with SPIRE 350\,$\mu$m contours overplotted. 
The white contour levels begin at 4\, Jy beam$^{-1}$ and increase in steps of 4\, Jy beam$^{-1}$.
The blue circles are PACS cores identified in \citet{Ragan2012b}, and green circles are candidate 70\,$\mu$m-dark cores. A circle is filled if it is recovered as a ``leaf'' with SABOCA, and it is left open if it is not. Cyan crosses are the positions of {\it all} \dendro~ leaves.
{\it Right:} SABOCA 350\,$\mu$m image is plotted in greyscale with red contours showing the ``leaf''  structures identified with \dendro~ and green contours show lower-level structures (``branches'') in emission, if any.  The scalebar is in units of Jy beam$^{-1}$.  See Section~\ref{sec:dendro} for details.
\label{fig:18223}}
\end{figure*}

\clearpage

\begin{figure*}
\begin{center}
\begin{tabular}{lr}
\includegraphics[width=0.46\linewidth]{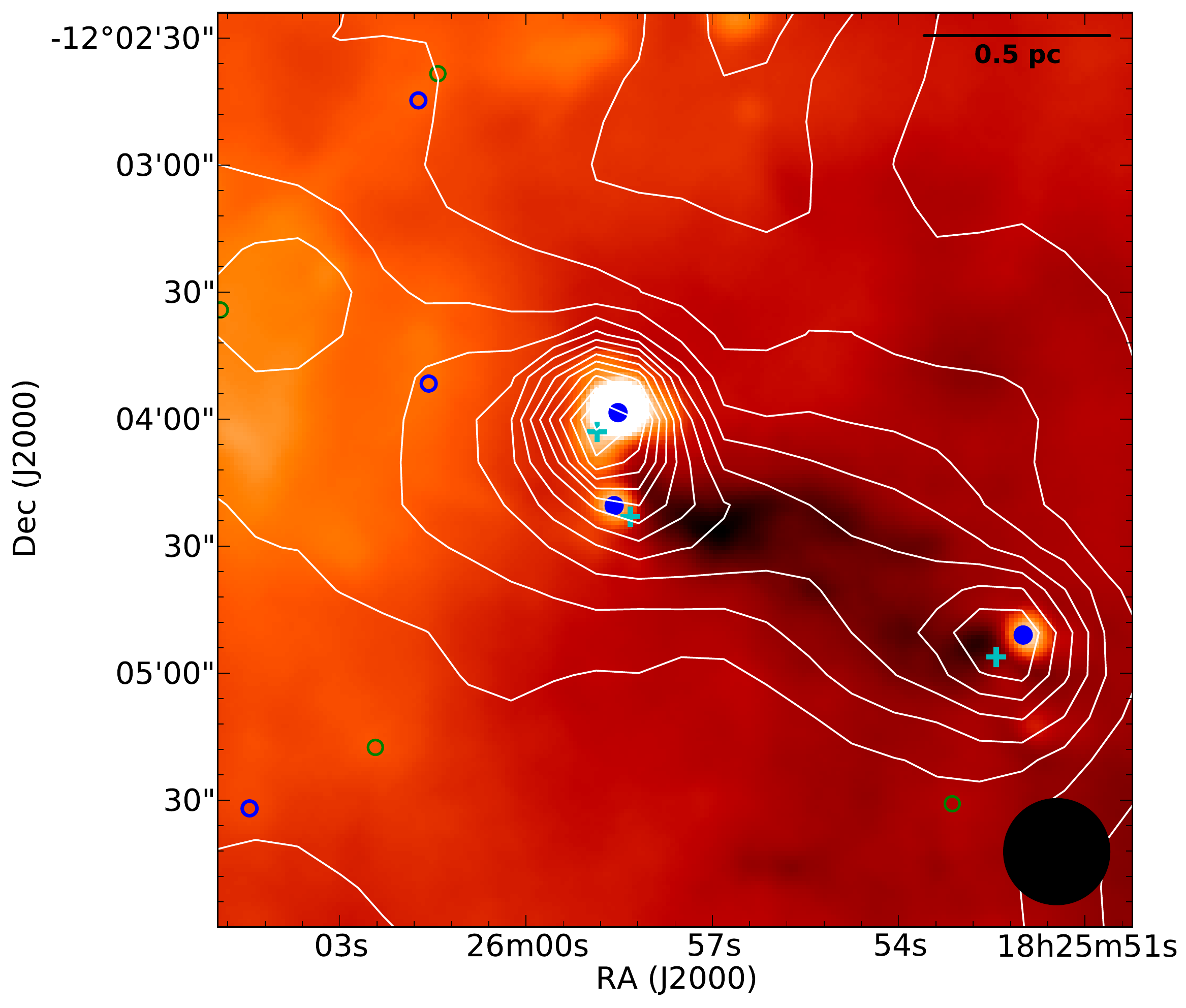} & \includegraphics[width=0.47\linewidth]{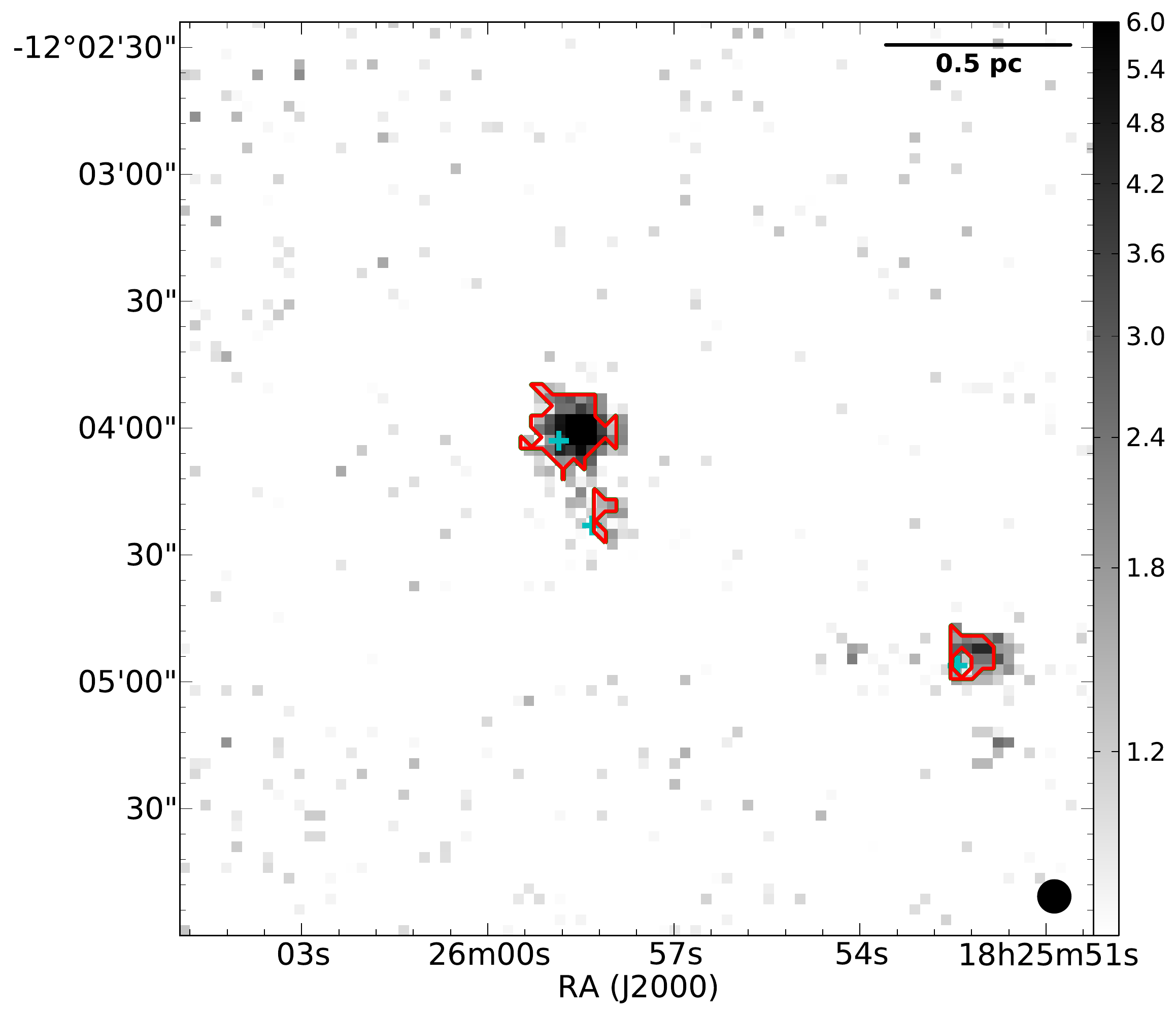}
\end{tabular}
\end{center}
\caption{IRDC019.30+0.07: {\it Left:} The image is the PACS 100\,$\mu$m with SPIRE 350\,$\mu$m contours overplotted. 
The white contour levels begin at 5\, Jy beam$^{-1}$ and increase in steps of 2\, Jy beam$^{-1}$.
The blue circles are PACS cores identified in \citet{Ragan2012b}, and green circles are candidate 70\,$\mu$m-dark cores. A circle is filled if it is recovered as a ``leaf'' with SABOCA, and it is left open if it is not. Cyan crosses are the positions of {\it all} \dendro~ leaves.
{\it Right:} SABOCA 350\,$\mu$m image is plotted in greyscale with red contours showing the ``leaf''  structures identified with \dendro~ and green contours show lower-level structures (``branches'') in emission, if any.  The scalebar is in units of Jy beam$^{-1}$.  See Section~\ref{sec:dendro} for details.
\label{fig:g1930}}
\end{figure*}


\begin{figure*}
\begin{center}
\begin{tabular}{lr}
\includegraphics[width=0.46\linewidth]{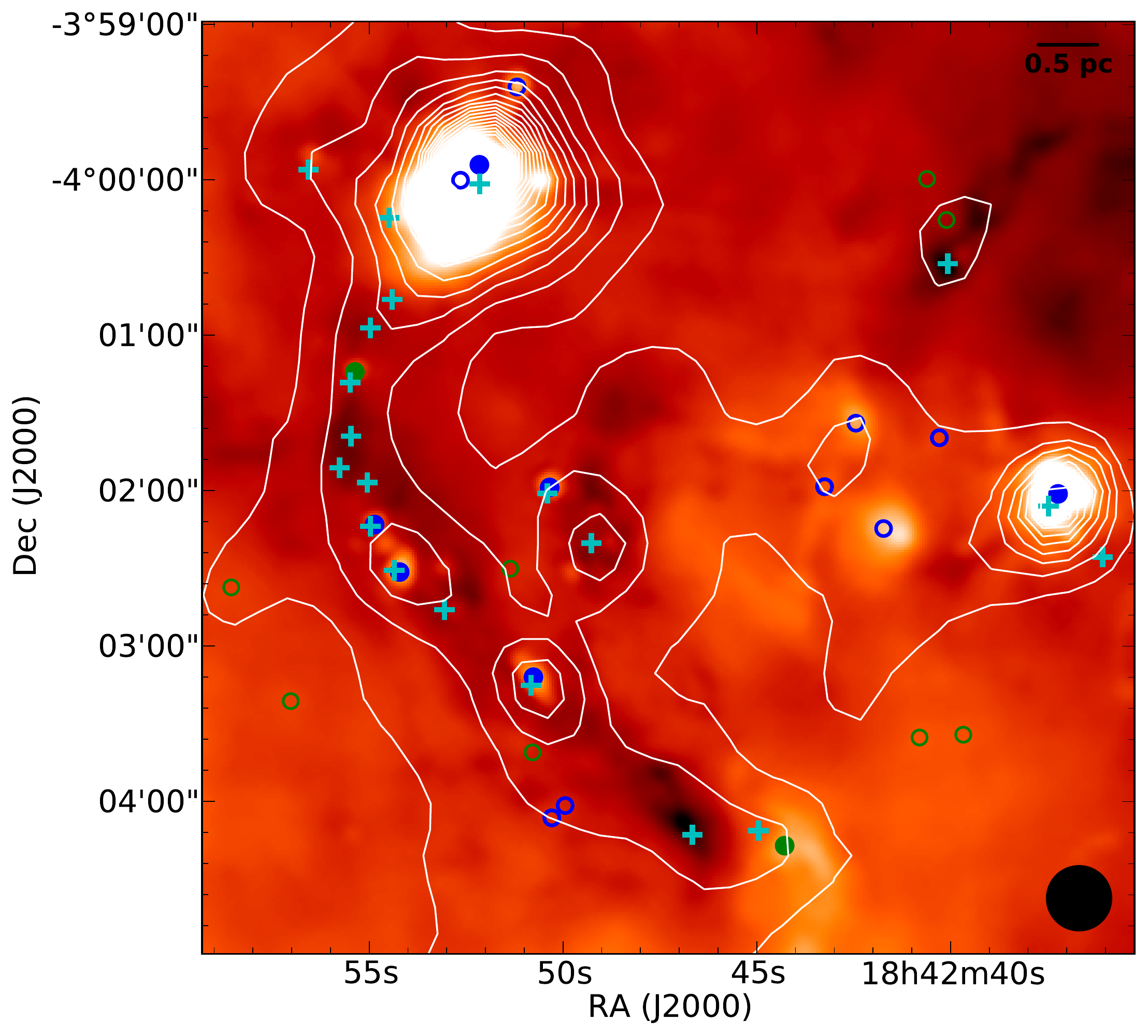} & \includegraphics[width=0.46\linewidth]{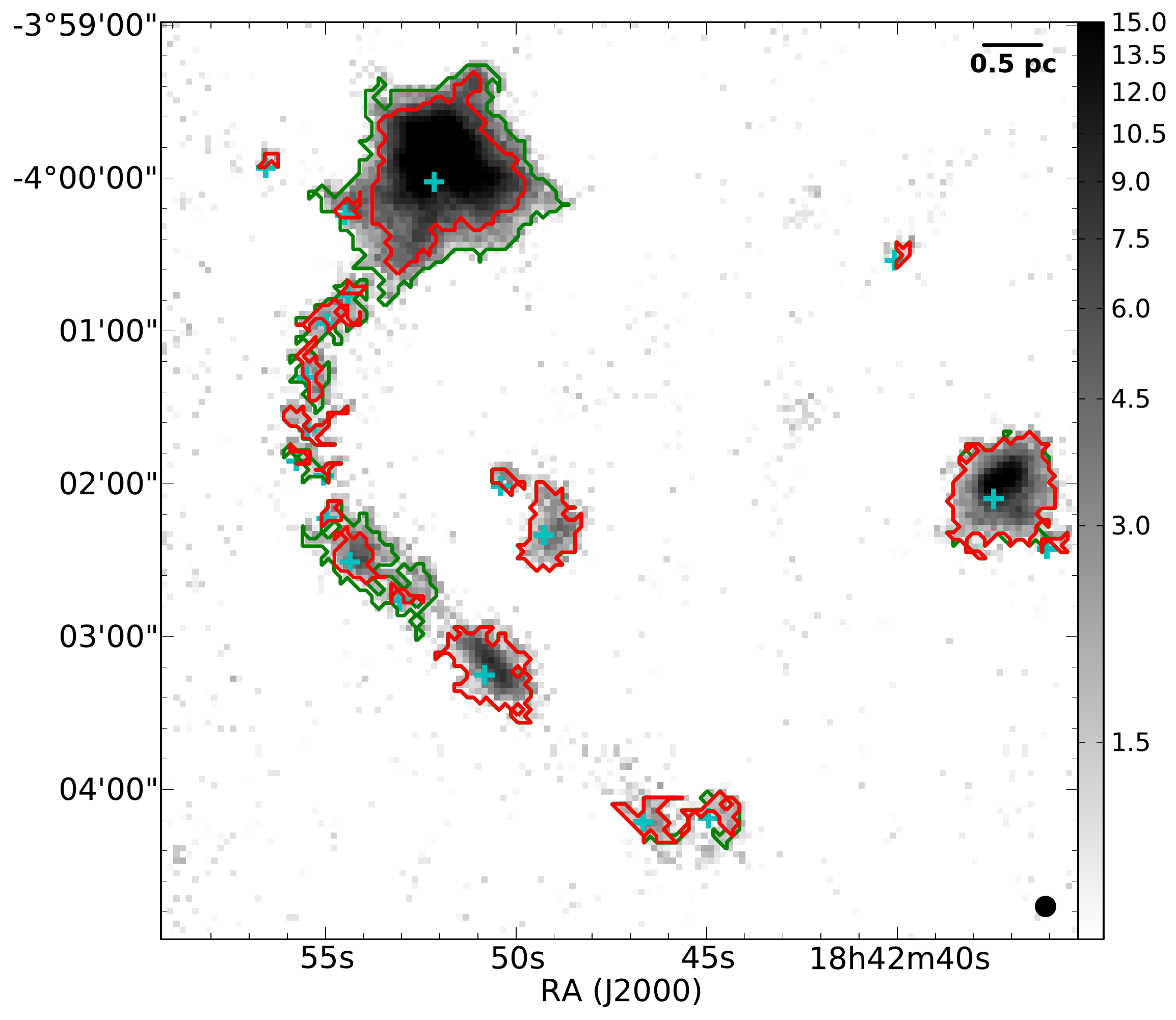}
\end{tabular}
\end{center}
\caption{IRDC028.34+0.06: {\it Left:} The image is the PACS 100\,$\mu$m with SPIRE 350\,$\mu$m contours overplotted. 
The white contour levels begin at 16\, Jy beam$^{-1}$ and increase in steps of 5\, Jy beam$^{-1}$. 
The blue circles are PACS cores identified in \citet{Ragan2012b}, and green circles are candidate 70\,$\mu$m-dark cores. A circle is filled if it is recovered as a ``leaf'' with SABOCA, and it is left open if it is not. Cyan crosses are the positions of {\it all} \dendro~ leaves.
{\it Right:} SABOCA 350\,$\mu$m image is plotted in greyscale with red contours showing the ``leaf''  structures identified with \dendro~ and green contours show lower-level structures (``branches'') in emission, if any.  The scalebar is in units of Jy beam$^{-1}$.  See Section~\ref{sec:dendro} for details.
\label{fig:g2834}}
\end{figure*}


\clearpage

\section{Dendrogram leaves}
\nopagebreak
\longtab{1}{
\begin{longtable}{lcccccrrrl}
\caption{Leaves from dendrogram decomposition of SABOCA 350\,$\mu$m image. \label{tab:newcores}} \\
\hline \hline 
Leaf & RA (J2000) & Dec (J2000) & $S_\mathrm{tot}^a$ & $S_{parent} $ & $< \mathrm{N}_{H_2} >^b$ & $r_{eff}$ & $M_\mathrm{350{\mu}m}^{20K}$ & $M_\mathrm{350{\mu}m}^{b}$ & Comment \\
number & ($^{h}$:$^{m}$:$^{s}$) & ($^{\circ}$:$^{'}$:$^{''}$) & (Jy) & (Jy beam$^{-1}$) & (cm$^{-2}$) & (pc) & ($\msun$) & ($\msun$) &  \\
\hline
\endfirsthead
\caption{continued.} \\
\hline 
Leaf & RA (J2000) & Dec (J2000) & $S_\mathrm{tot}^a$ & $S_{parent} $ & $< \mathrm{N}_{H_2} >^b$ & $r_{eff}$ & $M_\mathrm{350{\mu}m}^{20K}$ & $M_\mathrm{350{\mu}m}^{b}$ & Comment \\
number & ($^{h}$:$^{m}$:$^{s}$) & ($^{\circ}$:$^{'}$:$^{''}$) & (Jy) & (Jy beam$^{-1}$) & (cm$^{-2}$) & (pc) & ($\msun$) & ($\msun$) &  \\
\hline
\endhead
\hline
\endfoot
\hline
IRDC310.39 & & & & & & & & \\
\hline
 1$^c$ & 13:56:01.4 & -62:14:20 &    25.4 &   0.9 & 3.40E+22 &  0.41 &   223 &   307 & pacs \\ 
 2 & 13:56:04.2 & -62:14:02 &     0.8 &   0.9 &  1.52E+22 &  0.18 &     7 &    16 &   \\ 
\hline
IRDC316.72 & & & & & & & & \\
\hline
 1 & 14:44:04.8 & -59:41:47 &     2.8 &   0.0 & 1.53E+22 &  0.09 &     7 &    22 &   \\ 
 2 & 14:44:15.4 & -59:43:14 &     0.3 &   2.9 & 4.04E+22 &  0.06 &  1 &     1 &   \\ 
 3 & 14:44:16.1 & -59:43:25 &     0.3 &   3.5 & 4.99E+22 &  0.05 &  1 &     1 &   \\ 
 4 & 14:44:16.4 & -59:44:46 &     4.1 &   1.2 & 2.24E+22 &  0.16 &    11 &    34 &   \\ 
 5 & 14:44:17.1 & -59:43:33 &     0.5 &   3.6 & 4.99E+22 &  0.09 &     1 &     3 &   \\ 
 6 & 14:44:17.4 & -59:43:48 &     0.9 &   3.6 & 5.19E+22 &  0.09 &     2 &     3 & pacs \\ 
 7 & 14:44:19.4 & -59:44:20 &     9.0 &   3.1 & 5.12E+22 &  0.20 &    24 &    93 & cold \\ 
 8 & 14:44:21.1 & -59:44:41 &     0.9 &   3.1 & 4.59E+22 &  0.09 &     2 &     2 & pacs \\ 
 9 & 14:44:23.3 & -59:45:01 &     2.1 &   2.1 & 3.54E+22 &  0.11 &     6 &     8 & pacs \\ 
10 & 14:44:25.3 & -59:45:14 &     0.3 &   1.6 & 2.44E+22 &  0.06 &  1 &     1 &   \\ 
11 & 14:44:26.2 & -59:45:30 &     3.2 &   1.6 & 2.98E+22 &  0.13 &     9 &    26 &   \\ 
12 & 14:44:28.4 & -59:45:49 &     0.5 &   1.7 & 2.53E+22 &  0.09 &     1 &     3 &   \\ 
13 & 14:44:29.8 & -59:45:31 &     0.3 &   1.3 & 2.04E+22 &  0.06 &  1 &     2 &   \\ 
14 & 14:44:30.2 & -59:46:04 &     1.2 &   1.7 & 2.87E+22 &  0.09 &     3 &     8 &   \\ 
15 & 14:44:33.3 & -59:46:14 &     0.3 &   0.8 & 1.50E+22 &  0.06 &  1 &     2 &   \\ 
16 & 14:44:33.5 & -59:46:04 &     0.4 &   1.1 & 1.86E+22 &  0.06 &  1 &     1 & pacs \\ 
17 & 14:44:34.7 & -59:45:57 &     0.2 &   1.1 & 1.72E+22 &  0.05 &  1 &     1 &   \\ 
18 & 14:44:34.8 & -59:46:22 &     1.9 &   0.0 & 1.47E+22 &  0.08 &     5 &    14 & cold \\ 
19 & 14:44:35.6 & -59:45:35 &     2.2 &   0.8 & 1.91E+22 &  0.11 &     6 &    22 & cold \\ 
20 & 14:44:36.7 & -59:45:19 &     0.6 &   0.8 & 1.42E+22 &  0.09 &     2 &     4 &   \\ 
21 & 14:44:37.5 & -59:45:58 &     0.4 &   1.5 & 2.40E+22 &  0.06 &     1 &     2 &   \\ 
22$^c$ & 14:44:40.2 & -59:46:21 &    10.8 &   1.5 & 3.19E+22 &  0.20 &    29 &    60 & pacs \\ 
23 & 14:44:43.0 & -59:46:53 &     1.4 &   1.4 & 2.38E+22 &  0.10 &     4 &     9 &   \\ 
24 & 14:44:44.8 & -59:46:10 &     1.5 &   0.9 & 1.82E+22 &  0.11 &     4 &     8 & cold \\ 
25 & 14:44:46.7 & -59:46:38 &     0.6 &   1.8 & 2.91E+22 &  0.07 &     1 &     2 & cold \\ 
26 & 14:44:47.5 & -59:47:17 &     1.5 &   2.7 & 4.16E+22 &  0.10 &     4 &    10 &   \\ 
27 & 14:44:47.8 & -59:46:53 &     0.5 &   1.9 & 2.96E+22 &  0.07 &     1 &     3 &   \\ 
28 & 14:44:49.3 & -59:47:17 &     0.9 &   2.8 & 4.05E+22 &  0.09 &     2 &     5 &   \\ 
29 & 14:44:49.5 & -59:45:27 &     1.7 &   0.0 & 2.23E+22 &  0.06 &     4 &     8 & pacs \\ 
30 & 14:44:51.1 & -59:47:32 &     0.8 &   3.1 & 4.87E+22 &  0.06 &     2 &     4 &   \\ 
31 & 14:44:51.4 & -59:48:11 &     0.6 &   3.0 & 4.80E+22 &  0.05 &     1 &     3 &   \\ 
32 & 14:44:51.8 & -59:46:12 &     1.2 &   0.0 & 1.48E+22 &  0.06 &     3 &     7 & pacs \\ 
33 & 14:44:52.5 & -59:46:28 &     0.9 &   0.0 & 1.25E+22 &  0.06 &     3 &     6 &   \\ 
34 & 14:44:53.1 & -59:47:41 &     1.4 &   3.9 & 6.31E+22 &  0.07 &     4 &    10 &   \\ 
35 & 14:44:53.5 & -59:46:42 &     5.5 &   0.0 & 2.43E+22 &  0.10 &    15 &    23 & pacs \\ 
36 & 14:44:55.5 & -59:46:54 &     1.7 &   0.0 & 1.91E+22 &  0.06 &     5 &     8 & pacs \\ 
37 & 14:44:55.5 & -59:45:49 &     1.7 &   0.0 & 2.73E+22 &  0.05 &     4 &    11 &   \\ 
38 & 14:44:56.9 & -59:48:03 &    53.5 &   8.6 & 1.86E+23 &  0.18 &   142 &    90 & pacs \\ 
39 & 14:44:58.6 & -59:47:19 &     1.1 &   1.6 & 3.06E+22 &  0.07 &     3 &     7 &   \\ 
40 & 14:44:59.2 & -59:47:32 &     0.8 &   1.8 & 3.42E+22 &  0.06 &     2 &     4 &   \\ 
41 & 14:45:00.2 & -59:47:37 &     0.8 &   2.3 & 3.86E+22 &  0.06 &     2 &     5 &   \\ 
42 & 14:45:01.0 & -59:46:19 &     2.5 &   0.0 & 3.03E+22 &  0.06 &     7 &    17 & cold \\ 
\hline
IRDC320.27 & & & & & & & & \\
\hline
 1 & 15:07:32.4 & -57:53:30 &     1.1 &   0.0 & 1.44E+22 &  0.05 &     2 &     5 &   \\ 
 2 & 15:07:32.9 & -57:53:40 &     0.8 &   0.0 & 1.31E+22 &  0.04 &     1 &     3 &   \\ 
 3 & 15:07:55.5 & -57:54:34 &     0.6 &   1.3 & 2.41E+22 &  0.05 &     1 &     2 & cold \\ 
 4 & 15:07:57.6 & -57:54:26 &     1.3 &   1.3 & 2.43E+22 &  0.08 &     2 &     3 & pacs \\ 
\hline
IRDC321.73 & & & & & & & & \\
\hline
 1 & 15:18:00.7 & -57:21:14 &     1.8 &   0.0 & 1.38E+22 &  0.06 &     3 &     9 &   \\ 
 2 & 15:18:01.4 & -57:22:05 &     9.1 &   0.0 & 1.87E+22 &  0.13 &    17 &    34 & pacs \\ 
 3 & 15:18:05.8 & -57:24:34 &     1.0 &   0.0 & 1.24E+22 &  0.05 &     2 &     2 & pacs \\ 
 4 & 15:18:26.4 & -57:22:00 &    15.6 &   0.0 & 2.91E+22 &  0.13 &    29 &    43 & pacs \\ 
 5 & 15:18:29.3 & -57:21:56 &     2.8 &   0.0 & 1.53E+22 &  0.08 &     5 &     9 & pacs \\ 
\hline
IRDC004.36 & & & & & & & & \\
\hline
 1 & 17:55:41.7 & -25:13:56 &     0.8 &   0.0 & 9.23E+21 &  0.07 &     3 &     6 &   \\ 
 2 & 17:55:51.8 & -25:13:15 &     1.8 &   0.0 & 9.84E+21 &  0.11 &     7 &    18 &   \\ 
 3 & 17:55:53.9 & -25:13:05 &     0.3 &   0.8 & 1.45E+22 &  0.06 &     1 &     2 &   \\ 
 4 & 17:55:55.0 & -25:13:10 &     0.7 &   0.8 & 1.42E+22 &  0.10 &     3 &     6 &   \\ 
\hline
IRDC009.86 & & & & & & & & \\
\hline
 1 & 18:07:33.9 & -20:26:09 &     1.3 &   0.0 & 1.15E+22 &  0.07 &     3 &     8 &   \\ 
 2 & 18:07:34.0 & -20:26:21 &     1.4 &   0.0 & 1.30E+22 &  0.07 &     3 &     3 & pacs \\ 
\hline
IRDC011.11 & & & & & & & & \\
\hline
 1 & 18:09:58.9 & -19:27:43 &     2.2 &   0.0 & 2.09E+22 &  0.09 &     9 &    25 &   \\ 
 2 & 18:10:03.8 & -19:23:09 &     2.8 &   0.0 & 1.74E+22 &  0.11 &    11 &    33 &   \\ 
 3 & 18:10:05.0 & -19:23:11 &     2.9 &   0.0 & 1.45E+22 &  0.12 &    12 &    34 &   \\ 
 4 & 18:10:05.6 & -19:26:37 &     5.7 &   0.0 & 1.92E+22 &  0.14 &    23 &    36 & pacs \\ 
 5 & 18:10:06.9 & -19:27:45 &     1.6 &   1.4 & 2.33E+22 &  0.14 &     6 &    16 &   \\ 
 6 & 18:10:07.0 & -19:28:09 &     0.9 &   0.8 & 1.69E+22 &  0.11 &     4 &     9 &   \\ 
 7 & 18:10:07.4 & -19:29:15 &     3.0 &   3.3 & 5.99E+22 &  0.11 &    12 &    10 & pacs \\ 
 8 & 18:10:07.5 & -19:29:01 &     0.5 &   3.3 & 4.80E+22 &  0.08 &     2 &     4 &   \\ 
 9 & 18:10:07.5 & -19:28:02 &     1.3 &   1.2 & 2.51E+22 &  0.10 &     5 &    13 &   \\ 
10 & 18:10:07.8 & -19:27:51 &     0.3 &   1.4 & 2.17E+22 &  0.08 &     1 &     2 &   \\ 
11 & 18:10:08.3 & -19:28:16 &     0.4 &   1.1 & 1.90E+22 &  0.08 &     2 &     3 &   \\ 
12 & 18:10:13.2 & -19:24:22 &     2.8 &   0.0 & 1.55E+22 &  0.11 &    11 &    33 &   \\ 
13 & 18:10:18.1 & -19:24:38 &     2.0 &   0.0 & 1.43E+22 &  0.10 &     8 &    23 &   \\ 
14 & 18:10:19.3 & -19:24:24 &     2.1 &   0.0 & 1.44E+22 &  0.10 &     8 &    17 & pacs \\ 
15 & 18:10:24.6 & -19:23:05 &     1.0 &   0.0 & 1.30E+22 &  0.07 &     4 &     9 &   \\ 
16 & 18:10:25.0 & -19:17:45 &    12.6 &   0.0 & 2.05E+22 &  0.21 &    51 &   210 &   \\ 
17 & 18:10:27.0 & -19:18:11 &     1.5 &   0.0 & 1.54E+22 &  0.08 &     6 &     6 & pacs \\ 
18 & 18:10:28.5 & -19:22:35 &    48.5 &   1.6 & 5.10E+22 &  0.33 &   196 &   183 & pacs \\ 
19 & 18:10:30.2 & -19:14:43 &     4.2 &   0.0 & 1.62E+22 &  0.13 &    17 &    55 &   \\ 
20 & 18:10:31.1 & -19:22:27 &     0.2 &   1.6 & 2.39E+22 &  0.07 &  1 &     1 &   \\ 
21 & 18:10:32.8 & -19:22:28 &     0.7 &   1.3 & 2.23E+22 &  0.10 &     3 &     6 &   \\ 
22 & 18:10:32.8 & -19:22:13 &     0.2 &   1.5 & 2.22E+22 &  0.07 &  1 &     1 &   \\ 
23 & 18:10:33.3 & -19:21:56 &     0.4 &   1.7 & 2.61E+22 &  0.08 &     2 &     3 &   \\ 
24 & 18:10:33.6 & -19:22:07 &     0.3 &   1.7 & 2.58E+22 &  0.08 &     1 &     1 & pacs \\ 
25 & 18:10:33.9 & -19:21:47 &     0.7 &   1.5 & 2.54E+22 &  0.09 &     3 &     6 &   \\ 
26 & 18:10:34.2 & -19:21:36 &     0.2 &   1.4 & 2.15E+22 &  0.07 &  1 &     1 &   \\ 
27 & 18:10:34.4 & -19:20:55 &     6.8 &   0.0 & 1.63E+22 &  0.17 &    27 &   116 & cold \\ 
28 & 18:10:34.4 & -19:17:07 &     1.3 &   0.9 & 1.61E+22 &  0.14 &     5 &     6 & cold \\ 
29 & 18:10:34.8 & -19:16:51 &     0.5 &   1.0 & 1.70E+22 &  0.08 &     2 &     4 &   \\ 
30 & 18:10:34.8 & -19:17:26 &     1.1 &   0.0 & 1.76E+22 &  0.07 &     4 &    10 &   \\ 
31 & 18:10:35.0 & -19:16:39 &     0.4 &   1.0 & 1.82E+22 &  0.07 &     2 &     3 &   \\ 
32 & 18:10:35.5 & -19:17:00 &     1.1 &   0.0 & 1.76E+22 &  0.07 &     4 &    10 &   \\ 
33 & 18:10:38.3 & -19:19:53 &     1.2 &   0.0 & 1.43E+22 &  0.08 &     5 &     3 & pacs \\ 
34 & 18:10:40.1 & -19:11:35 &     1.6 &   0.0 & 1.62E+22 &  0.08 &     6 &    16 &   \\ 
35 & 18:10:40.3 & -19:15:44 &     5.7 &   0.0 & 1.50E+22 &  0.16 &    23 &    78 &   \\ 
\hline
IRDC015.05 & & & & & & & & \\
\hline
 1 & 18:17:32.0 & -15:46:40 &     0.6 &   1.6 & 3.09E+22 &  0.06 &     2 &     4 &   \\ 
 2 & 18:17:32.2 & -15:46:26 &     1.8 &   1.6 & 3.15E+22 &  0.10 &     5 &    14 &   \\ 
 3 & 18:17:38.2 & -15:51:46 &     7.0 &   0.0 & 3.33E+22 &  0.11 &    21 &    77 &   \\ 
 4 & 18:17:39.0 & -15:48:45 &     0.4 &   1.9 & 3.15E+22 &  0.06 &     1 &     2 &   \\ 
 5 & 18:17:39.3 & -15:49:00 &     6.0 &   1.9 & 3.71E+22 &  0.16 &    18 &    63 &   \\ 
 6 & 18:17:40.3 & -15:49:44 &     4.1 &   0.0 & 2.60E+22 &  0.09 &    13 &    41 &   \\ 
 7 & 18:17:51.5 & -15:51:58 &     3.7 &   0.0 & 2.58E+22 &  0.09 &    11 &    35 &   \\ 
 8 & 18:17:52.5 & -15:49:56 &     2.6 &   0.0 & 2.47E+22 &  0.07 &     8 &    23 &   \\ 
\hline
IRDC18223 & & & & & & & & \\
\hline
 1 & 18:25:06.5 & -12:46:11 &     0.2 &   0.7 & 1.16E+22 &  0.08 &  1 &     1 &   \\ 
 2 & 18:25:07.4 & -12:45:56 &     3.4 &   1.1 & 2.32E+22 &  0.17 &    15 &    17 & pacs \\ 
 3 & 18:25:07.4 & -12:47:53 &     4.0 &   0.0 & 1.44E+22 &  0.14 &    17 &    25 & pacs \\ 
 4 & 18:25:08.7 & -12:45:24 &    25.3 &   1.1 & 3.52E+22 &  0.30 &   108 &   143 & pacs \\ 
 5 & 18:25:09.5 & -12:44:52 &     0.2 &   1.0 & 1.64E+22 &  0.07 &  1 &     1 &   \\ 
 6 & 18:25:09.9 & -12:44:14 &    14.3 &   1.0 & 2.42E+22 &  0.31 &    61 &    99 & pacs \\ 
 7 & 18:25:10.8 & -12:42:30 &    64.2 &   2.5 & 7.59E+22 &  0.33 &   274 &   254 & pacs \\ 
 8 & 18:25:11.3 & -12:41:48 &     0.7 &   2.5 & 3.76E+22 &  0.10 &     3 &     2 & pacs \\ 
 9 & 18:25:12.2 & -12:41:08 &     0.7 &   0.0 & 1.12E+22 &  0.07 &     3 &     3 & cold \\ 
10 & 18:25:17.7 & -12:48:27 &     2.2 &   0.0 & 1.46E+22 &  0.11 &     9 &    14 & pacs \\ 
\hline
IRDC019.30 & & & & & & & & \\
\hline
 1 & 18:25:52.4 & -12:04:56 &     4.5 &   0.0 & 2.83E+22 &  0.07 &     9 &    13 & pacs \\ 
 2 & 18:25:58.3 & -12:04:23 &     1.7 &   0.0 & 2.03E+22 &  0.05 &     3 &     4 & pacs \\ 
 3 & 18:25:58.9 & -12:04:03 &    17.9 &   0.0 & 4.31E+22 &  0.12 &    35 &    36 & pacs \\ 
\hline
IRDC028.34 & & & & & & & & \\
\hline
 1 & 18:42:36.1 & -04:02:25 &     0.8 &   1.3 & 2.56E+22 &  0.11 &     6 &    13 &   \\ 
 2 & 18:42:37.5 & -04:02:05 &    85.9 &   1.3 & 6.18E+22 &  0.48 &   611 &   782 & pacs \\ 
 3 & 18:42:40.1 & -04:00:32 &     1.6 &   0.0 & 2.10E+22 &  0.10 &    11 &    29 &   \\ 
 4 & 18:42:45.0 & -04:04:11 &     2.2 &   1.3 & 2.43E+22 &  0.18 &    15 &    22 & cold \\ 
 5 & 18:42:46.7 & -04:04:12 &     3.9 &   1.3 & 2.70E+22 &  0.21 &    27 &    87 &   \\ 
 6 & 18:42:49.3 & -04:02:20 &    22.6 &   0.0 & 3.06E+22 &  0.30 &   161 &   769 &   \\ 
 7 & 18:42:50.4 & -04:02:01 &     3.7 &   0.0 & 2.88E+22 &  0.13 &    26 &    37 & pacs \\ 
 8 & 18:42:50.8 & -04:03:15 &    48.3 &   0.0 & 4.35E+22 &  0.37 &   344 &  1369 & pacs \\ 
 9$^c$ & 18:42:52.1 & -04:00:01 &   387.3 &   3.9 & 1.60E+23 &  0.66 &  2755 &  5784 & pacs \\ 
10 & 18:42:53.1 & -04:02:45 &     0.6 &   2.3 & 3.64E+22 &  0.11 &     4 &     9 &   \\ 
11 & 18:42:54.4 & -04:02:30 &     6.8 &   2.3 & 4.89E+22 &  0.21 &    49 &   110 & pacs \\ 
12 & 18:42:54.4 & -04:00:46 &     0.5 &   1.8 & 3.08E+22 &  0.09 &     3 &     7 &   \\ 
13 & 18:42:54.5 & -04:00:14 &     1.2 &   3.9 & 6.16E+22 &  0.12 &     9 &    21 &   \\ 
14 & 18:42:55.0 & -04:00:57 &     2.1 &   1.8 & 3.31E+22 &  0.17 &    15 &    41 &   \\ 
15 & 18:42:55.0 & -04:02:13 &     0.8 &   1.4 & 2.54E+22 &  0.11 &     6 &     5 & pacs \\ 
16 & 18:42:55.0 & -04:01:56 &     0.3 &   1.5 & 2.43E+22 &  0.09 &     2 &     4 &   \\ 
17 & 18:42:55.5 & -04:01:38 &     6.2 &   0.0 & 2.43E+22 &  0.18 &    44 &   155 &   \\ 
18 & 18:42:55.5 & -04:01:18 &     1.9 &   1.8 & 3.27E+22 &  0.16 &    13 &    19 & cold \\ 
19 & 18:42:55.8 & -04:01:51 &     0.4 &   1.5 & 2.52E+22 &  0.09 &     3 &     5 &   \\ 
20 & 18:42:56.6 & -03:59:56 &     1.6 &   0.0 & 2.31E+22 &  0.09 &    11 &    29 &   \\ 
\hline

\end{longtable}


\tablefoottext{a}{Integrated flux contribution from \dendro~leaf (with parent branch/trunk flux subtracted).}

\tablefoottext{b}{Assuming $T_\mathrm{dust}$ of the best fit (see Table~\ref{tab:coreprops}) or (if no {\em Herschel} counterpart) the upper limit temperature.}

\tablefoottext{c}{Leaf overlaps with multiple {\em Herschel} cores (see Section~\ref{ss:individual} for details).}

}

\end{document}